\documentclass{sig-alternate_mod}

\newtheorem{theorem}{Theorem}
\usepackage{amsmath,url,cite,times}
\usepackage{amsfonts}
\usepackage{graphicx}
\usepackage{subfigure,algorithm}

\usepackage{etoolbox}
\makeatletter
\patchcmd{\maketitle}{\@copyrightspace}{}{}{}
\makeatother

\begin{document}

\title{b-Bit Minwise Hashing in Practice: Large-Scale Batch and Online Learning and Using GPUs for Fast Preprocessing with Simple Hash Functions\vspace{-0.1in}}
\numberofauthors{3}
\author{
\alignauthor
Ping Li\\
\affaddr{Dept. of Statistical Science}\\
       \affaddr{Cornell University}\\
             \affaddr{Ithaca, NY 14853}\\
       \email{pingli@cornell.edu}
\alignauthor
%\and
Anshumali Shrivastava\\
       \affaddr{Dept. of Computer Science}\\
       \affaddr{Cornell University}\\
              \affaddr{Ithaca, NY 14853}\\
       \email{anshu@cs.cornell.edu}
\alignauthor
Arnd Christian K\"{o}nig\\
       \affaddr{Microsoft Research}\\
       \affaddr{Microsoft Corporation}\\
              \affaddr{Redmond, WA 98052}\\
       \email{chrisko@microsoft.com}
}

\maketitle

\begin{abstract}

\vspace{-0.05in}

Minwise %\footnote{This paper is an updated version of arXiv:1108.3072.}
hashing is a standard technique in the context of search for approximating set similarities. The recent work~\cite{Proc:HashLearning_NIPS11} demonstrated  a potential  use of $b$-bit minwise hashing~\cite{Article:Li_Konig_CACM11} for batch learning on large data. However, several critical issues must be tackled before one can apply b-bit minwise hashing
to the volumes of data often used industrial applications, especially in the context of search.

\vspace{0.05in}

\noindent (b-bit) Minwise hashing requires an expensive preprocessing step that computes $k$ (e.g., $500$) minimal values after applying the corresponding permutations for each data vector. Note that the required $k$ is often substantially larger for classification tasks than for duplicate detections (which mainly concern highly similar pairs) . We developed a parallelization scheme using GPUs and observed that the preprocessing time can be reduced by a factor of $20\sim 80$ and becomes substantially smaller than the data loading time. Reducing the preprocessing time is highly beneficial in practice, e.g., for duplicate Web page detection (where minwise hashing is a major step in the crawling pipeline) or for increasing the testing speed of online classifiers.

\vspace{0.05in}

\noindent  One major advantage of $b$-bit minwise hashing is that it can
substantially reduce the amount of memory required for
batch learning. However, as online algorithms become increasingly
popular for large-scale learning in the context of search, it is not clear if $b$-bit minwise
yields significant improvements for them. This paper demonstrates that $b$-bit minwise hashing
provides an effective data size/dimension reduction
scheme and hence it can dramatically  reduce the data loading time for each epoch of the online training
process. This is significant because online learning often requires
many (e.g., 10 to 100) epochs to reach a sufficient accuracy.

\vspace{0.05in}

\noindent Another critical issue is that for very large data sets it becomes impossible to store a (fully) random permutation matrix, due to its space requirements. Our paper is the first study to demonstrate that $b$-bit minwise hashing implemented using simple hash functions, e.g., the 2-universal (2U)  and  4-universal (4U) hash families, can produce very similar learning results as using fully random permutations.  Experiments on datasets of up to 200GB are presented.

\end{abstract}

%\vspace{-0.05in}
\section{Introduction}

Minwise hashing~\cite{Proc:Broder,Proc:Broder_WWW97,Proc:Broder_STOC98} is a standard technique
for efficiently computing  set similarities in the context of search, with further
applications in the context of content matching for online advertising~\cite{Proc:Pandey_WWW09}, detection of redundancy in enterprise file systems~\cite{Article:Forman09}, syntactic similarity algorithms for enterprise information management~\cite{Proc:Cherkasova_KDD09}, Web spam~\cite{Article:Urvoy08}, etc. The recent development of $b$-bit minwise hashing~\cite{Article:Li_Konig_CACM11} provided a substantial improvement in the estimation accuracy and speed by proposing a new estimator that stores only the lowest $b$ bits of each hashed value.
More recently,\cite{Proc:HashLearning_NIPS11} proposed the use of $b$-bit minwise hashing in the context of learning algorithms such as SVM or logistic regression on large binary data (which is typical in Web classification tasks). $b$-bit minwise hashing can enable scalable learning where otherwise massive (and expensive) parallel architectures would have been required, at negligible reduction in learning quality. In \cite{Proc:HashLearning_NIPS11}, experiments showed this for the {\em webspam} dataset which has 16 million features with a total disk size of 24GB in standard LibSVM format.

However,    several crucial issues  must be tackled before one can apply $b$-bit minwise hashing to  industrial applications. To understand these issues, we begin with a review of the method.

\subsection{A Review of $b$-Bit Minwise Hashing}

 Minwise hashing mainly focuses on binary (0/1) data, which can be viewed as sets. Consider  sets $S_1,   S_2 \subseteq \Omega = \{0, 1, 2, ..., D-1\}$, minwise hashing  applies a random permutation $\pi: \Omega\rightarrow\Omega$ on $S_1$ and $S_2$  and uses the following collision probability
 \begin{align}
\mathbf{Pr}\left(\text{min}({\pi}(S_1)) = \text{min}({\pi}(S_2)) \right) = \frac{|S_1
  \cap S_2|}{|S_1 \cup S_2|}=R
\end{align}
to estimate  $R$, which is the resemblance between $S_1$ and $S_2$. With $k$ permutations: $\pi_1$, ..., $\pi_k$, one can estimate $R$ without bias:
 \begin{align}\label{eqn_RM}
&\hat{R}_{M} = \frac{1}{k}\sum_{j=1}^{k}1\{z_1= z_2\}\\\notag
&z_1 = {\min}({\pi_j}(S_1)), \ \ z_2 = {\min}({\pi_j}(S_2)).
\end{align}
A common practice is to store each hashed value, e.g., ${\min}({\pi}(S_1))$, using 64 bits~\cite{Proc:Fetterly_WWW03}. The storage (and computational) cost  is prohibitive in  industrial applications~\cite{Proc:Manku_WWW07}. The recent work of $b$-bit minwise hashing~\cite{Article:Li_Konig_CACM11} provides a  simple solution  by storing only the lowest $b$ bits of each hashed value. For convenience, we define
%\begin{align}\notag
%e_{1,i} = i\text{th lowest bit of }z_1, \hspace{0.5in} e_{2,i} = i\text{th lowest bit of }z_2.
%\end{align}
\begin{align}\notag
z_1^{(b)} = \text{the lowest $b$ bits of }z_1, \hspace{0.3in} z_2^{(b)} = \text{the lowest $b$-bits of }z_2.
\end{align}
%For example, if $z_1 = 7$ (111 in binary), then $z_1^{(1)} = 1$ and $z_1^{(2)}=3$.

\begin{theorem}\cite{Article:Li_Konig_CACM11}\label{The_basic}
Assume $D$ is large.
\begin{align}\label{eqn_basic}
&P_b=\mathbf{Pr}\left(z_1^{(b)} = z_2^{(b)}\right) = C_{1,b} + \left(1-C_{2,b}\right) R,%\\\notag
%&r_1 = \frac{f_1}{D}, \hspace{0.1in} r_2 = \frac{f_2}{D}, \ \ f_1 = |S_1|,\  \ f_2 =|S_2|,\\\notag
%&C_{1,b} = A_{1,b} \frac{r_2}{r_1+r_2} + A_{2,b}\frac{r_1}{r_1+r_2},\\\notag
%&C_{2,b} = A_{1,b} \frac{r_1}{r_1+r_2} + A_{2,b}\frac{r_2}{r_1+r_2},\\\notag
%&A_{1,b} = \frac{r_1\left[1-r_1\right]^{2^b-1}}{1-\left[1-r_1\right]^{2^b}},\hspace{0.5in}
%A_{2,b} = \frac{r_2\left[1-r_2\right]^{2^b-1}}{1-\left[1-r_2\right]^{2^b}}.\Box
\end{align}
where $C_{1,b}$ and $C_{2,b}$ are functions of ($D, |S_1|, |S_2|, |S_1\cap S_2|$). $\Box$
\end{theorem}

\vspace{0.05in}

Based on Theorem~\ref{The_basic},  we can  estimate $P_b$ (and $R$) from $k$ independent permutations $\pi_1, \pi_2, ..., \pi_k$:
\begin{align}\label{eqn_R_b}
&\hat{R}_b = \frac{\hat{P}_b - C_{1,b}}{1-C_{2,b}},\hspace{0.3in} \hat{P}_{b} = \frac{1}{k}\sum_{j=1}^{k}1\left\{z_{1,\pi_j}^{(b)} = z_{2,\pi_j}^{(b)}\right\},%\\%\notag
%\end{align}\begin{align}
%&\text{Var}\left(\hat{R}_b\right)% = \frac{\text{Var}\left(\hat{P}_b\right)}{\left[1-C_{2,b}\right]^2}\\\label{eqn_Var_b}
%=\frac{1}{k}\frac{\left[C_{1,b}+(1-C_{2,b})R\right]\left[1-C_{1,b}-(1-C_{2,b})R\right]}{\left[1-C_{2,b}\right]^2}
\end{align}
The estimator $\hat{P}_b$ is an inner product between two vectors in $2^b\times k$ dimensions with exactly $k$ 1's, because
\begin{align}
1\left\{z_1^{(b)} = z_2^{(b)}\right\} = \sum_{t=0}^{2^b-1}1\{z_1^{(b)} = t\}\times 1\{z_2^{(b)} = t\}
\end{align}
This  provides a  practical strategy for using $b$-bit minwise hashing for large-scale learning. That is, each original data vector is transformed into a new data point consisting of $k$ $b$-bit integers, which is expanded into a $2^b\times k$-length binary vector at the run-time.

These days, many machine learning applications, especially in the context of search, are faced with  large and inherently high-dimensional datasets. For example, \cite{GoogleBlog} discusses training datasets with (on average) $n=10^{11}$ items and $D=10^9$ distinct features.  \cite{Proc:Weinberger_ICML2009} experimented with a dataset of potentially $D=16$ trillion ($1.6\times10^{13}$) unique features. Effective algorithms for  data/feature reductions will be highly beneficial for these  industry applications.

\subsection{Linear Learning Algorithms}

Clearly, $b$-bit minwise hashing can  approximate both linear and nonlinear kernels (if they are functions of the inner products). We focus on linear learning because  many high-dimensional  datasets used in the context of search are naturally  suitable for linear  algorithms. Realistically, for industrial applications, ``{\em almost all the big impact algorithms operate in pseudo-linear or better time}''~\cite{Url:Lanford_Learning0511}.

Linear algorithms such as linear SVM and logistic regression have become very powerful and extremely popular. Representative software packages include SVM$^\text{perf}$~\cite{Proc:Joachims_KDD06}, Pegasos~\cite{Proc:Shalev-Shwartz_ICML07}, Bottou's SGD SVM~\cite{URL:Bottou_SGD}, and LIBLINEAR~\cite{Article:Fan_JMLR08}.

Given a dataset $\{(\mathbf{x}_i, y_i)\}_{i=1}^n$, $\mathbf{x}_i\in\mathbb{R}^{D}$, $y_i\in\{-1,1\}$, the $L_2$-regularized linear SVM solves the following optimization  problem:
\begin{align}\label{eqn_SVM}
\min_{\mathbf{w}}\ \ \frac{1}{2}\mathbf{w^Tw} + C \sum_{i=1}^n \max \left\{1 - y_i\mathbf{w^Tx_i},\ 0\right\},
\end{align}
and the $L_2$-regularized logistic regression solves a  similar problem:
\begin{align}\label{eqn_logit}
\min_{\mathbf{w}}\ \ \frac{1}{2}\mathbf{w^Tw} + C \sum_{i=1}^n \log \left(1 + e^{-y_i\mathbf{w^Tx_i}}\right).
\end{align}
Here $C>0$ is an important penalty parameter.

\vspace{0.05in}

Next, we elaborate on 3 major issues one must tackle in order to apply $b$-bit minwise hashing to large-scale industrial applications.

\subsection{Issue 1: Expensive Preprocessing}

($b$-bit) Minwise hashing requires a very expensive  preprocessing step (substantially more costly than loading the data) in order to compute $k$ (e.g., $k=500$) minimal values (after permutation) for each data vector. Note that in prior studies for duplicate detection~\cite{Proc:Broder}, $k$ was usually not too large (i.e., 200), mainly because duplicate detection concerns highly similar pairs (e.g., $R>0.5$). With $b$-bit minwise hashing, we have to use larger $k$ values according to the analysis in~\cite{Article:Li_Konig_CACM11} even in the context of duplicate detections. Note that classification tasks are quite different from duplicate detections. For example, in our most recent experiments on image classifications~\cite{Report:Li_EM}, even $k=2000$ did not appear to be sufficient.

Consider, for example, the task of computing $b$-bit minwise hash signatures for the task of Web page duplicate detection. While parallelizing this task  is conceptually simple (as the signatures of different pages can be computed independently) it still comes at the cost of using additional hardware and electricity. Thus, any improvements in the speed of signature computation may be directly reflected in the cost of the required infrastructure.

For machine learning research and applications, this expensive preprocessing step can be a significant issue in scenarios where
(either due to changing data distributions or features) models are frequently re-trained. In  user-facing applications, the testing time performance can be severely affected by the preprocessing step if the (new) incoming data have not been previously processed.

%\vspace{0.05in}

This paper studies how to speed up the execution of the signature computation through the use of graphical processing units (GPUs). GPUs offer, compared to current  CPUs, higher instruction parallelism and very low latency access to the internal GPU memory, but comparatively slow latencies when accessing the main memory~\cite{Kim:2010:FFA:1807167.1807206}. As a result, many data processing algorithms (especially such with random memory access patterns) do not benefit significantly when implemented using a GPU. However, the characteristics of the minwise hashing algorithm make it very well suited for execution using a GPU.  The algorithm  accesses each set in its entirety, which allows for the use of main memory pre-fetching to reduce access latencies. Moreover, since we compute $k$ different hash minima for each item in a set, the algorithm can make good use of the high degree of parallelism in GPUs. This is especially true for $b$-bit minwise hashing, which, compared to the original algorithm, typically increases the number of hash functions (and minima) to be computed by a factor of 3 or more. Also, note that any improvements to the speed of ($b$-bit) minwise hashing are directly applicable to large-scale instances of the other applications of minwise hashing mentioned previously.

\subsection{Issue 2: Online Learning}

Online learning has become increasingly
popular in the context of web search and advertising~\cite{Ciaramita:2008:OLC:1367497.1367529,Bilenko:2005:APN:1106326.1106331,
L'Huillier:2010:OPC:1809400.1809421,Sculley:2007:ROS:1277741.1277813}
as it only requires loading one feature vector at
a time and thus avoids the overhead of storing a potentially very large dataset in memory or the complexity and cost of parallel learning architectures. In this paper, we demonstrate that $b$-bit minwise hashing can also be highly beneficial for online learning   because reducing the data size substantially decreases the  loading time, which usually dominates the training cost, especially when many training epochs are needed.  At the same time, the resulting reduction in the overall learning accuracy that we see in the experiments is negligible. Moreover, $b$-bit minwise hashing can also serve as an effective dimensionality reduction scheme here, which can be important in the context of web search, as (i) machine learning models based on text $n$-gram features generally have very large numbers of features, and thus require significant storage and (ii) there are
typically a number of different models deployed in user-facing search servers, all of which compete for the available main memory space.

There is another  benefit.  Machine learning researchers have been actively developing good (and sometimes complex) online algorithms to minimize the need for loading the data many times (e.g.,~\cite{Report:Wu_arXiv11}). If the data loading time is no longer a major bottleneck, then perhaps very simple online algorithms may be sufficient in practice.

\subsection{Issue 3: Massive Permutation Matrix}

When the data dimension ($D$) is not too large, e.g.,  millions, the implementation of $b$-bit minwise hashing for learning is  straightforward. Basically, we can assume a ``fully random permutation matrix'' of size $D\times k$, which defines  $k$ permutation mappings. This is actually how researchers use (e.g., Matlab) simulations to verify the theoretical results assuming perfectly random permutations.

Unfortunately, when the dimension is on the order of billions (let alone $2^{64}$), it becomes impractical (or too expensive) to store such a permutation matrix. Thus, we have to resort to simple hash functions  such as various forms of 2-universal (2U) hashing (e.g.,~\cite{Dietzfelbinger:1996:UHK:646511.695324}). Now the question is how reliable those hash functions are in the context of learning with $b$-bit minwise hashing.

There were prior studies on the impact of limited randomness on the estimation accuracy of (64-bit) minwise hashing, e.g.,~\cite{Indyk:2001:SAM:370968.370980, Patrascu:2010:KRL:1880918.1880996}. However, no prior studies reported how the learning accuracies were affected by the use of simple hash functions for $b$-bit minwise hashing. This study provides the empirical support that, as long as the data are reasonably sparse (as virtually always the case in the context of search), using 2U/4U hash functions results in negligible reduction of learning accuracies (unless $b=1$ and $k$ is very small).

One  limitation of GPUs is that they have fairly limited memory~\cite{Url:NVIDIA_CUDA}. Thus, it becomes even more beneficial if we can reliably replace a massive permutation matrix with simple hash functions.

\section{Simple Hash Functions}\label{sec_hash}

As previously discussed, in large-scale industry practice, it is often infeasible to assume perfect random permutations. For example, when $D=2^{30}$ (about 1 billion) and $k=500$, a matrix of $D\times k$ integers (4-byte each) would require $>2000$GB of storage.

To overcome the difficulty in achieving perfect permutations, the common practice is to use the so-called {\em universal hashing}~\cite{Proc:Carter_STOC77}. One standard 2-universal (2U) hash function is, for $j = 1$ to $k$,
\begin{align}\label{eqn_h_2U}
h_j^{(2U)}(t) = \left\{a_{1,j} + a_{2,j}\ t\ \text{  mod  }\ p\right\}\ \text{ mod } \ D,
\end{align}
where $p>D$ is a prime number and $a_{1,j}$, $a_{2,j}$ are chosen uniformly from $\{0, 1, ..., p-1\}$. To increase  randomness, one can also use
 the following 4-universal (4U) hash function:
\begin{align}\label{eqn_h_4U}
h_j^{(4U)}(t) = \left\{\sum_{i=1}^4 a_{i,j} t^{i-1} \ \text{  mod  }\ p\right\}\ \text{ mod } \ D,
\end{align}
where  the $a_{i,j}$ ($i=1,2,3,4$) are chosen uniformly from $\{0, 1, ..., p-1\}$. The storage cost for retaining the $a_{i,j}$'s is minimal, compared to storing a permutation matrix. In theory, the 4U hash function is (in the worst-case) more random than the 2U hash function.

Now, to compute the minwise hashes for a given feature vector (e.g., a parsed document represented as a list of 1-grams, 2-grams, and 3-grams, where each $n$-gram can be viewed as a binary feature), we iterate over all non-zero features;
any non-zero location $t$ in the original feature vector is mapped to its new location $h_j(t)$; we then iterate over all mapped locations to find their minimum, which will be the $j$th hashed value for that feature vector.

\section{GPU for Fast Preprocessing}\label{sec_GPU}

In this section we will describe and analyze the use of graphics processors (GPUs) for fast computation of minwise hashes. We will first sketch the relevant properties of GPUs in general and then describe in how far
minwise hashing computation is suited for execution on this architecture. Subsequently, we will describe our implementation and analyze the resulting performance improvements over a CPU-based implementation.

\subsection{Introduction}

The use of GPUs as general-purpose coprocessors is relatively recent and primarily due to their high computational power at comparatively low cost. In comparison with commodity CPUs, GPUs offer significantly increased computation speed and memory bandwidth. However, since GPUs have been designed for graphics processing, the programming model (which includes massively parallel Single-Instruction-Multiple-Data (SIMD) processing and limited bus speeds for data transfers to/from main memory) is not suitable for arbitrary data processing applications~\cite{He:2009:RQC:1620585.1620588}. GPUs consist of a number of SIMD multiprocessors. At each clock cycle, all processors in a multiprocessor execute identical instructions, but on different parts of the data. Thus, GPUs can leverage spatial locality in data access and group accesses to consecutive memory addresses into a single access; this is referred to as {\em coalesced~access}.

%GPUs are known to be energy-efficient~\cite{Proc:Huang_IPDPS09}.

\subsection{Our Approach}

In light of the properties of GPU processing, our GPU algorithm to compute $b$-bit minwise hashes proceeds in \textbf{3 distinct phases}:
First, we read in chunks of  10K sets from disk into main memory and write these to the GPU memory. Then, we compute the hash values and the corresponding minima by applying all $k$ hash functions to the data currently in the GPU and retaining,  for each hash function and  set, the corresponding minima. Finally, we write out the resulting  minima back to main memory and repeat the process.

This batch-style computation has a number of advantages.  Because we transfer larger blocks of data, the main memory latency is reduced through the use of main memory pre-fetching. Moreover, because the computation within the GPU itself scans through consecutive blocks of data in the GPU-internal memory (as opposed to random memory access patterns), performing the same computation (with a different hash function) for each set entry $k$ times, we can take full advantage of coalesced access and the massive parallelism inherent in the GPU architecture.

Because GPUs are known to have fairly limited memory capacity, it becomes even more impractical to store a fully random permutation matrix; and hence it is crucial to utilize simple hash functions. We implemented both 2U and 4U hash functions introduced in Section~\ref{sec_hash}. However, because the modulo operations in the definitions of the 2U/4U hash functions are expensive especially for GPUs~\cite{Url:NVIDIA_CUDA}, we have used the following tricks to avoid them and make our approach (more) suitable for GPU-based execution.

\subsection{Avoid Modulo Operations in 2U Hashing}

To avoid the modulo operations in 2U hashing, we adopt a common trick~\cite{Article:Dietzfelbinger97}. Here, for simplicity, we assume $D = 2^s<2^{32}$ (note that $D=2^{30}$ corresponds to about a billion features). It is known that the following hash function is  essentially 2U~\cite{Article:Dietzfelbinger97}:
\begin{align}\label{eqn_h_s}
h_j^{(s)}(t) = \left\{a_{1,j} + a_{2,j}\ t\ \text{  mod  }\ 2^{32}\right\}\ \text{ mod } \ 2^s,
\end{align}
where $a_{1,j}$ is chosen uniformly from $\{0,1,...,2^{32}-1\}$ and $a_{2,j}$ uniformly from $\{1, 3, ..., 2^{32}-1\}$ (i.e., $a_{2,j}$ is odd).  This scheme is much faster because we can effectively leverage the integer overflow mechanism and the efficient bit-shift operation. In this paper, we  always implement 2U hash using $h_j^{(s)}$.

\subsection{Avoid Modulo Operations in 4U Hashing}\label{sec_h_4U-bit}

It is slightly  tricky to avoid the modulo operations in evaluating 4U hash functions. Assuming $D<p=2^{31}-1$ (a prime number), we provide the $C\#$ code to compute $v\ mod\ p$ with $p=2^{31}-1$:
{\small\begin{verbatim}
private static ulong BitMod(ulong v)
{
    ulong p = 2147483647; // p = 2^31-1
    v = (v >> 31) + (v & p);
    if (v >= 2 * p)
       v = (v >> 31) + (v & p);
    if (v >= p)
       return v - p;
    else
       return v;
}
\end{verbatim}}
To better understand the code, consider
\begin{align}\notag
&v \ mod \ p = x, \hspace{0.25in}\text{and  }\hspace{0.25in}\    v \ mod \ 2^{31} = y\\\notag
\Longrightarrow& v = p\times Z + x  = 2^{31} \times S + y \\\notag
\Longrightarrow& x = 2^{31}(S - Z) + Z + y
\end{align}
for two integers $S$ and $Z$. $S$ and $y$ can be efficiently evaluated using bit operations: $S = v>>31$ and $y = v\ \&\ p$.

A recent paper~\cite{Proc:Thorup_ALENEX10} implemented a similar trick for $p=2^{61}-1$, which was simpler than ours because with $p=2^{61}-1$ there is no need to check the condition ``if (v >= 2 * p)''. We find the case of $p=2^{31}-1$  useful in machine learning practice because it suffices for  datasets with less than a billion features. Note that a large value of $p$ potentially increases the dimensionality  of the hashed data.

\subsection{Experiments: Datasets}

Table~\ref{tab_data} summarizes the two datasets used in this evaluation: {\em webspam} and {\em rcv1}.  The {\em webspam} dataset was used in the recent paper~\cite{Proc:HashLearning_NIPS11}. Since the {\em webspam} dataset (24 GB in LibSVM format) may be too small compared to
datasets used in industrial practice, in this paper we also present an empirical study on the {\em expanded rcv1} dataset~\cite{URL:Bottou_SGD}, which we generated by using the original features + all pairwise combinations (products) of features + 1/30 of 3-way combinations (products) of features. Note that, for {\em rcv1}, we did not include the original test set in~\cite{URL:Bottou_SGD}, which has only 20242 examples. To ensure reliable test results, we randomly split our  expanded {\em rcv1} dataset into two halves, for training and testing.\vspace{-0.05in}

\begin{table}[h]
\caption{Data information\vspace{-0.15in}}
\begin{center}{\scriptsize
\begin{tabular}{l l l l l}
\hline \hline
Dataset &  $n$ &$D$ & \# Avg Nonzeros  &Train / Test \\\hline
Webspam (24 GB) &350000  &16609143 & 3728 &$80\%$ / $20\%$ \\
Rcv1 (200 GB) &781265  &1010017424  &12062 &$50\%$ / $50\%$\\
\hline\hline
\end{tabular}
}
\end{center}
\label{tab_data}\vspace{-0.2in}
\end{table}

\subsection{Experiments: Platform}

 The GPU platform we use in our experiments is the NVIDIA Tesla C2050, which has 15 Simultaneous Multiprocessors (SMs), each with 2 groups of 16
scalar processors (hence 2 sets of 16-element wide SIMD units). The peak (single precision) GFlops of this GPU are 1030, with a peak memory bandwidth of 144 GB/s. In comparison, the numbers for a
Intel Xeon processor X5670 (Westmere) processor are 278 GFlops and 60 GB/s.

\subsection{Experiments: CPU Results}

We use the setting of $k=500$ for these experiments. Table~\ref{tab_CPU} shows the overhead of the CPU-based implementation, broken down into the time required to load the data into memory  and the
time for the minwise hashing computation.  For 2U, we always use the 2U hash function (\ref{eqn_h_s}). For 4U (Mod), we use the 4U hash function (\ref{eqn_h_4U}) which requires  the modulo operation. For 4U (Bit), we use the implementation in Section~\ref{sec_h_4U-bit}, which converted the modulo operation into bit operations.  Note that for {\em rcv1} dataset, we only report the experimental results for 2U hashing. %\vspace{-0.05in}

 \begin{table}[h!]
\caption{The data loading and preprocessing (for $k=500$ permutations) times (in seconds). Note that we measured the data loading times of LIBLINEAR which used a plain text data format. The data loading times could be reduced by a factor of 5 or so when the data were converted into binary. In other words, the (relative) preprocessing costs of minwise hashing would be even much more expensive if we optimized the data loading procedure of LIBLINEAR. This further explains why reducing the cost by using GPUs could be so beneficial. \vspace{-0.15in}}
\begin{center}{\scriptsize
\begin{tabular}{l l l l l l l}
\hline \hline
Dataset & Loading &Permu & 2U & 4U (Mod) & 4U (Bit) \\\hline
Webspam &$9.7\times10^2$&$6.1\times10^3$  &$4.1\times 10^3$ &$4.4\times 10^{4}$ &$1.4\times10^4$ \\
Rcv1  &$1.0\times10^4$ &--  &$3.0\times10^4$  &-- &--\\
\hline\hline
\end{tabular}
}
\end{center}
\label{tab_CPU}\vspace{-0.1in}
\end{table}

Table~\ref{tab_CPU} shows that the preprocessing using CPUs (even for 2U) can be very expensive, substantially more  than data-loading. 4U hashing with modulo operations can take an order of magnitude more time than 2U hashing. As expected, the cost for 4U hashing can be substantially reduced if modulo operations are avoided.

Note that for {\em webspam} dataset (with only 16 million features), using permutations is actually slightly faster than the algebra required for 2U hash functions. The main constraint here is the storage space. The permutations are generated once and then stored in main memory. This makes them impractical for use with larger feature sets such as the {\em rcv1} data (with about a billion features)

\subsection{Experiments: GPU results}

The total overhead for the GPU-based processing for batch size = 10K is summarized in Table~\ref{tab_GPU}, demonstrating the substantial time reduction compared to the CPU-based processing in Table~\ref{tab_CPU}. For example, the cost of 2U processing on the {\em webspam} dataset is reduced from 4100 seconds to 51 seconds, a 80-fold reduction.  We also observe improvements of similar magnitude for 4U processing (both modulo and bit versions) on {\em webspam}. For the {\em rcv1} dataset, the time reduction of the GPU-based implementation is about 20-fold, compared to the CPU-based implementation.

For both datasets, the costs for the GPU-based preprocessing become substantially smaller than the data loading time. Thus, while achieving further reduction of the preprocessing cost is still interesting, it becomes less practically significant because we have to load the data once in the learning process. \vspace{-0.1in}

\begin{table}[h!]
\caption{The data loading and preprocessing (for $k=500$ permutations) times (in seconds) for using GPUs. \vspace{-0.05in} }
\begin{center}{\scriptsize
\begin{tabular}{l l  l l l l}
\hline \hline
Dataset & Loading & GPU 2U & GPU 4U (Mod) & GPU 4U (Bit) \\\hline
Webspam &$9.7\times10^2$  &$51$ &$5.2\times 10^{2}$ &$1.2\times10^2$ \\
Rcv1  &$1.0\times10^4$  &$1.4\times10^3$  &$1.5\times10^4$ &$3.2\times10^3$ \\
\hline\hline
\end{tabular}
}
\end{center}
\label{tab_GPU}\vspace{-0.1in}
\end{table}

\begin{figure}[h!]
\begin{center}
\mbox{
\includegraphics[width=1.7in]{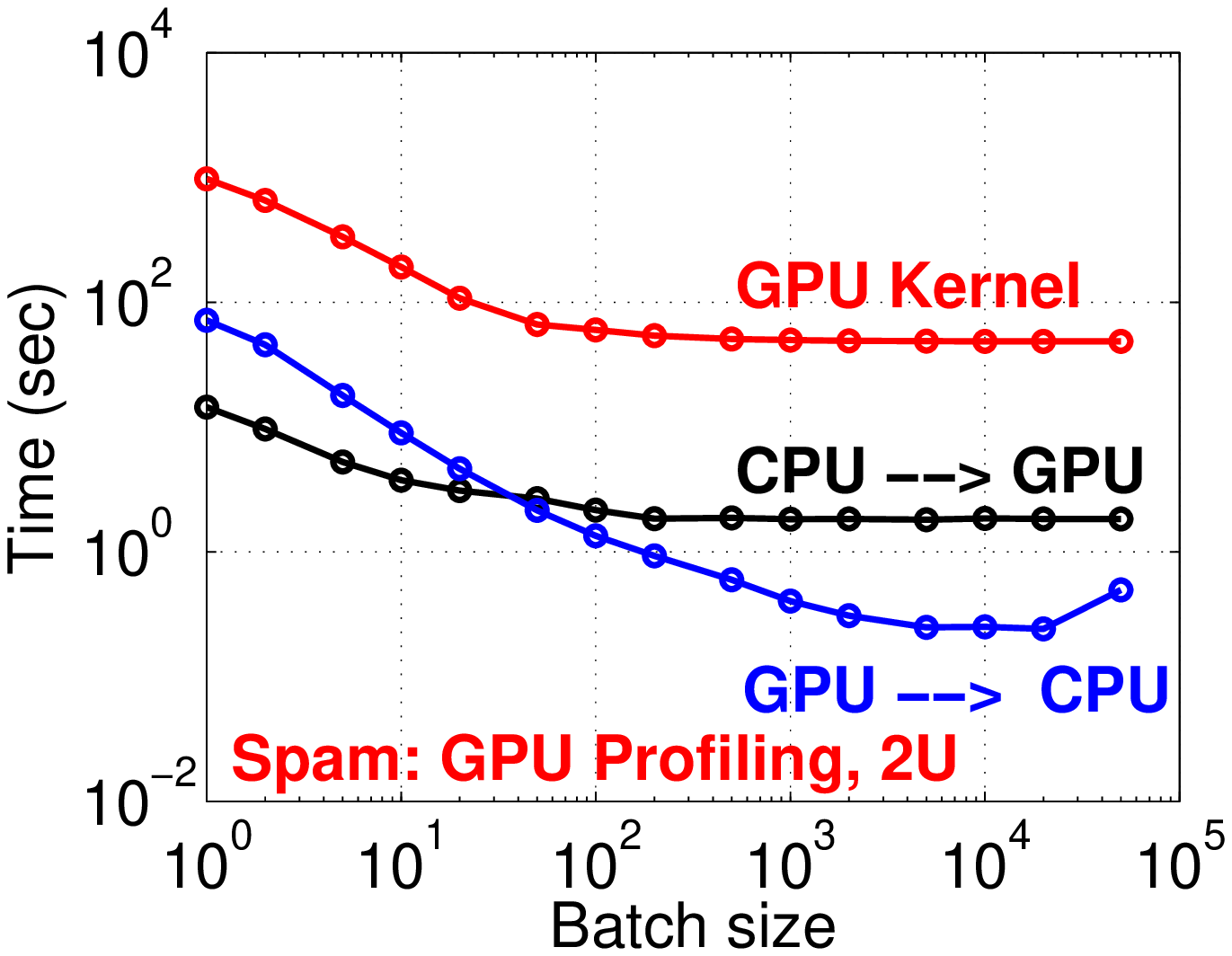}\hspace{-0.1in}
\includegraphics[width=1.7in]{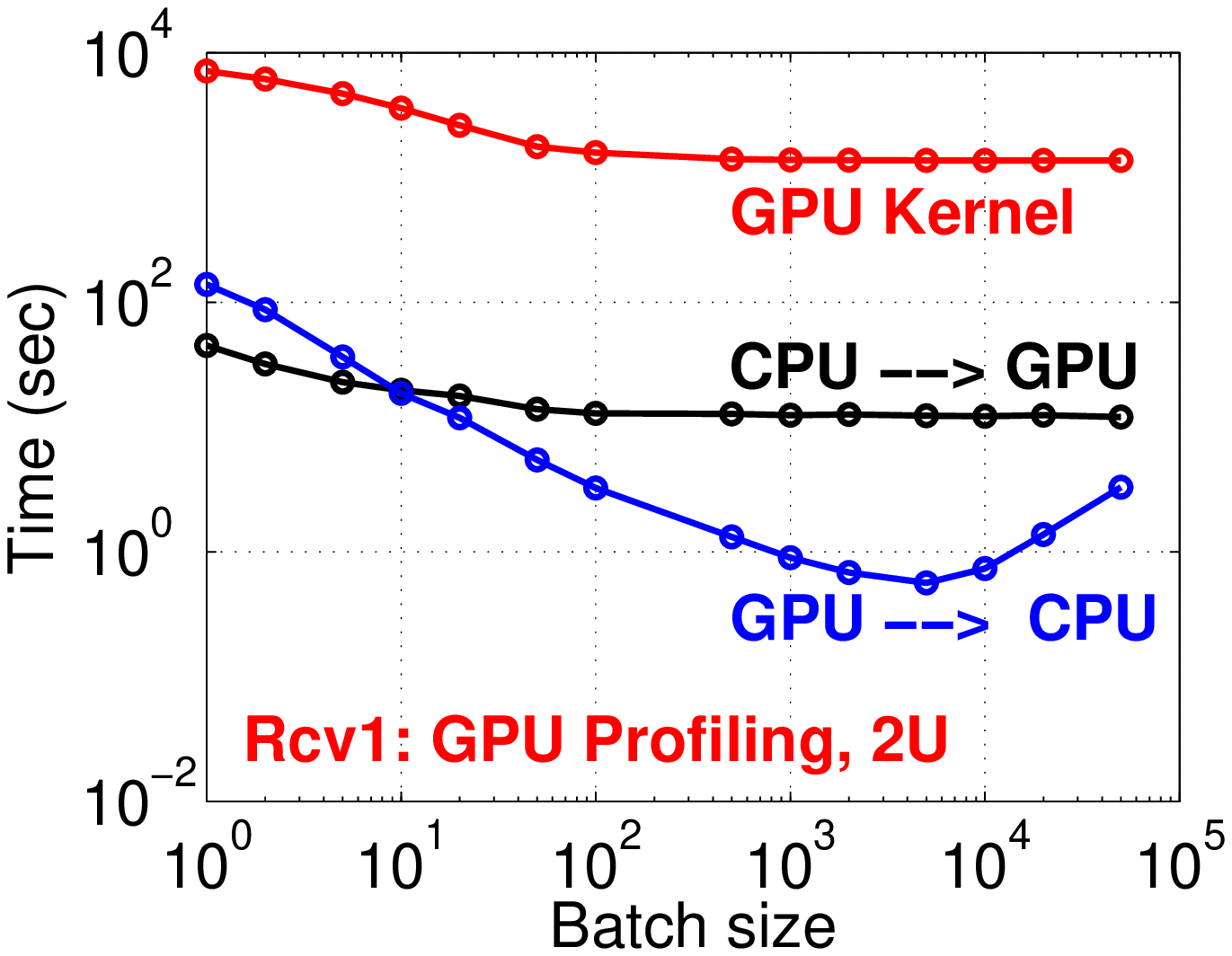}}
\end{center}
\vspace{-0.25in}
\caption{2U. Overhead of the three phases of the GPU-based implementation using 2U hash functions, for both {\em webspam} (left panel) and {\em rcv1} (right panel) datasets.}\label{fig_GPU_2U}\vspace{-0.1in}
\end{figure}

\begin{figure}[h!]
\begin{center}
\mbox{
\includegraphics[width=1.7in]{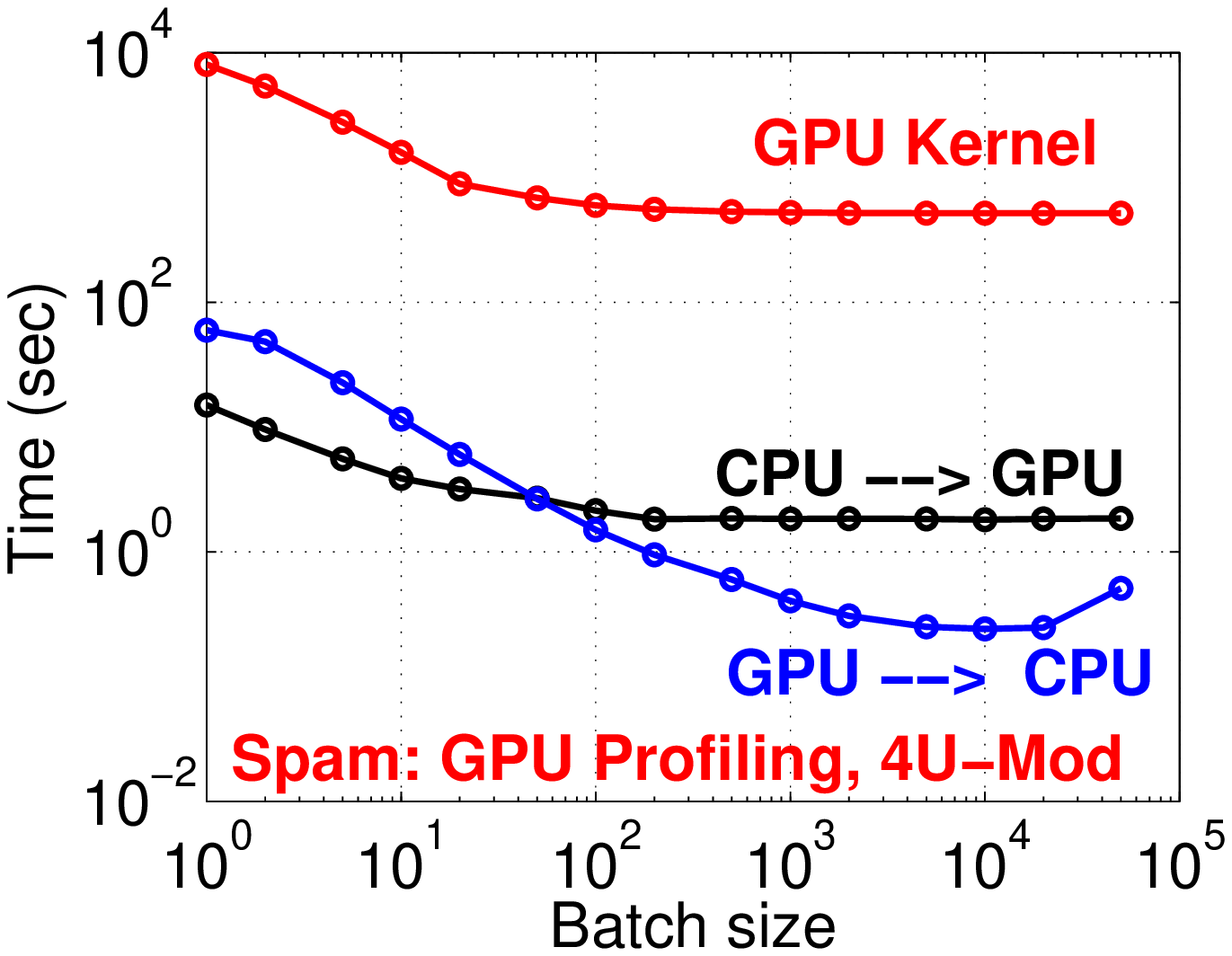}\hspace{-0.1in}
\includegraphics[width=1.7in]{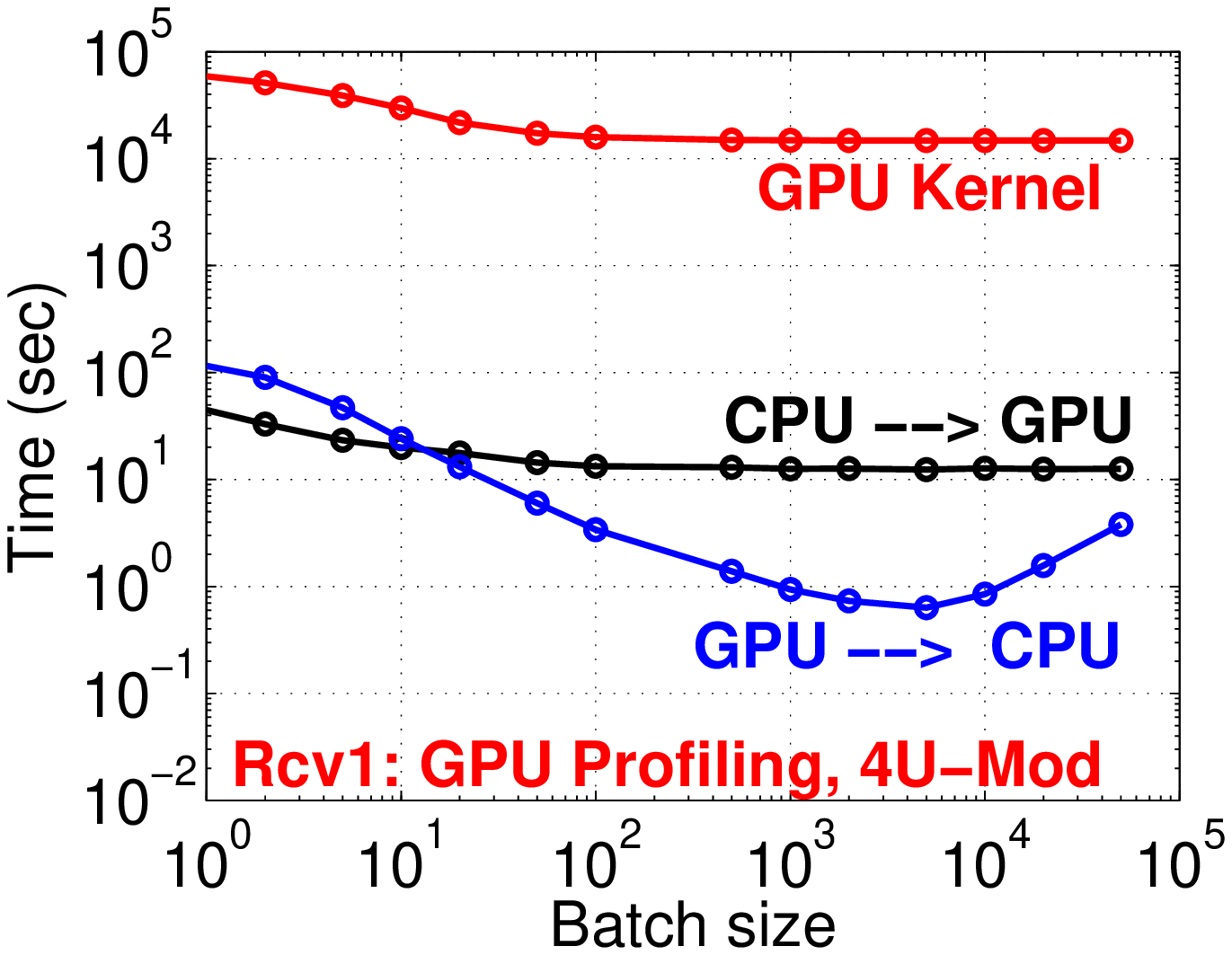}}
\end{center}
\vspace{-0.25in}
\caption{4U-Mod. Overhead of the three phases of the GPU-based implementation using 4U hash functions with modulo operations, for both datasets. }\label{fig_GPU_4U_Mod}\vspace{-0.1in}
\end{figure}

\begin{figure}[h!]
\begin{center}
\mbox{
\includegraphics[width=1.7in]{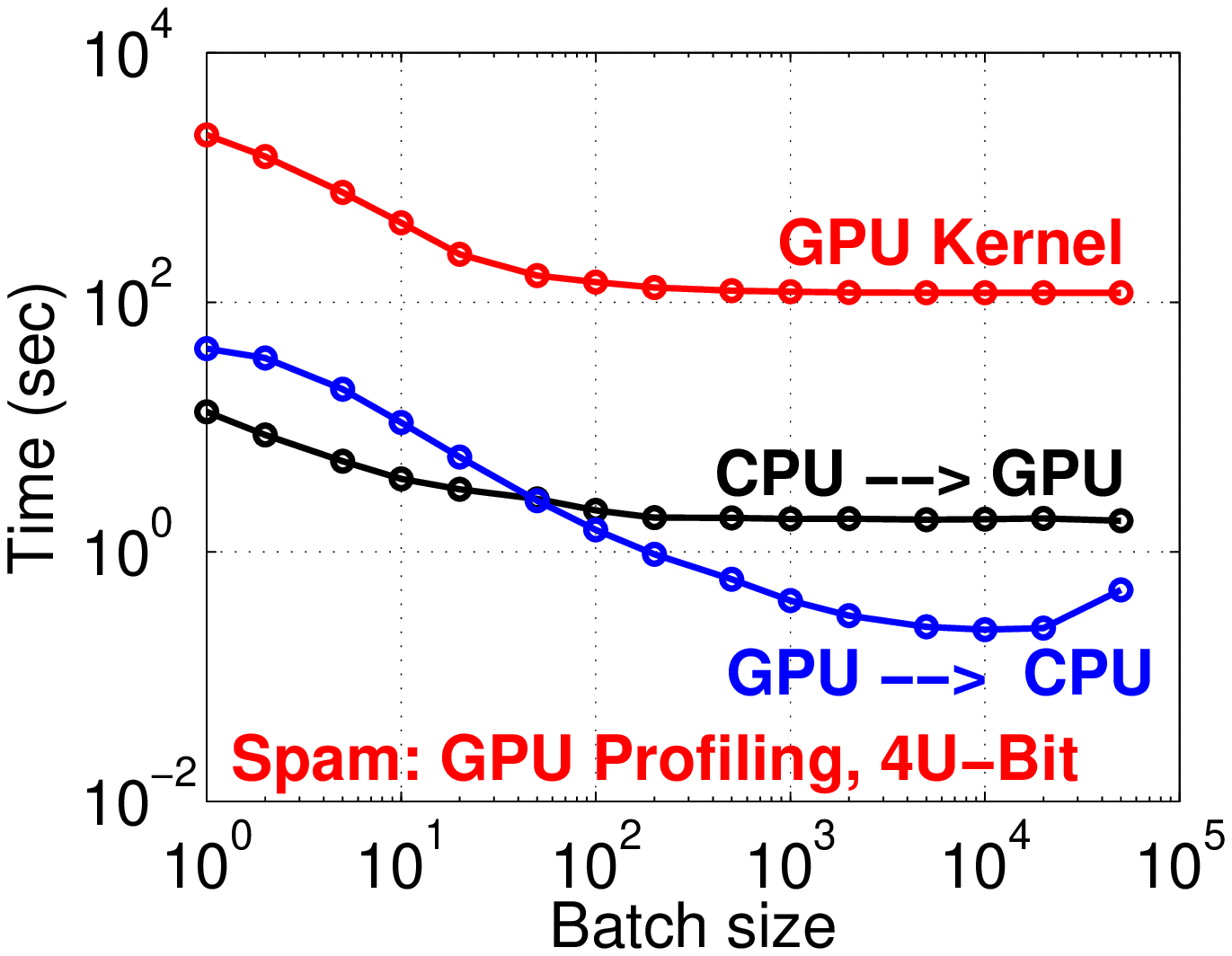}\hspace{-0.1in}
\includegraphics[width=1.7in]{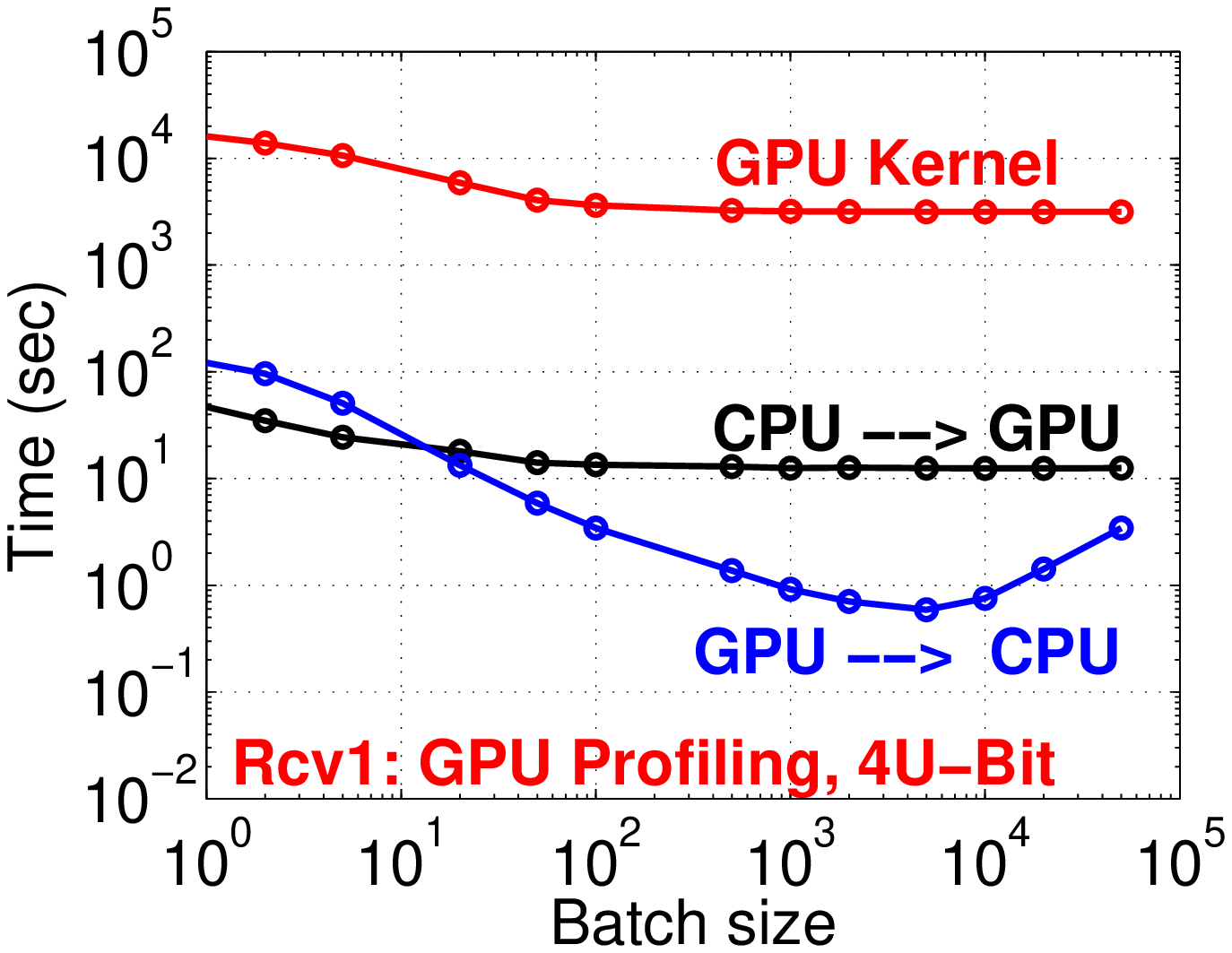}}
\end{center}
\vspace{-0.25in}
\caption{4U-Bit. Overhead of the three phases of the GPU-based implementation using 4U hash functions without modulo operations (Section~\ref{sec_h_4U-bit}), for both datasets. }\label{fig_GPU_4U_Bit}
\end{figure}

Figures~\ref{fig_GPU_2U} to ~\ref{fig_GPU_4U_Bit} provide the breakdowns of the overhead for the GPU-based implementations, using 2U hashing, 4U hashing with modulo operations, and 4U hashing without modulo operations, respectively. As shown in the figures, we separate the overhead into three components:  (i) time spent transferring the data from main memory to the GPU (``\textbf{CPU} $\rightarrow$ \textbf{GPU}''), (ii) the actual computation (``\textbf{GPU Kernel}'') and (iii) transferring the $k$ minima back to main memory (``\textbf{GPU} $\rightarrow$ \textbf{CPU}'').

Recall we adopt a batch-based GPU implementation by reading chunks of sets from disk into main memory and write them to the GPU memory. If the chunk size is not chosen appropriately, it may affect the GPU performance. Thus, we vary this parameter and report the performance for chuck sizes ranging from 1 to 50000.

We can see from Figures~\ref{fig_GPU_2U} to ~\ref{fig_GPU_4U_Bit} that the overall cost is not significantly affected by the batch size, as long as it is reasonably large (e.g., in this case $>100$). This nice property may  simplify the design because practitioners will not have  to  try out many batch sizes. Note that the time spent in transferring the data to the GPU is not affected significantly by the batch size, but the speed at which data is transferred back does vary significantly with this parameter. However, for any setting of the batch size does it hold that the time spent transferring data is about two orders of magnitude smaller than the time spent on the actual processing within the GPU. This is the key to the large speed-up over CPU implementations we see.

\section{Validation of the Use of 2U/4U Hash Functions for Learning}

For large-scale industrial applications, because storing a fully random permutation matrix is not practical, we have to resort to simple hash functions such as 2U or 4U hash families. However, before we can recommend them to practitioners, we must first validate on a smaller dataset that using such hash functions will not hurt the learning performance. To the best of our knowledge, this section is the first empirical study of the impact of hashing functions on machine learning with $b$-bit minwise hashing.

In addition, Appendix A  provides another set of experiments for estimating  resemblances using $b$-bit minwise hashing with simple hash functions. Those experiments demonstrate that, as long as the data are not too dense, using 2U hash will produce very similar estimates  as using fully random permutations. That set of experiments may help understand the experimental results in this section. Note that both  datasets listed in Table~\ref{tab_data} are extremely sparse. \\

The {\em webspam} dataset is small enough (24GB and 16 million features) that we can conduct experiments using a permutation matrix.
We chose LIBLINEAR as the underlying learning procedure. All experiments were conducted on  workstations with Xeon(R) CPU (W5590@3.33GHz) and 48GB RAM, on a Windows 7 System.

\subsection{Experimental Results}

We experimented with both 2U and 4U hash schemes for  training  linear SVM and logistic regression. We tried out 30 values for the regularization parameter $C$ in the interval $[10^{-3},100]$. We experimented with 11 $k$ values  from $k=10$ to $k=500$, and for 7 $b$ values: $b=1, 2, 4, 6, 8, 10, 16$. Each experiment was repeated 50 times. The total number of training experiments turns out to be
\begin{align}\notag
2\times 2\times 30\times 11\times 7 \times 50= 462000.
\end{align}

To maximize the repeatability, whenever page space allows, we always would like to present the detailed experimental results for all the parameters instead of, for example, only reporting the best results or the cross-validated results. In this subsection, we only present the  results for {\em webspam} dataset using linear SVM because the results for logistic regression lead to the same conclusion.

\begin{figure}[h!]

\begin{center}
\mbox{
\includegraphics[width=1.6in]{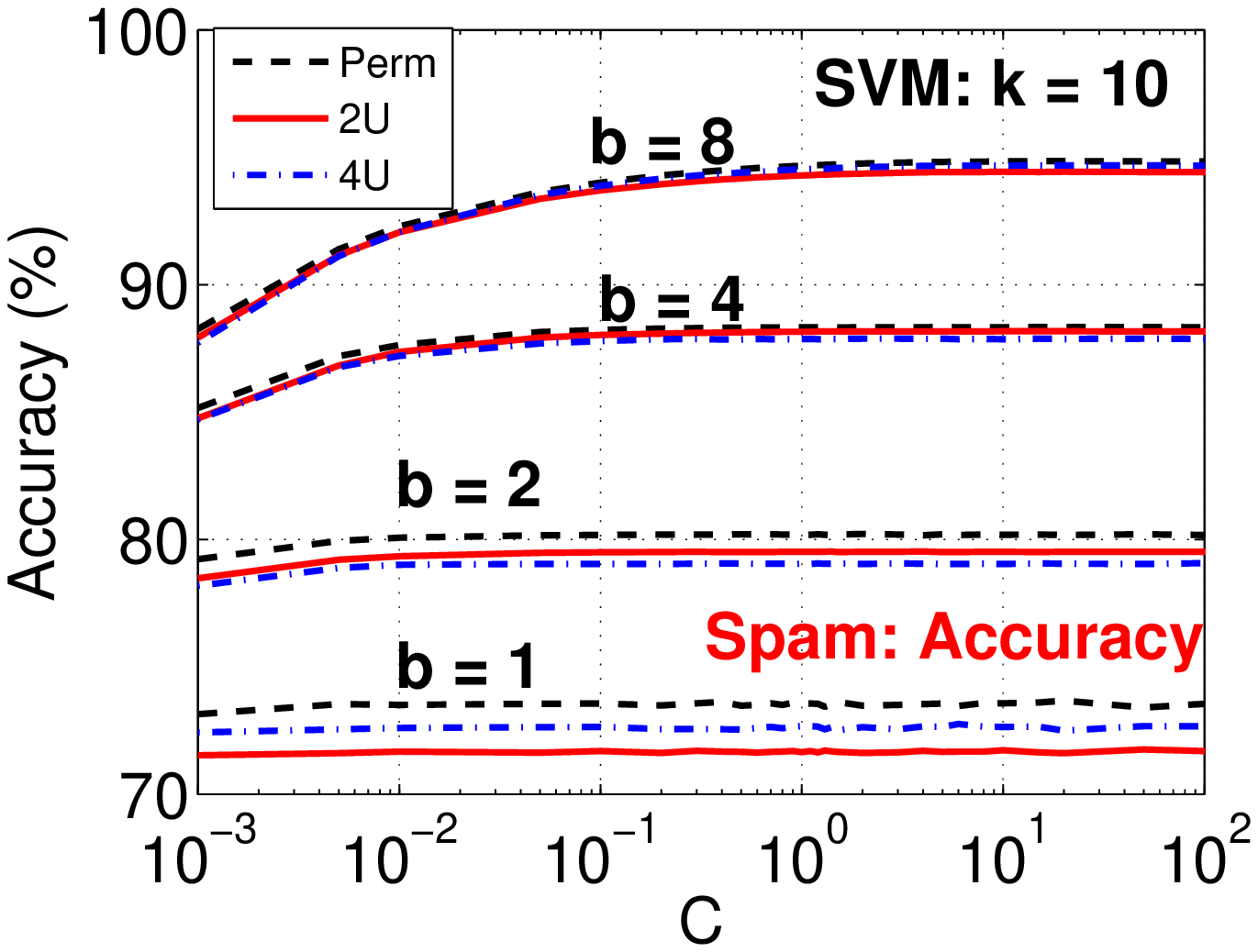}
%\includegraphics[width=1.6in]{fig/spam80_accuracy_k20_2u4u.eps}}

%\vspace{-0.1in}

%\mbox{
%\includegraphics[width=1.6in]{fig/spam80_accuracy_k30_2u4u.eps}
\includegraphics[width=1.6in]{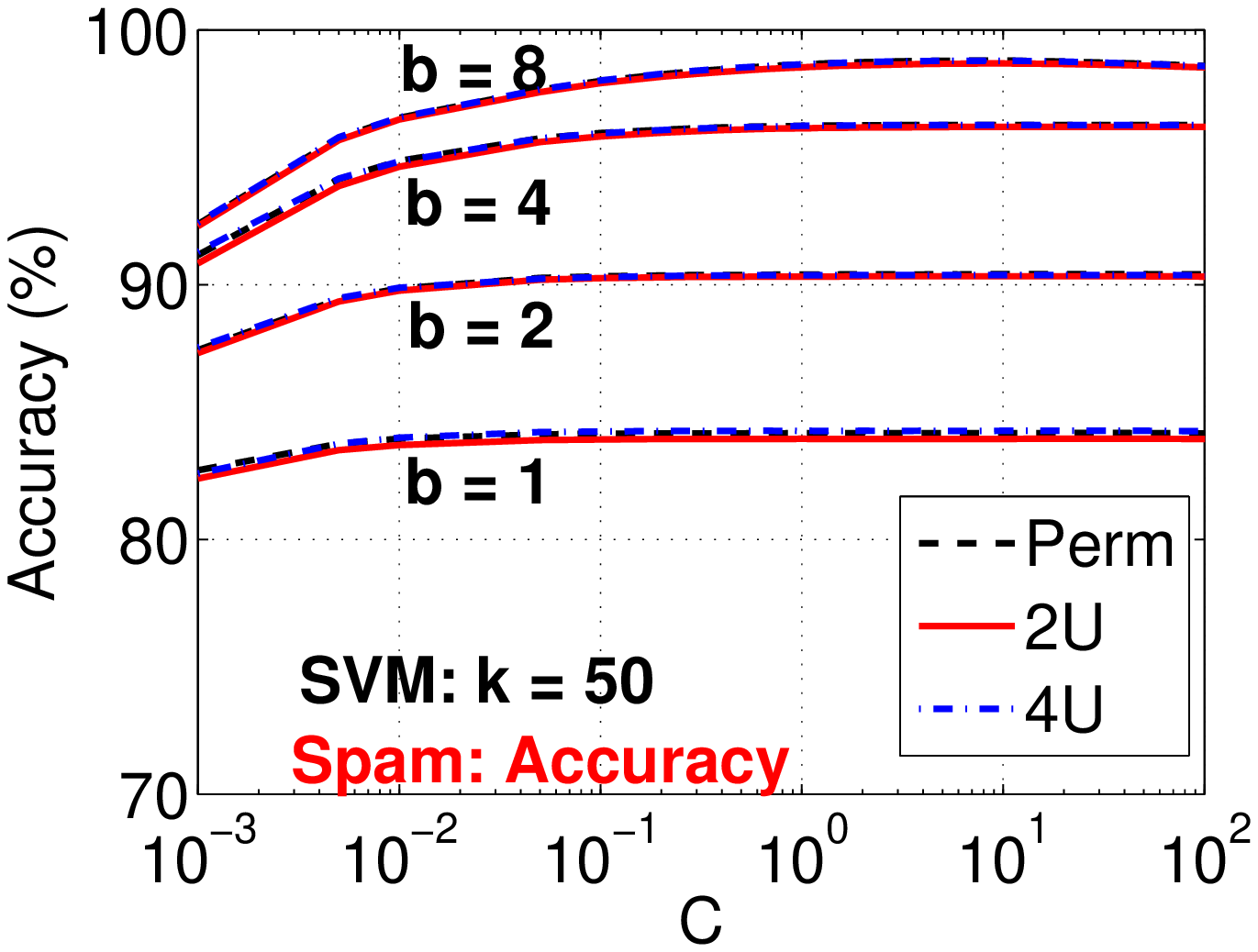}}

\vspace{-0.1in}

\mbox{
\includegraphics[width=1.6in]{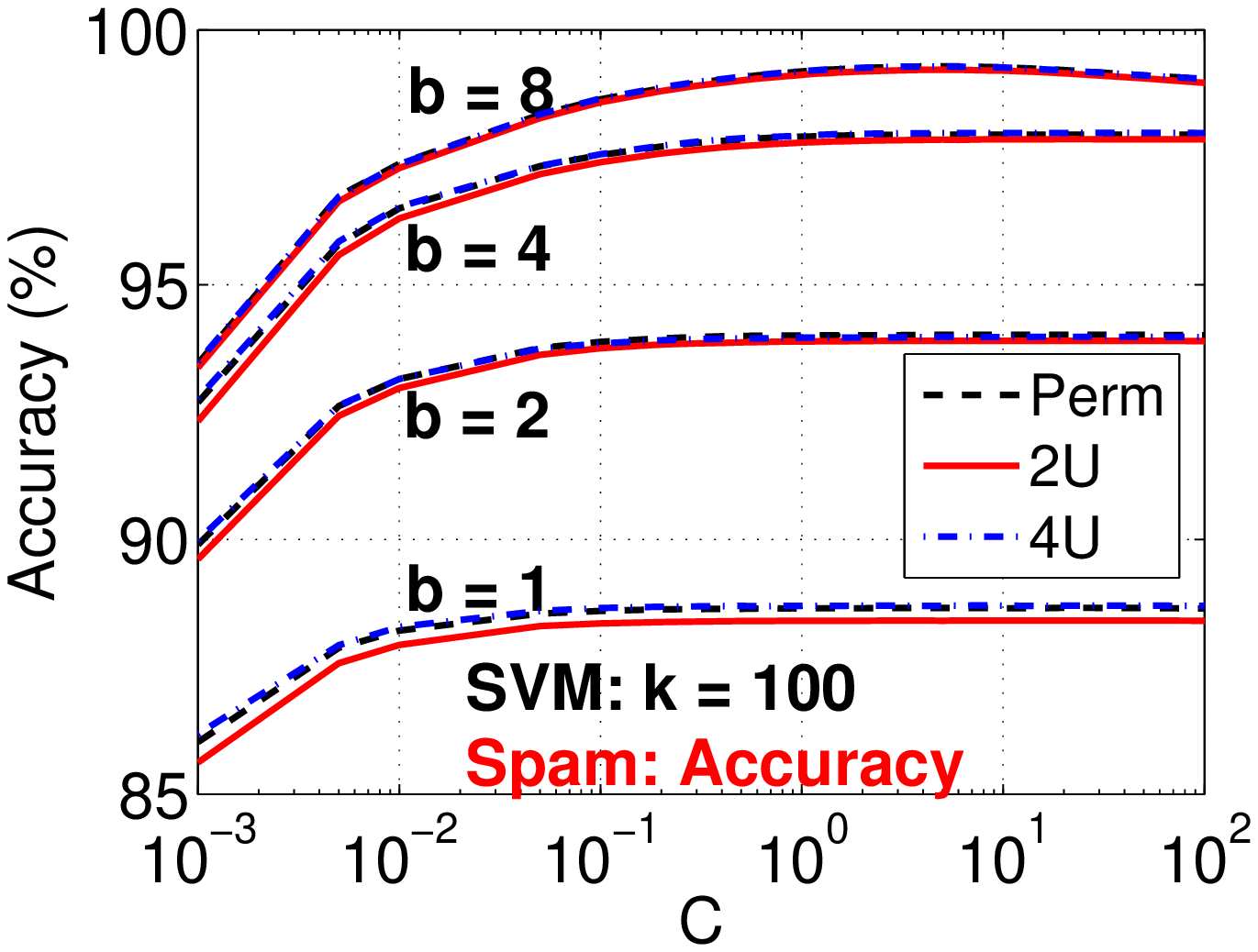}
\includegraphics[width=1.6in]{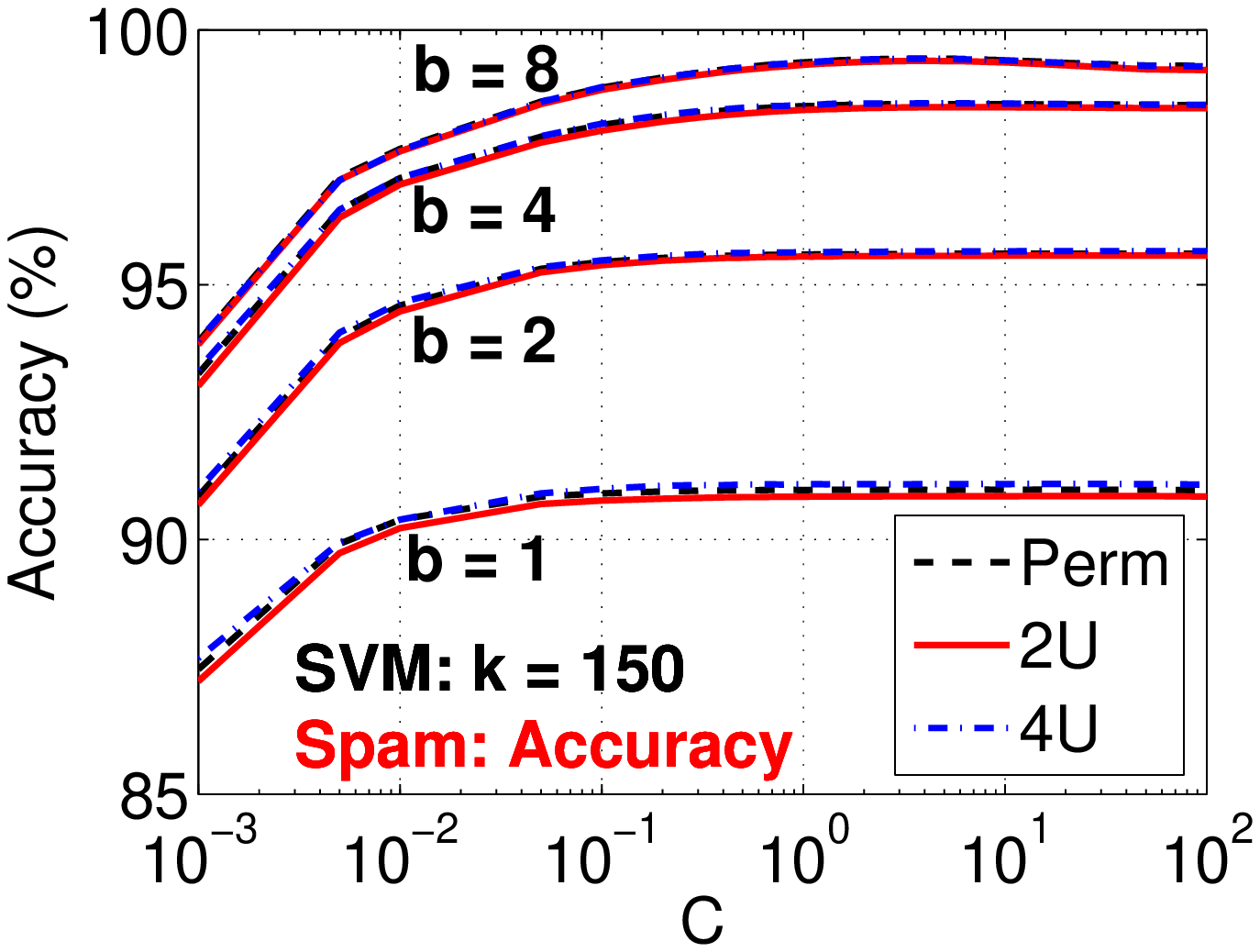}}

\vspace{-0.1in}

\mbox{
\includegraphics[width=1.6in]{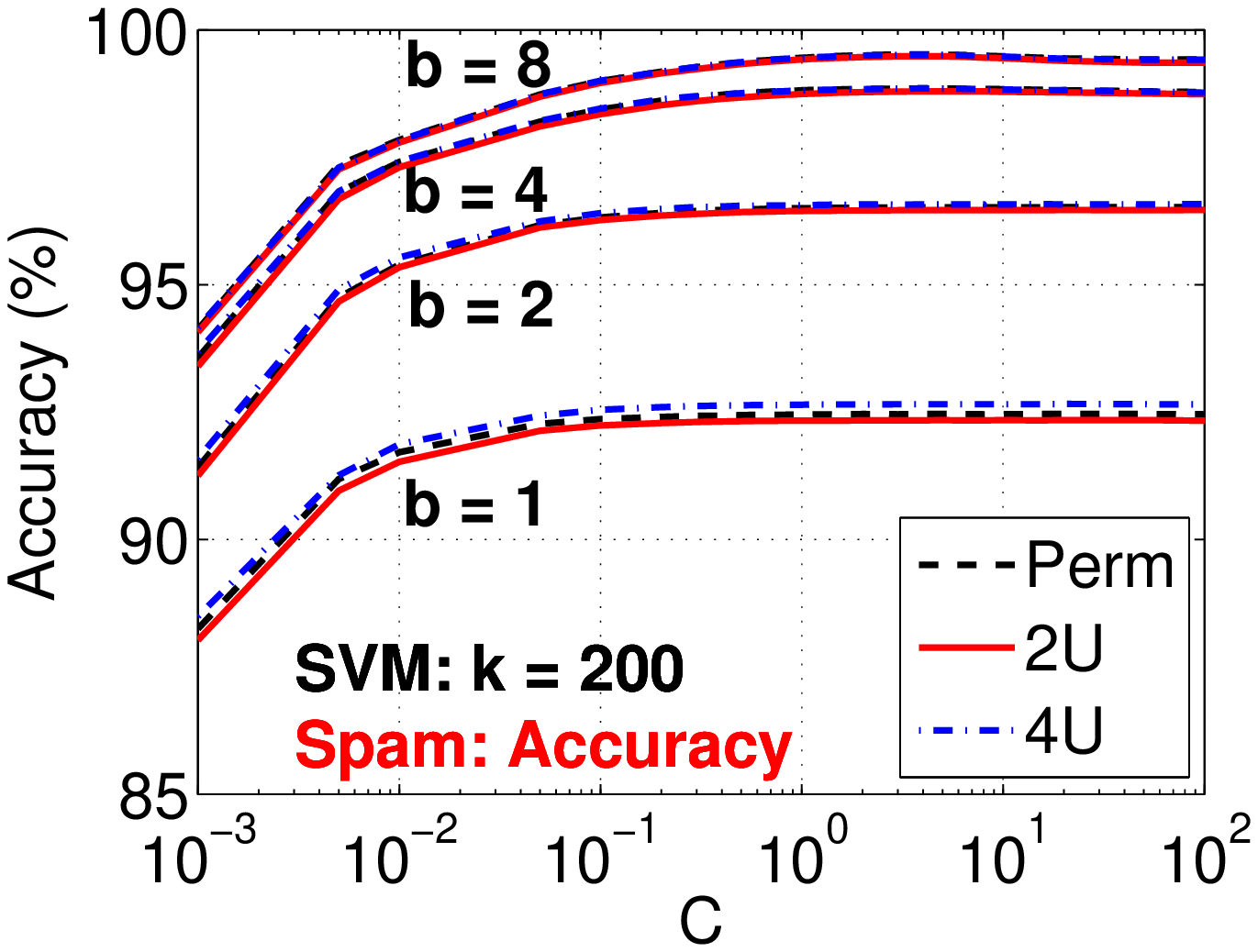}
\includegraphics[width=1.6in]{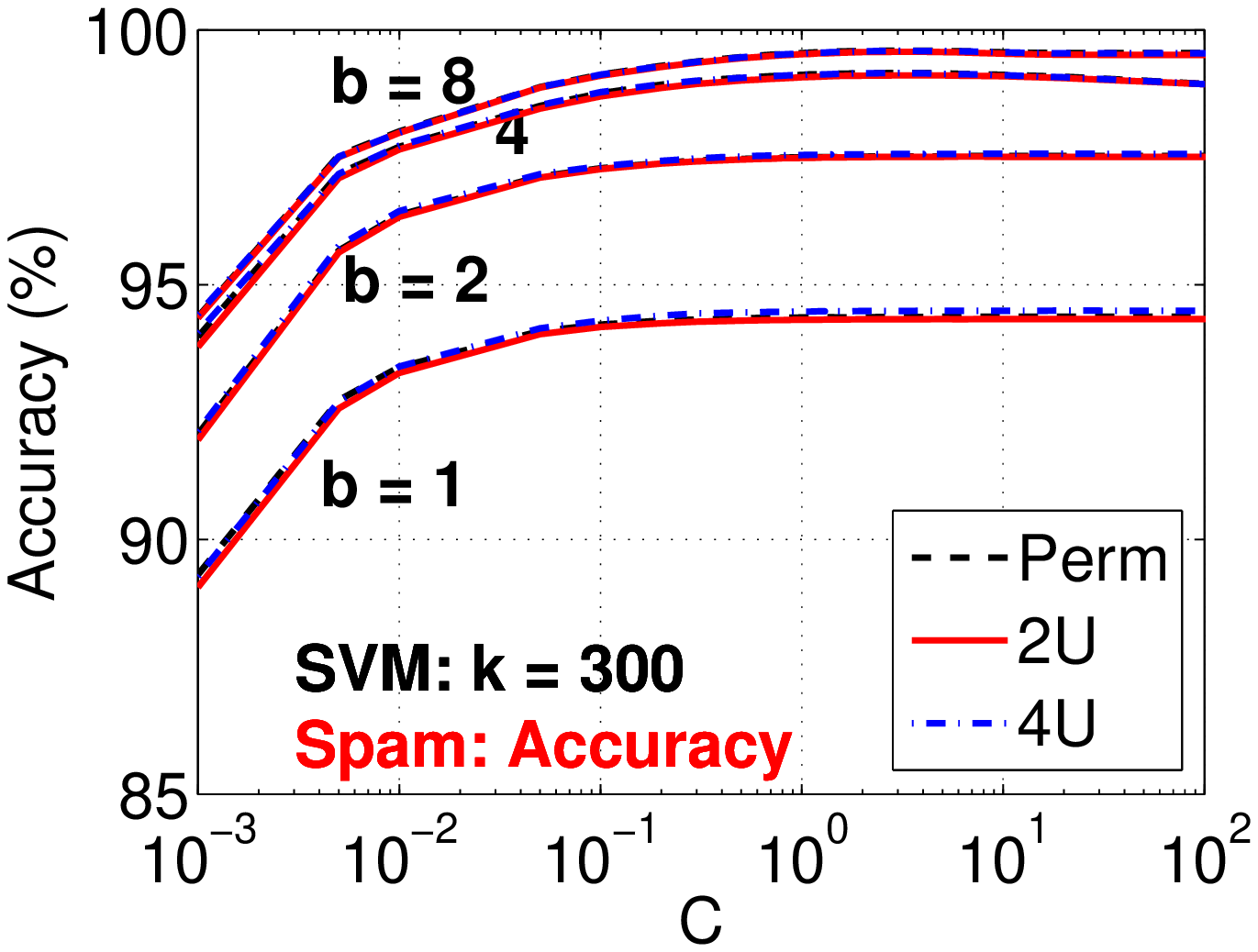}}

\vspace{-0.1in}

\mbox{
\includegraphics[width=1.6in]{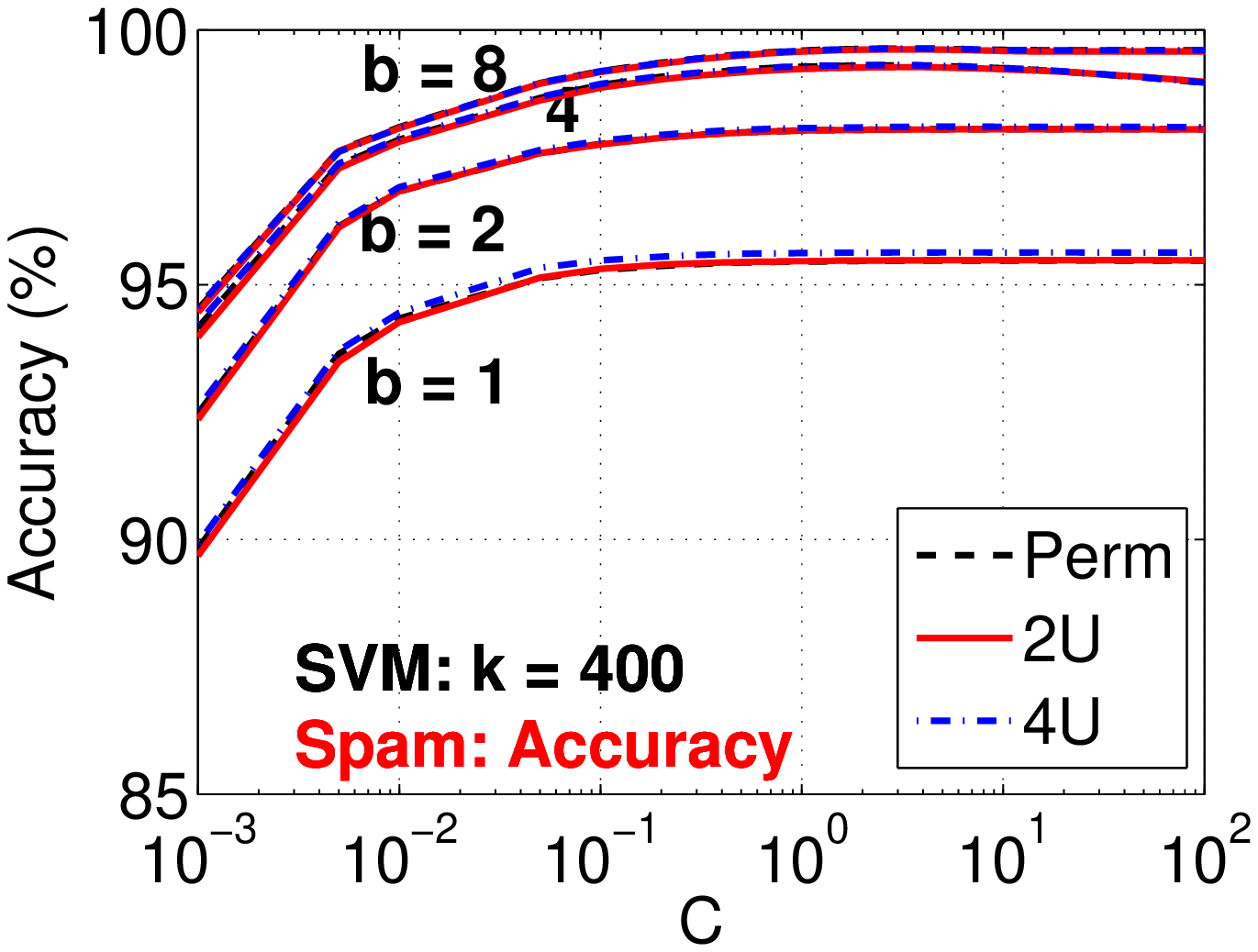}
\includegraphics[width=1.6in]{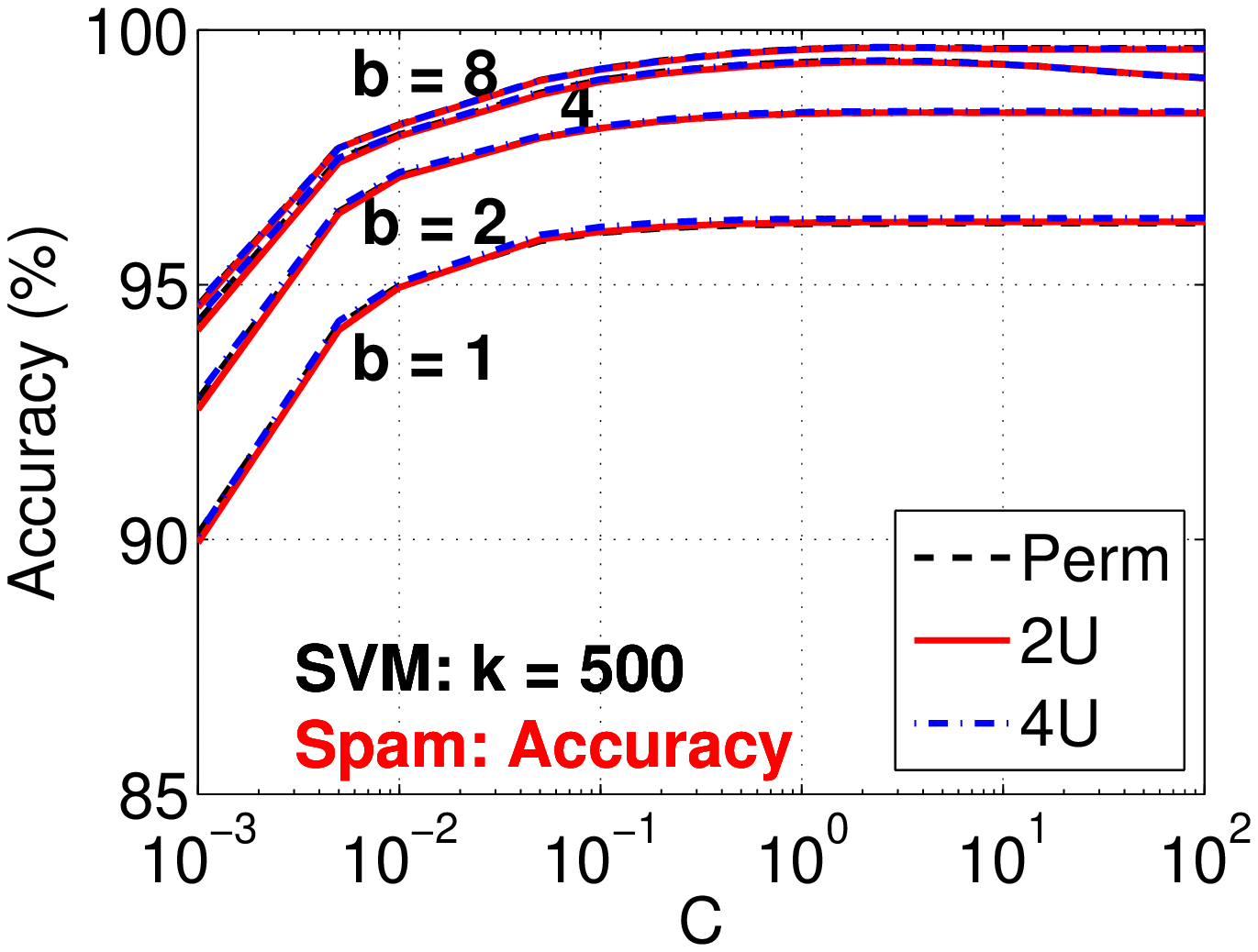}}

\vspace{-0.3in}
\end{center}

\caption{Linear SVM test accuracies on {\em webspam}, using three hashing schemes: permutations, 2U, and 4U.  Both 2U and 4U perform well as the curves essentially overlap except when $b=1$ or small $k$. It appears that 4U is only slightly better than 2U. }\label{fig_spam_acc_2u4u}\vspace{-0.1in}
\end{figure}

Figure~\ref{fig_spam_acc_2u4u} presents the SVM test accuracies (averaged over 50 runs).   For the test cases that correspond to the most likely parameter settings used in practice (e.g., $k\geq 200$ and $b\geq 4$), we can see that the results from the three hashing schemes (full permutations, 2U, and 4U) are essentially identical. Overall, it appears that 4U is slightly better than 2U when $b=1$ or $k$ is very small.

This set of experiments can provide us with a strong experimental evidence that the simple and highly efficient 2U hashing scheme may be sufficient in practice, when used in the context of large-scale machine learning using $b$-bit minwise hashing.

\subsection{Experimental Results on VW Algorithm}

The application domains of $b$-bit minwise hashing are limited to binary data, whereas other hashing algorithms such as random projections and Vowpal Wabbit (VW) hashing algorithm~\cite{Article:Shi_JMLR09,Proc:Weinberger_ICML2009} are not restricted to binary data. It is shown in~\cite{Proc:HashLearning_NIPS11} that  VW and random projections have essentially the same variances as reported in~\cite{Proc:Li_Hastie_Church_KDD06}. In this study, we also provide some experiments on VW hashing algorithm because we expect that practitioners will find applications which are more suitable for VW than $b$-bit minwise hashing. Since there won't be space to explain the details of VW algorithm, interested readers please refer to~\cite{Article:Shi_JMLR09,Proc:Weinberger_ICML2009,Proc:HashLearning_NIPS11}.

\begin{figure}[h!]
\begin{center}
\mbox{
\includegraphics[width=1.6in]{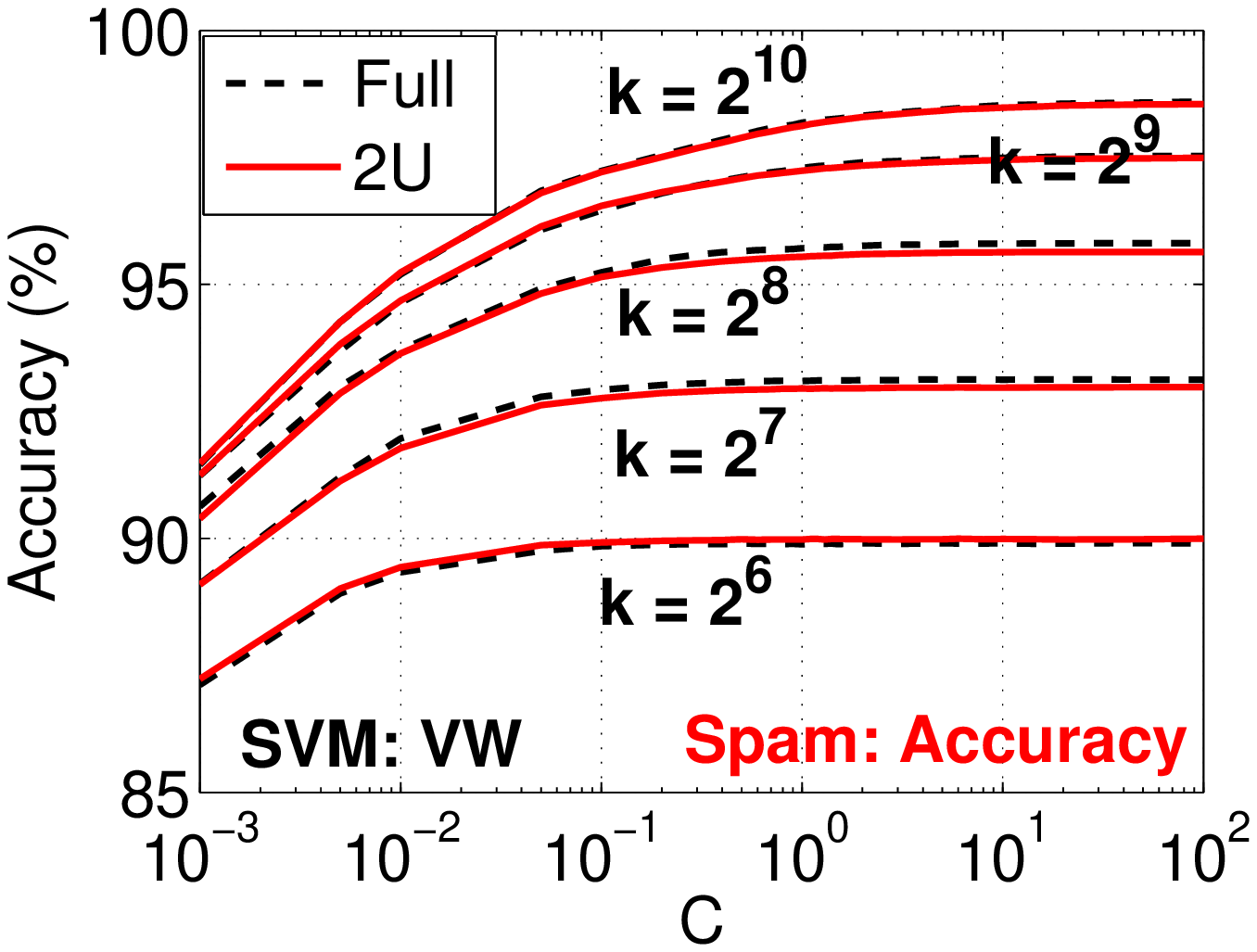}
\includegraphics[width=1.6in]{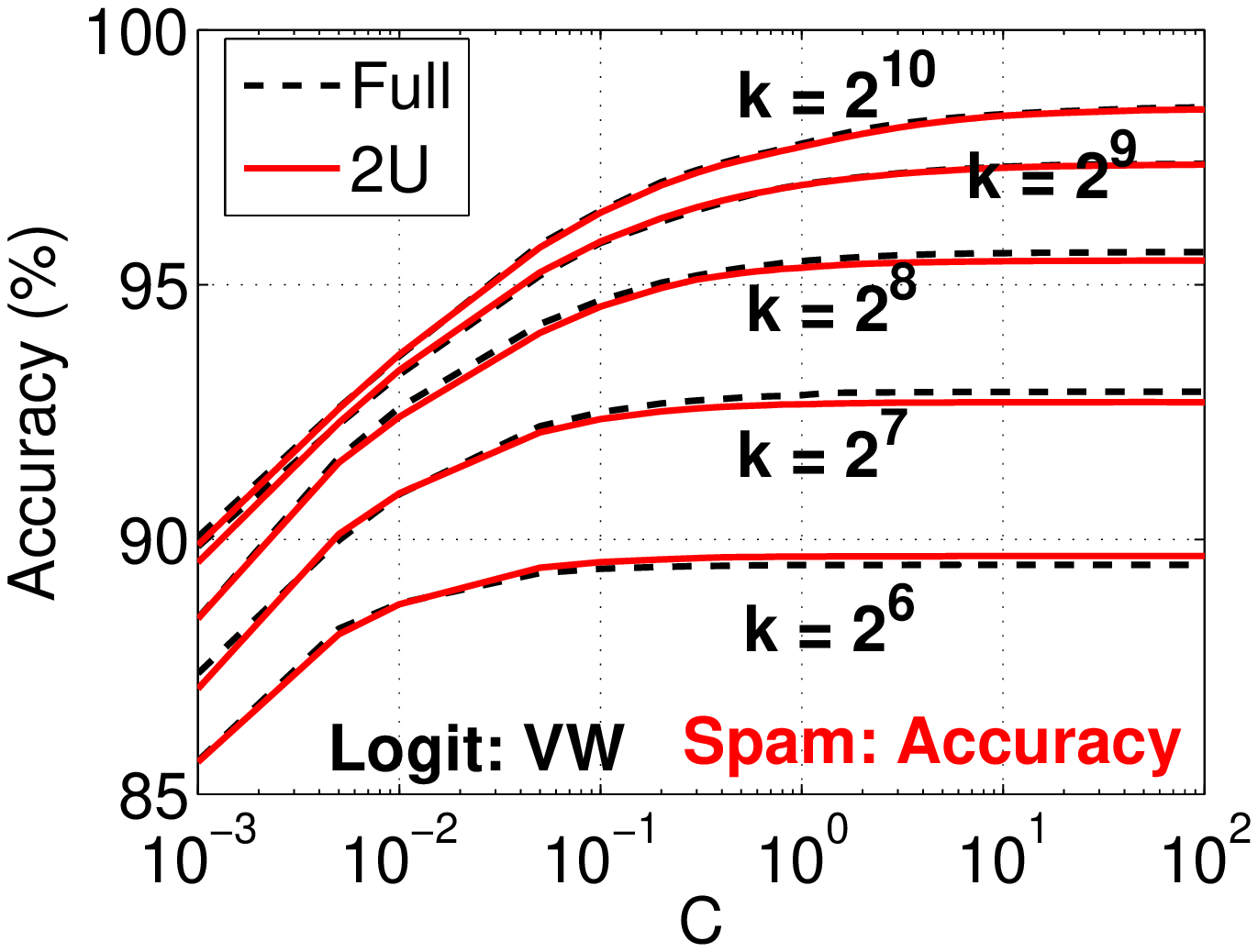}
}
\end{center}
\vspace{-0.3in}
\caption{Linear SVM (left panel) and logistic regression (right panel) test accuracies on {\em webspam} using full randomness (``FULL'') as well as 2U hash scheme (``2U''). We can see that the test accuracies are not affected much by using 2U hash. }\label{fig_VW_2u}
\end{figure}

In particular, in this subsection, we present a comparison between the accuracy of VW when using a fully random implementation and when using 2U
hash functions. Figure~\ref{fig_VW_2u} present the experimental results for both linear SVM and logistic regression. As we can see, the two graphs are nearly identical, meaning that the performance 2U hashing seen earlier is not limited to the context of learning with $b$-bit minwise hashing.

\section{Learning on Rcv1 Data (200GB)}

Compared to {\em webspam}, the size of the expanded {\em rcv1} dataset may be more close to the training data sizes used in industrial applications. We report the experiments on linear SVM and logistic regression, as well as the comparisons with the VW hash algorithm.

\subsection{Experiments on Linear SVM }
\begin{figure}[h!]
\begin{center}

\mbox{
\includegraphics[width=1.6in]{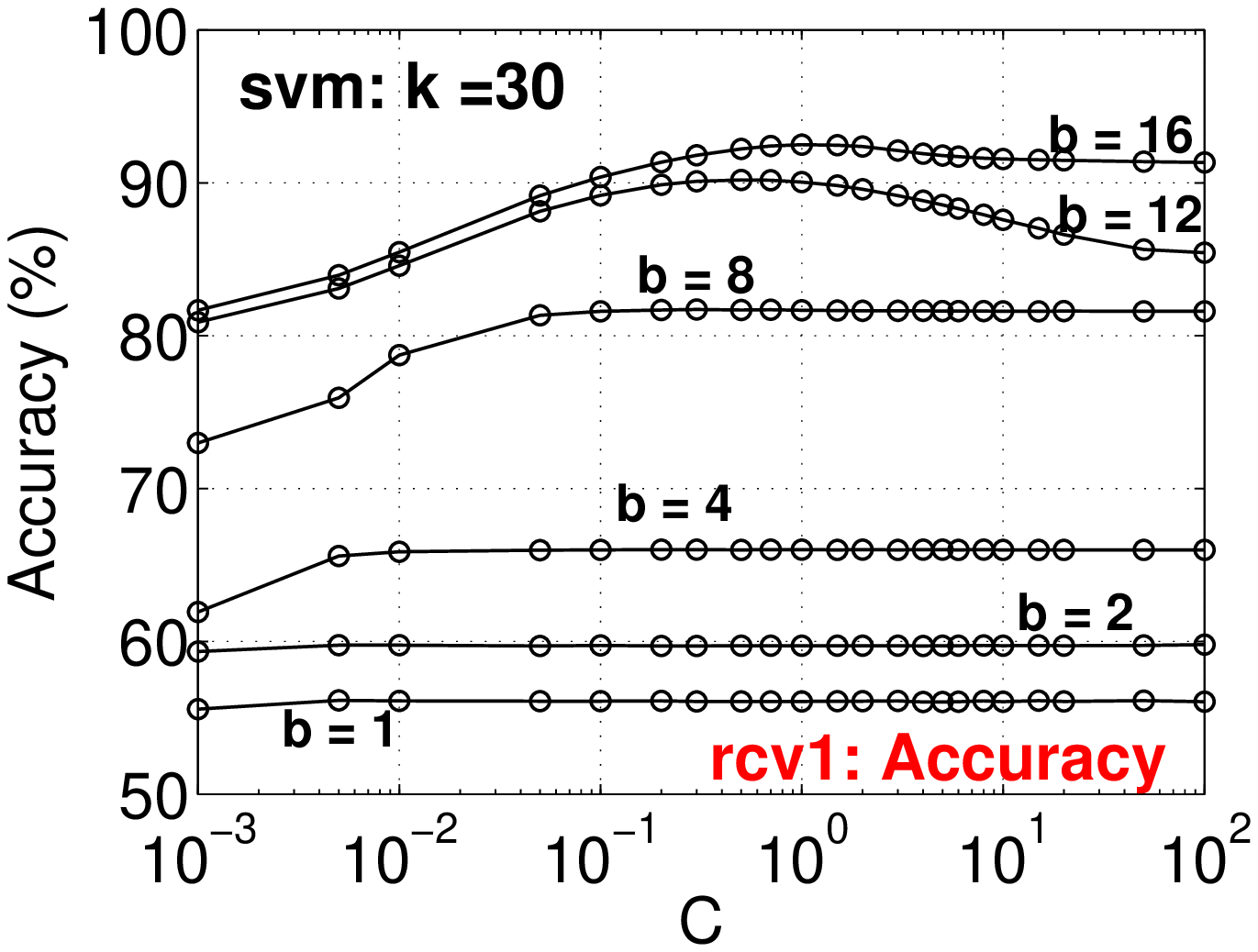}\hspace{-0.1in}
\includegraphics[width=1.6in]{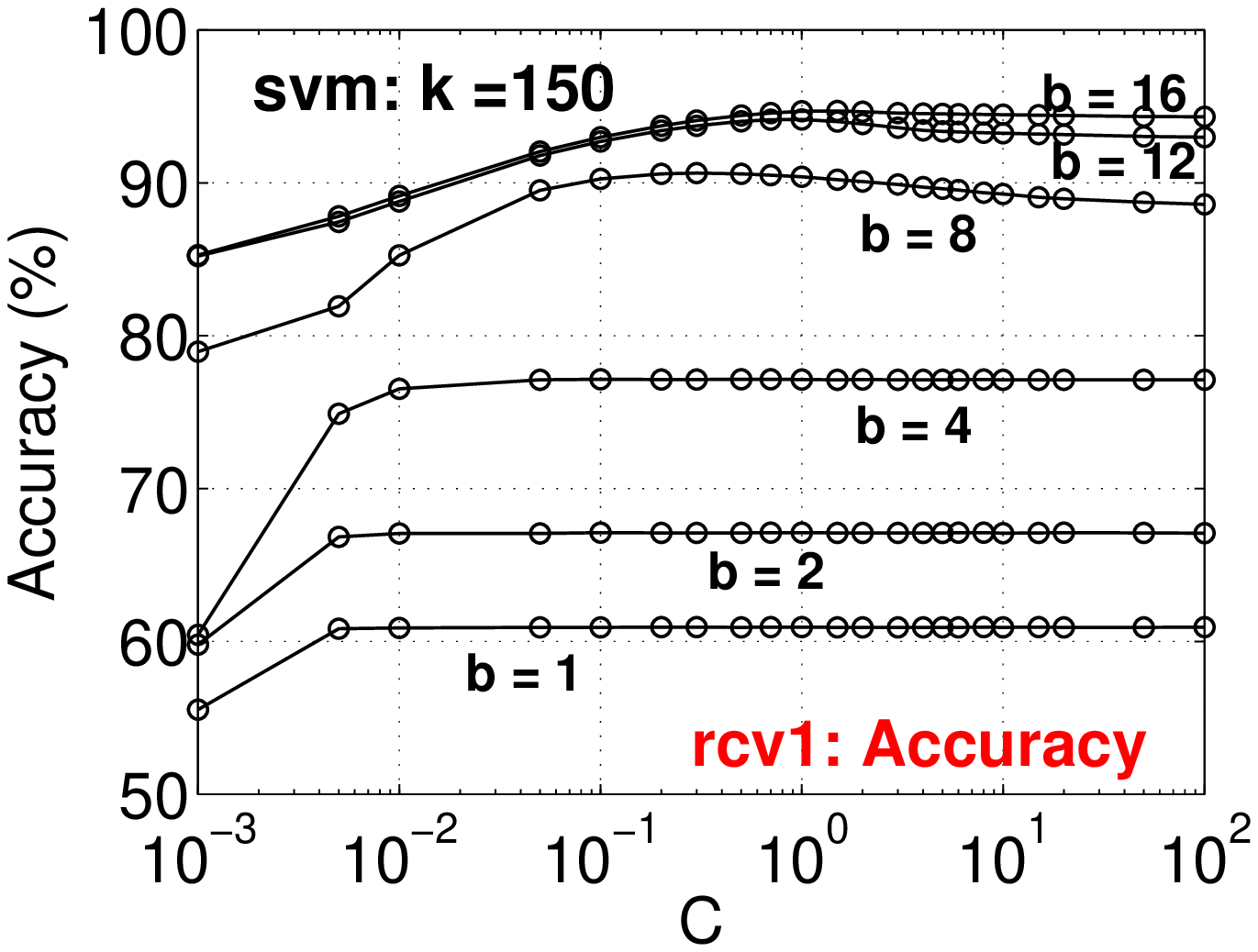}
}
\vspace{-0.1in}

\mbox{
\includegraphics[width=1.6in]{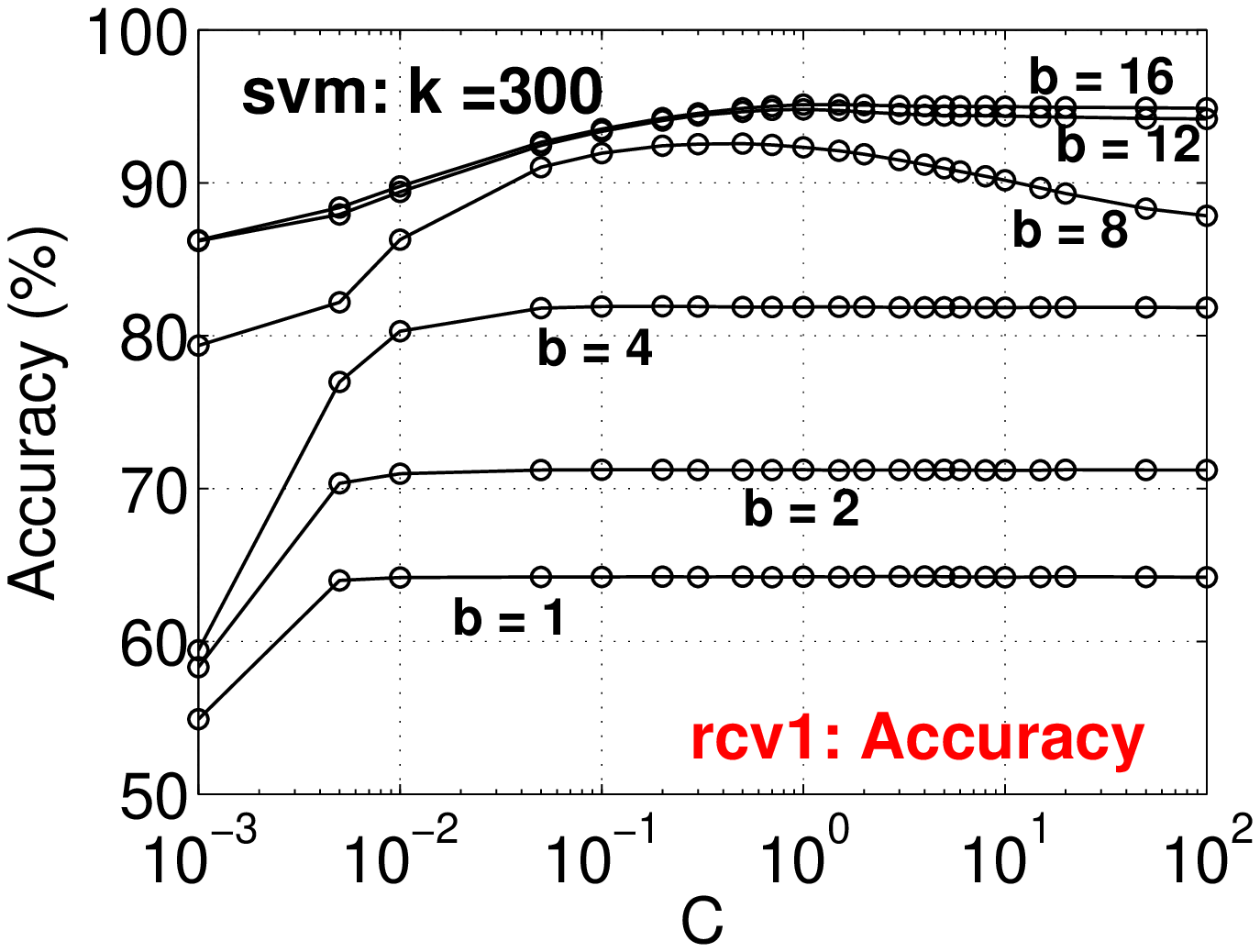}
\includegraphics[width=1.6in]{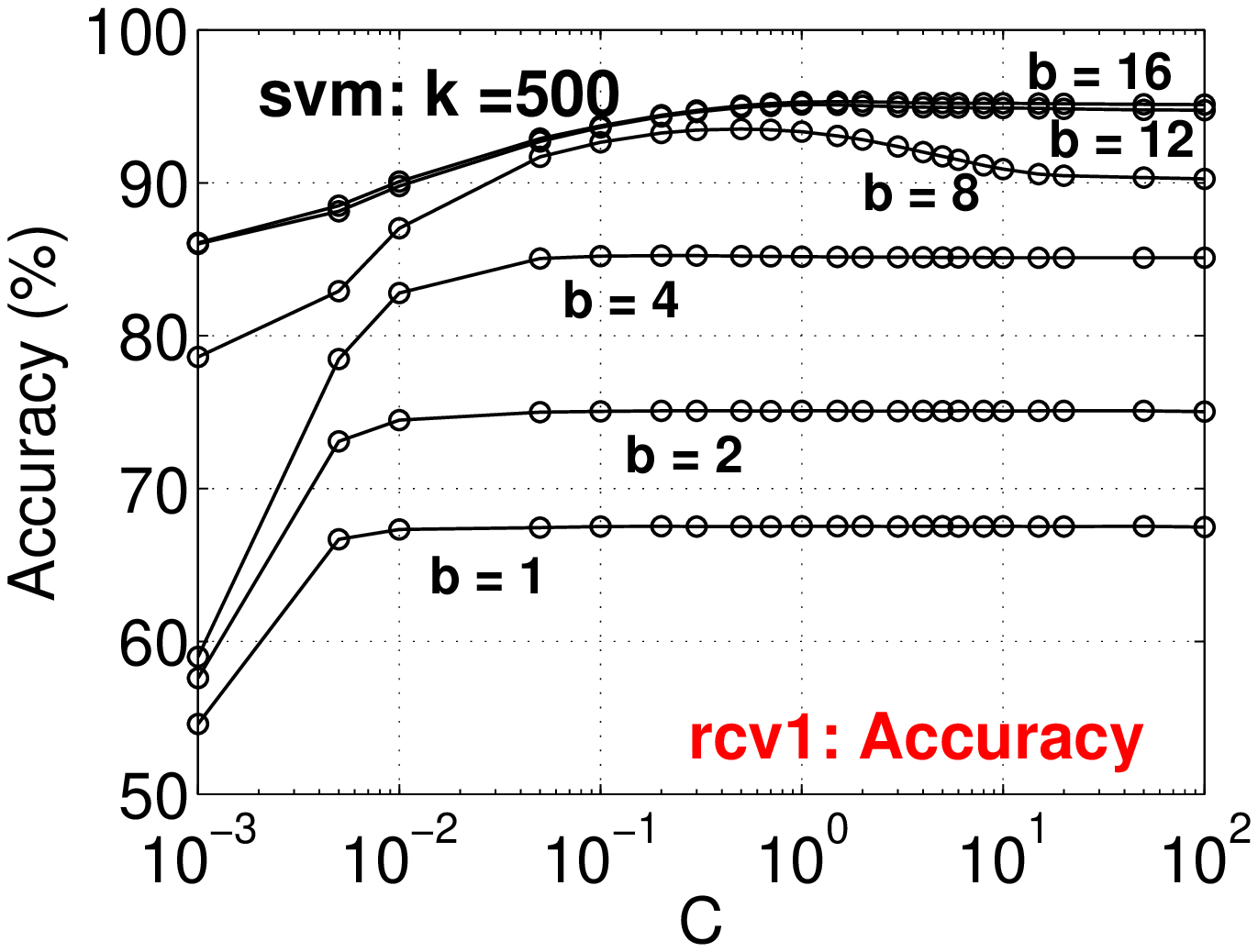}}

\end{center}
\vspace{-0.3in}

\caption{\textbf{Linear SVM test accuracy on rcv1}.   }\label{fig_rcv1_acc}\vspace{-0.2in}

\end{figure}

\begin{figure}[h!]
\begin{center}
\mbox{
\includegraphics[width=1.6in]{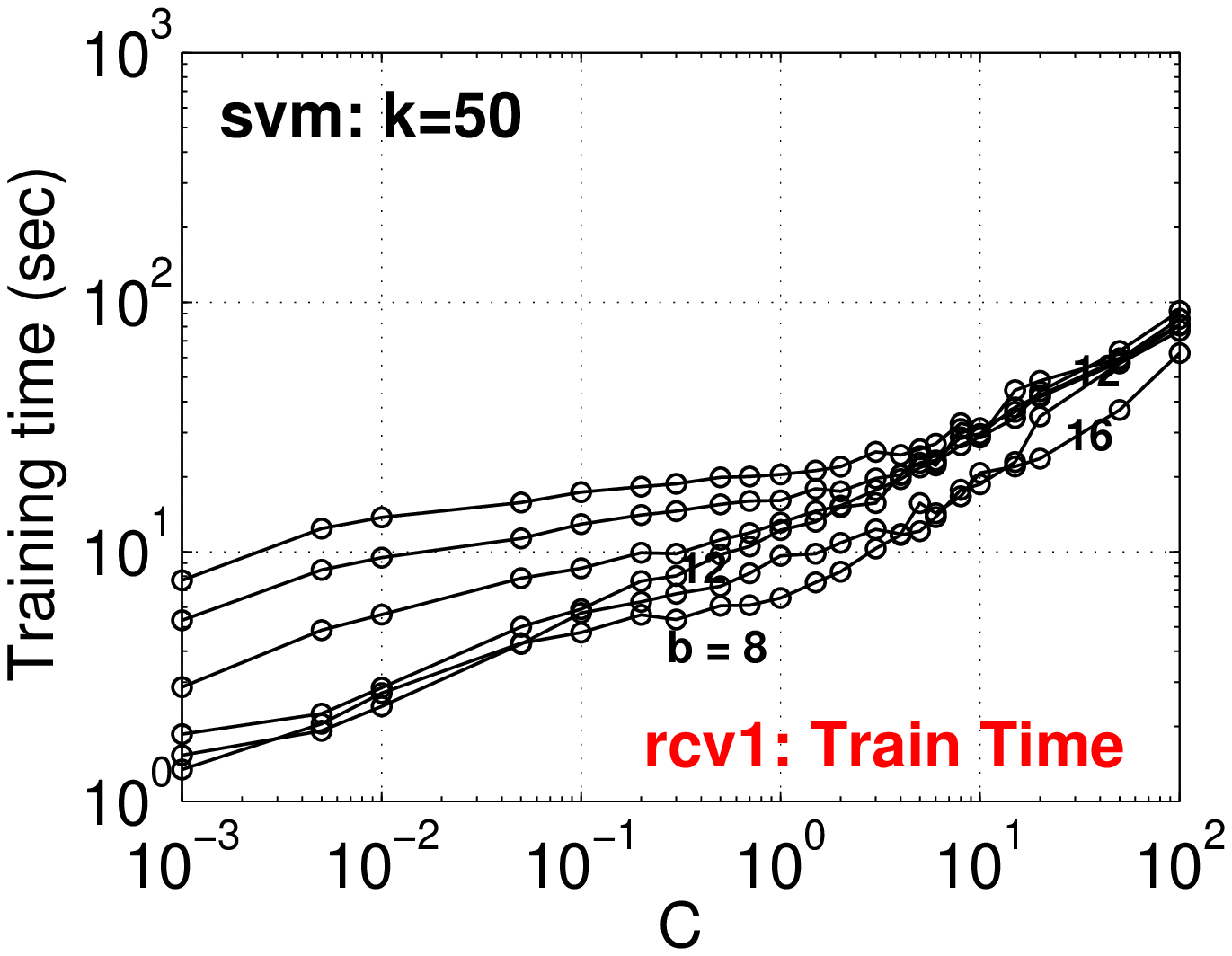}
%\includegraphics[width=1.6in]{fig/rcv1_train_k200_2u.eps}
%\includegraphics[width=1.6in]{fig/rcv1_train_k150_2u.eps}}

%\vspace{-0.1in}

%\mbox{
%\includegraphics[width=1.6in]{fig/rcv1_train_k300_2u.eps}
%\includegraphics[width=1.6in]{fig/rcv1_train_k300_2u.eps}\hspace{-0.1in}
%\includegraphics[width=1.6in]{fig/rcv1_train_k400_2u.eps}\hspace{-0.1in}
\includegraphics[width=1.6in]{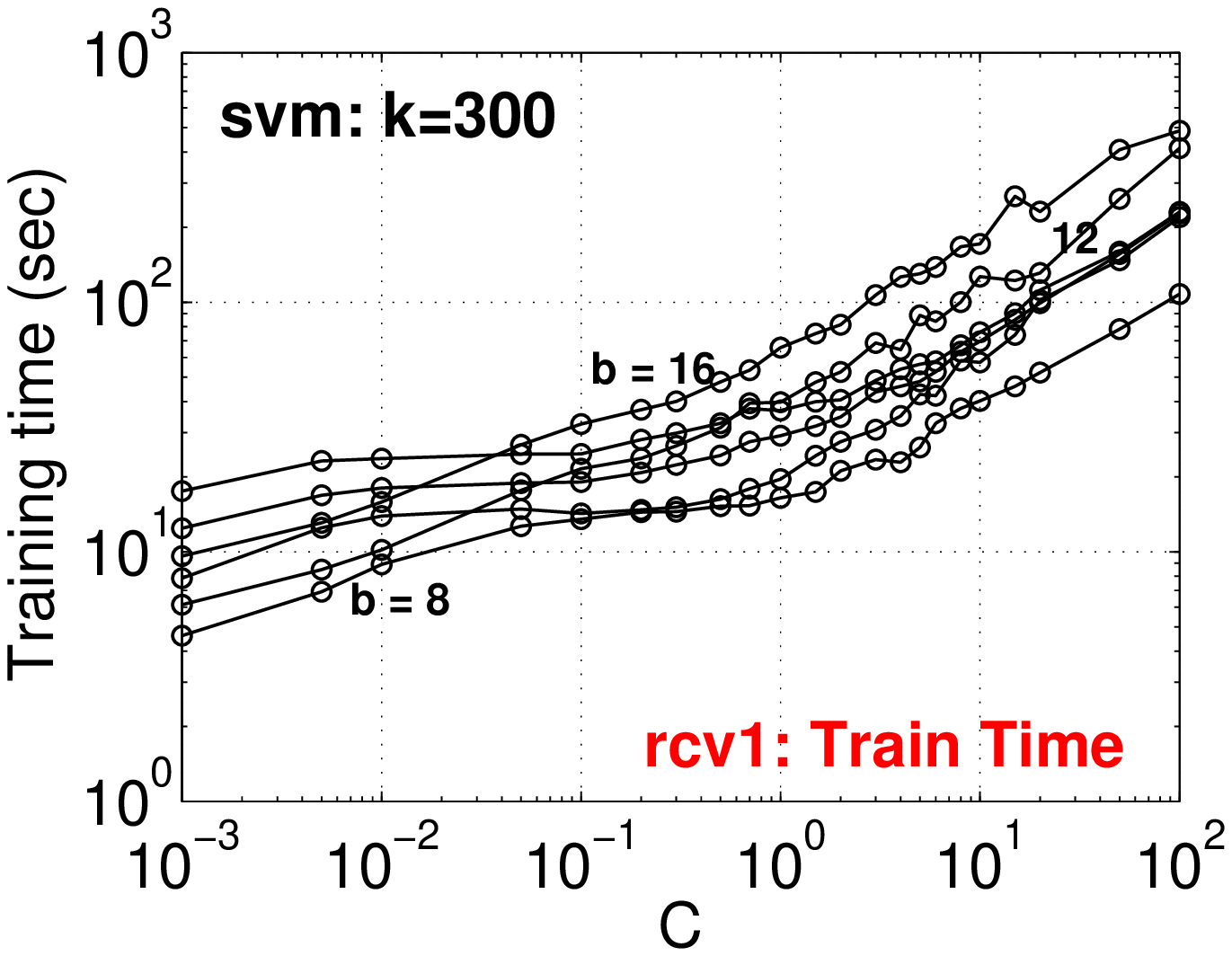}}

\end{center}
\vspace{-0.3in}

\caption{\textbf{Linear SVM training time on rcv1}.}\label{fig_rcv1_train}\vspace{-0.1in}
\end{figure}

Figure~\ref{fig_rcv1_acc} and Figure~\ref{fig_rcv1_train} respectively provide the test accuracies and train times, for training linear SVM. We can not report the baseline because the original dataset exceeds the memory capacity. Using merely $k=30$ and $b=12$, our method can achieve $>90\%$ test accuracies. With  $k\geq 300$, we can achieve $>95\%$ test accuracies.

%\clearpage

\subsection{Experiments on Logistic Regression}
\begin{figure}[h!]
\begin{center}
\mbox{
\includegraphics[width=1.6in]{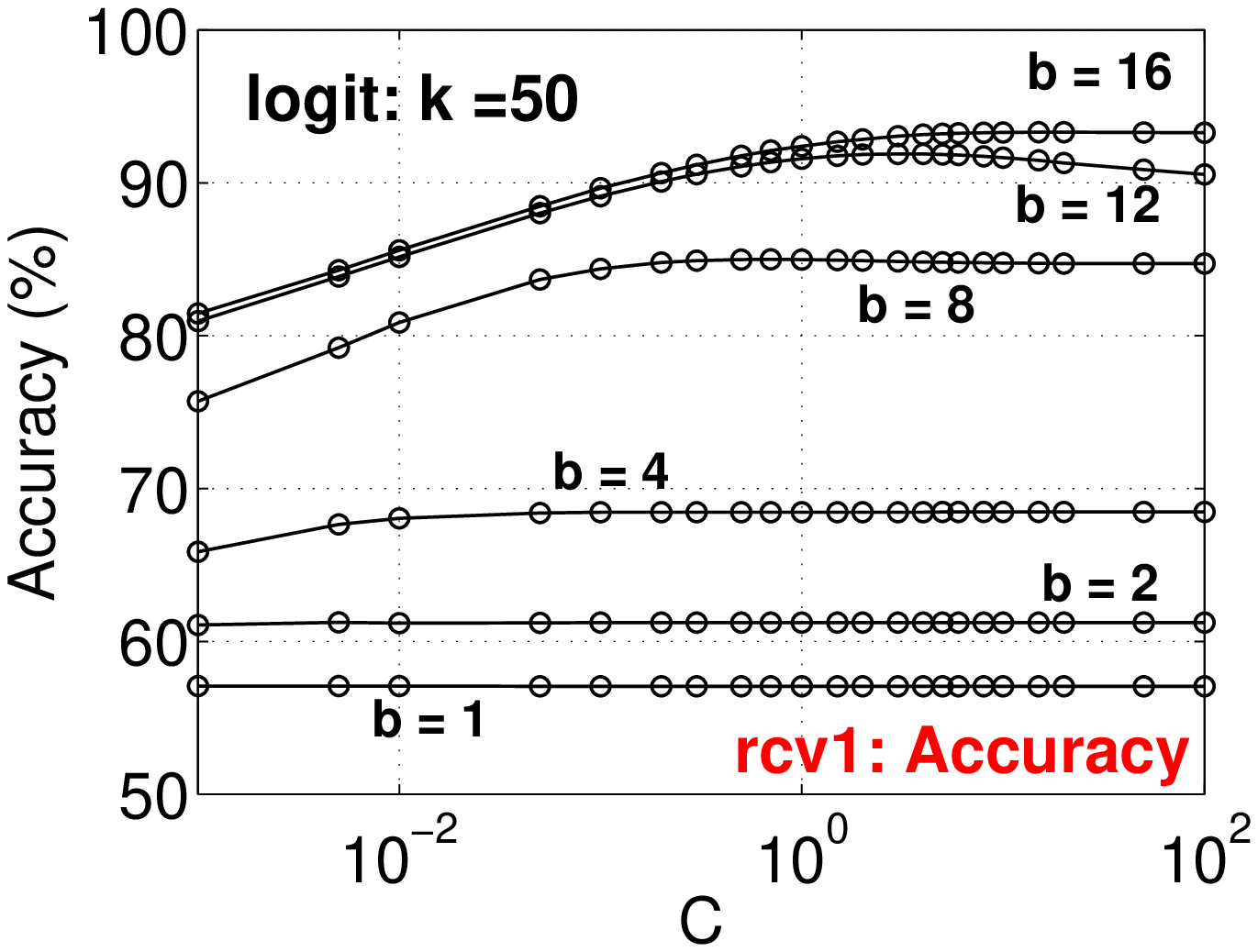}
\includegraphics[width=1.6in]{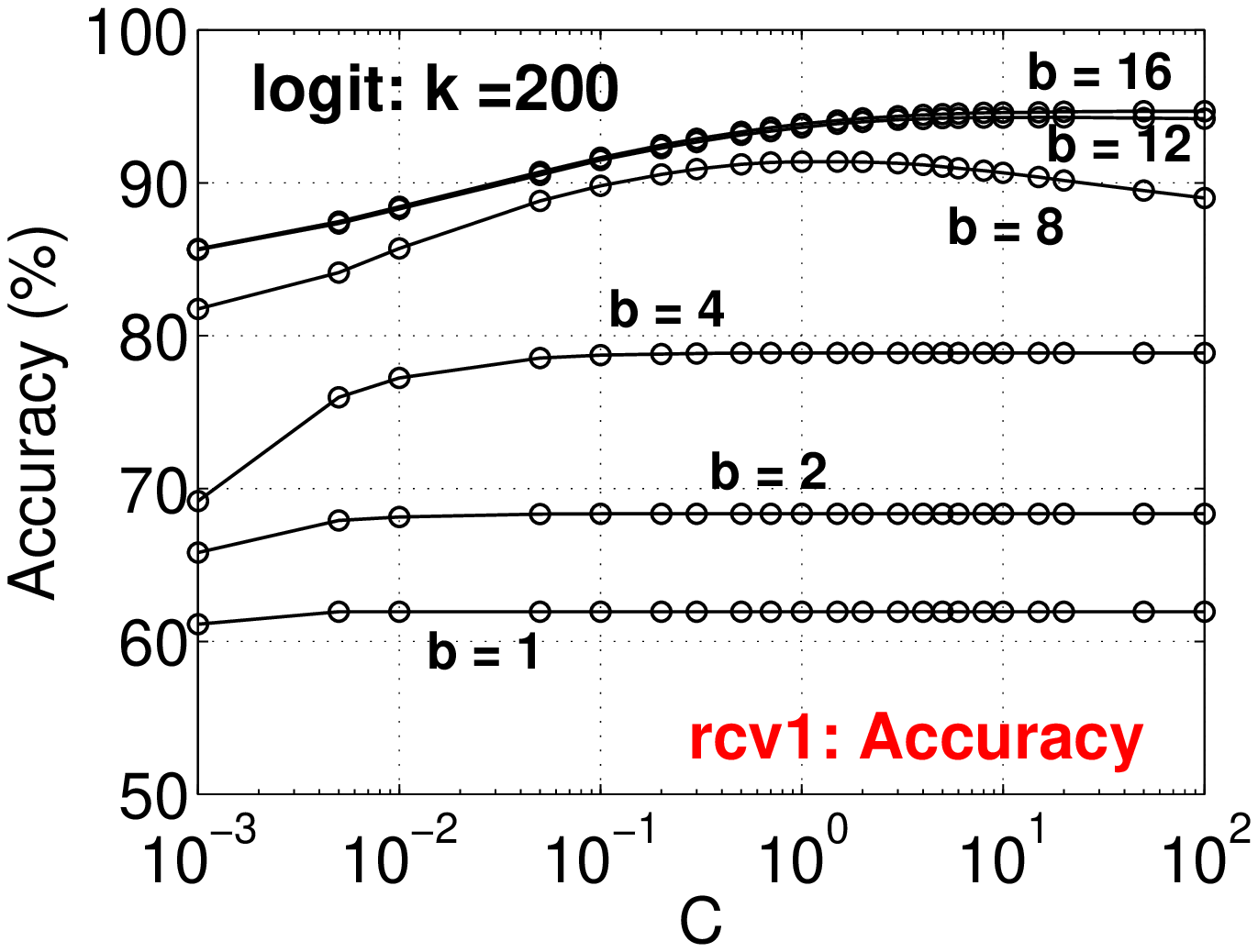}
}

\vspace{-0.1in}

\mbox{
\includegraphics[width=1.6in]{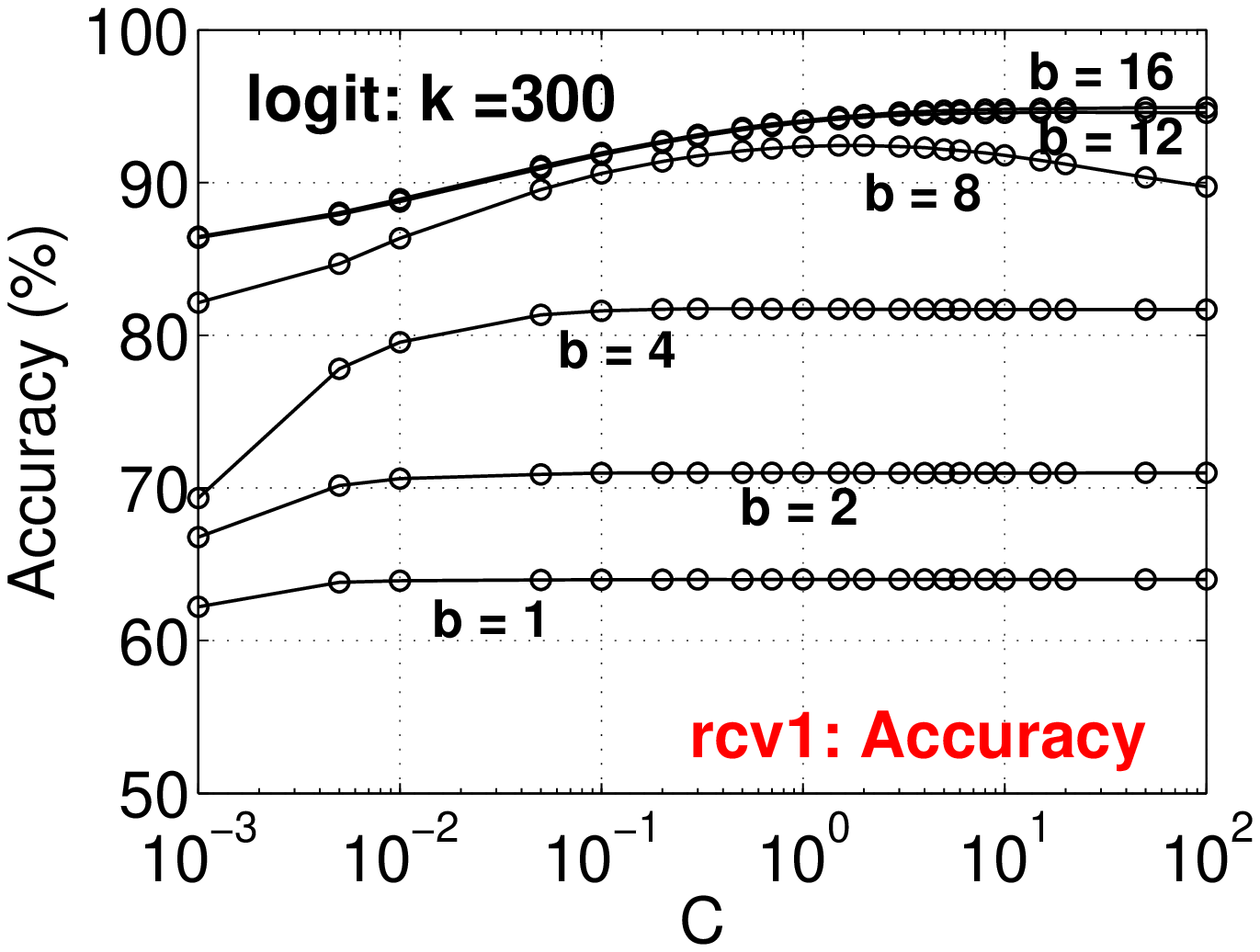}
\includegraphics[width=1.6in]{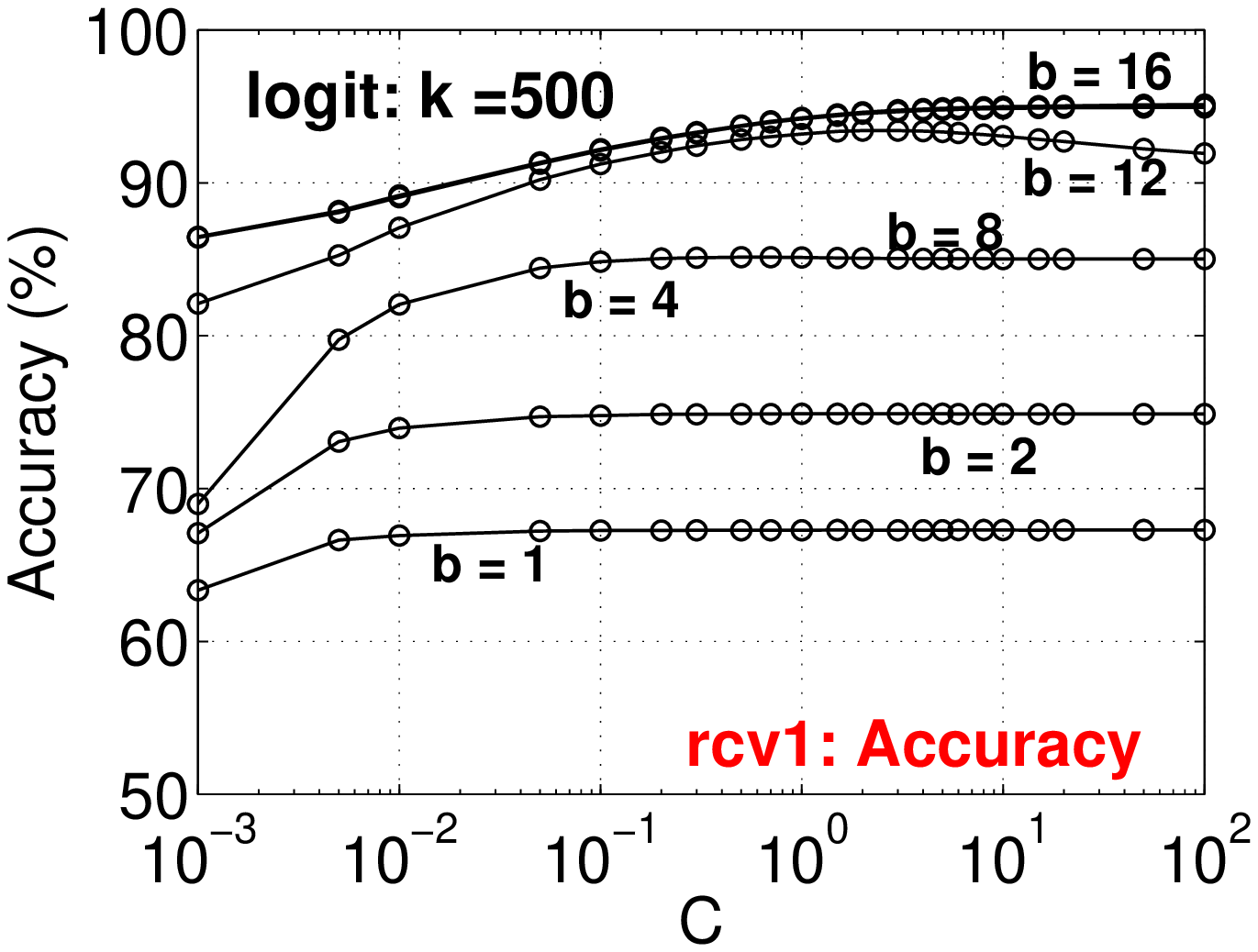}}

\vspace{-0.25in}

\end{center}
\caption{\textbf{Logistic regression test accuracy on rcv1}.  }\label{fig_rcv1_acc_logit}\vspace{-0.15in}
\end{figure}

\begin{figure}[h!]
\begin{center}
\mbox{
\includegraphics[width=1.6in]{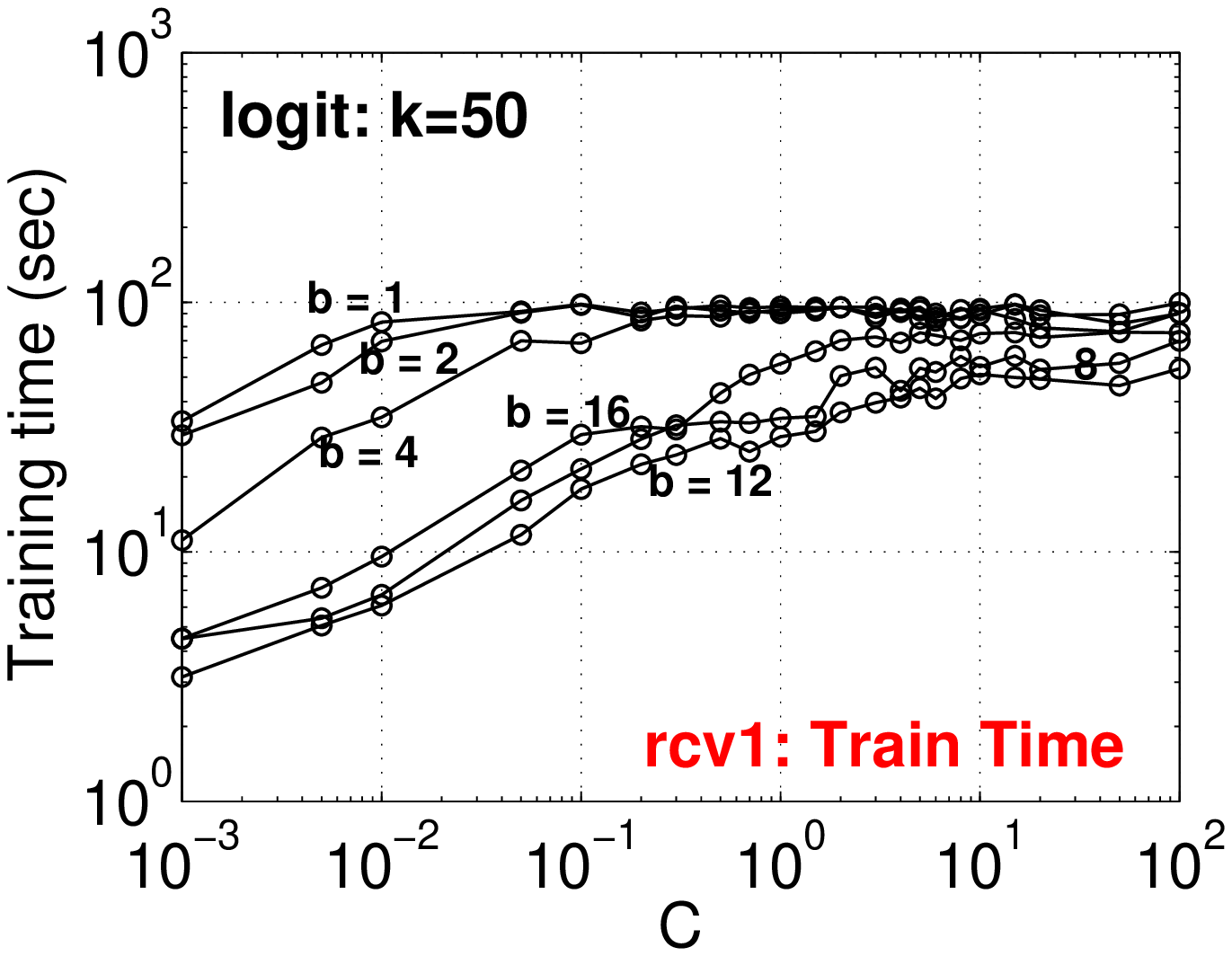}
%\includegraphics[width=1.6in]{fig/logit_rcv1_train_k100_2u.eps}\hspace{-0.1in}
%\includegraphics[width=1.6in]{fig/logit_rcv1_train_k200_2u.eps}}

%\vspace{-0.1in}

%\mbox{
%\includegraphics[width=1.6in]{fig/logit_rcv1_train_k300_2u.eps}
%\includegraphics[width=1.6in]{fig/logit_rcv1_train_k300_2u.eps}\hspace{-0.1in}
%\includegraphics[width=1.6in]{fig/logit_rcv1_train_k400_2u.eps}\hspace{-0.1in}
\includegraphics[width=1.6in]{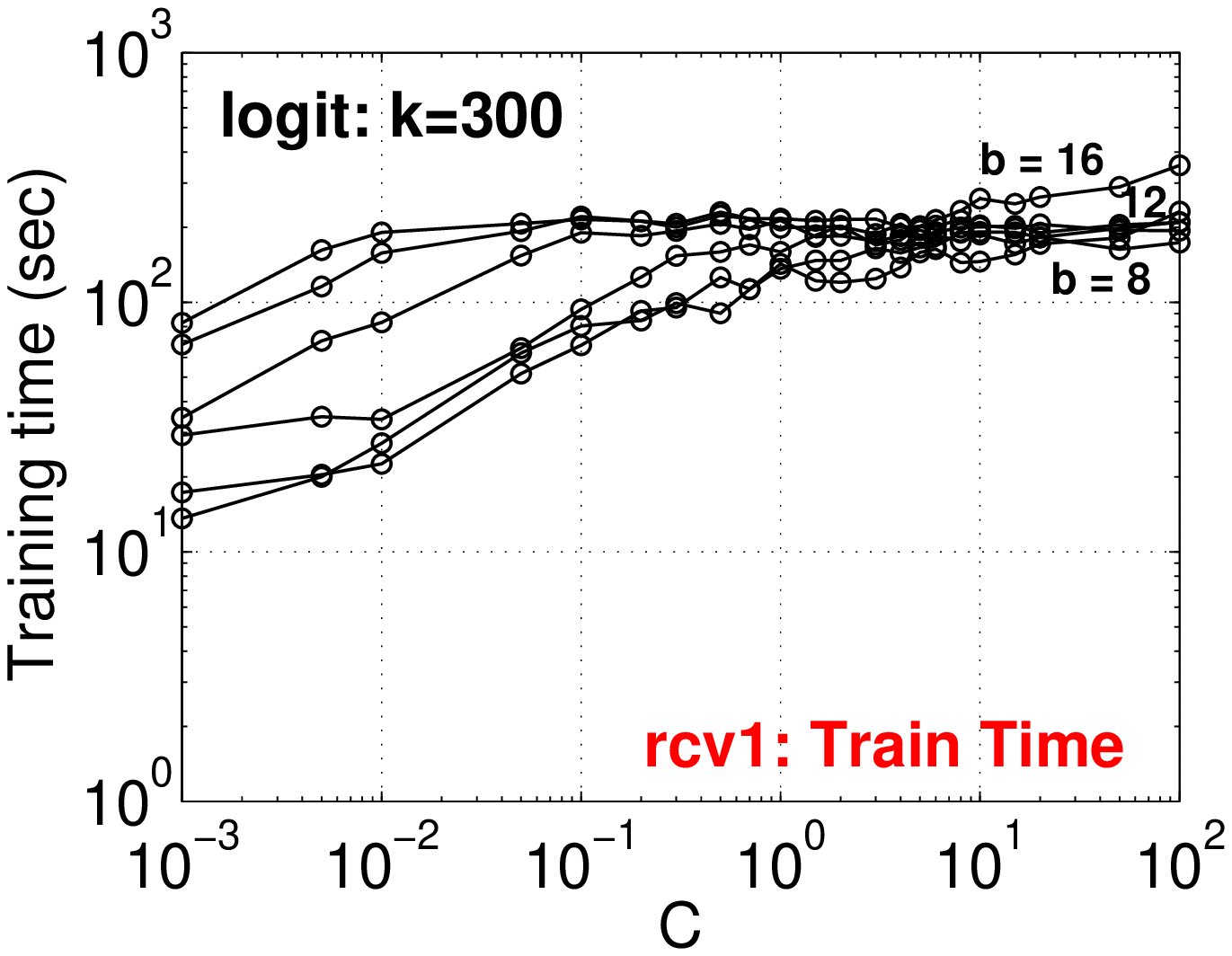}}
\end{center}
\vspace{-0.25in}

\caption{\textbf{Logistic regression training time on rcv1}.}\label{fig_rcv1_train_logit}
\end{figure}

Figure~\ref{fig_rcv1_acc_logit} and Figure~\ref{fig_rcv1_train_logit} respectively present the test accuracies and training times for training logistic regression. Again,  using merely $k=30$ and $b=12$, our method can achieve $>90\%$ test accuracies. With  $k\geq 300$, we can achieve $>95\%$ test accuracies.

To help understand the significance of these results, next we provide a comparison study with the  VW hashing algorithm~\cite{Article:Shi_JMLR09,Proc:Weinberger_ICML2009}.

\subsection{Comparisons with VW  Algorithm}

The Vowpal Wabbit (VW)  algorithm~\cite{Article:Shi_JMLR09,Proc:Weinberger_ICML2009} is an influential hashing method for data/dimension reduction. Since \cite{Proc:HashLearning_NIPS11} only compared $b$-bit minwise hashing with VW on a small dataset, it is more informative to conduct a comparison of the two algorithms on this much larger dataset (200GB). We experimented with VW using $k = 2^5$ to  $2^{14}$ hash bins (sample size). Note that $2^{14} = 16384$. It is difficult to train LIBLINEAR with $k=2^{15}$ because the training size of the hashed  data by VW is  close to 48 GB when $k=2^{15}$.

\begin{figure}[h!]
\begin{center}
\mbox{
\includegraphics[width=1.6in]{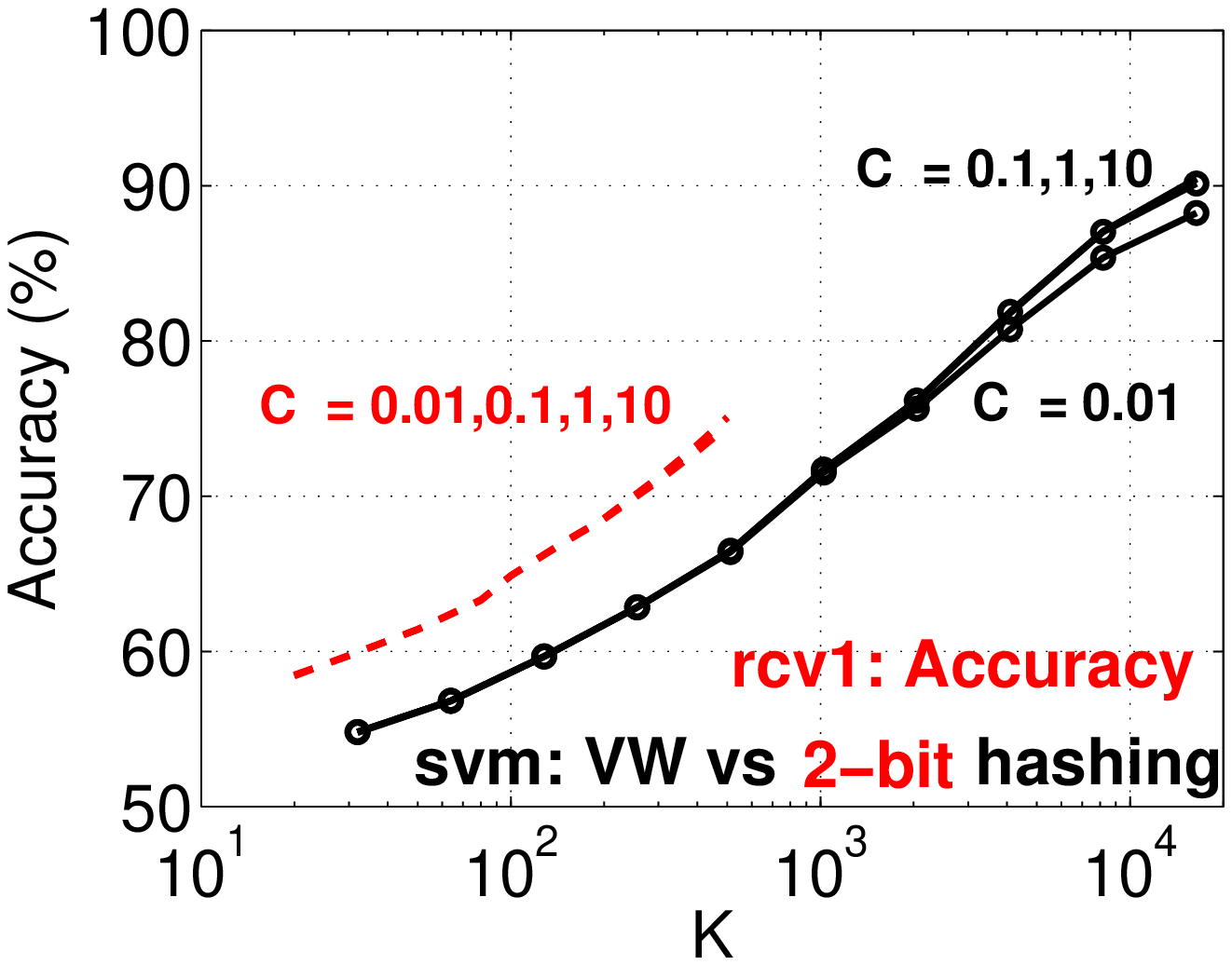}%\hspace{0.2in}
\includegraphics[width=1.6in]{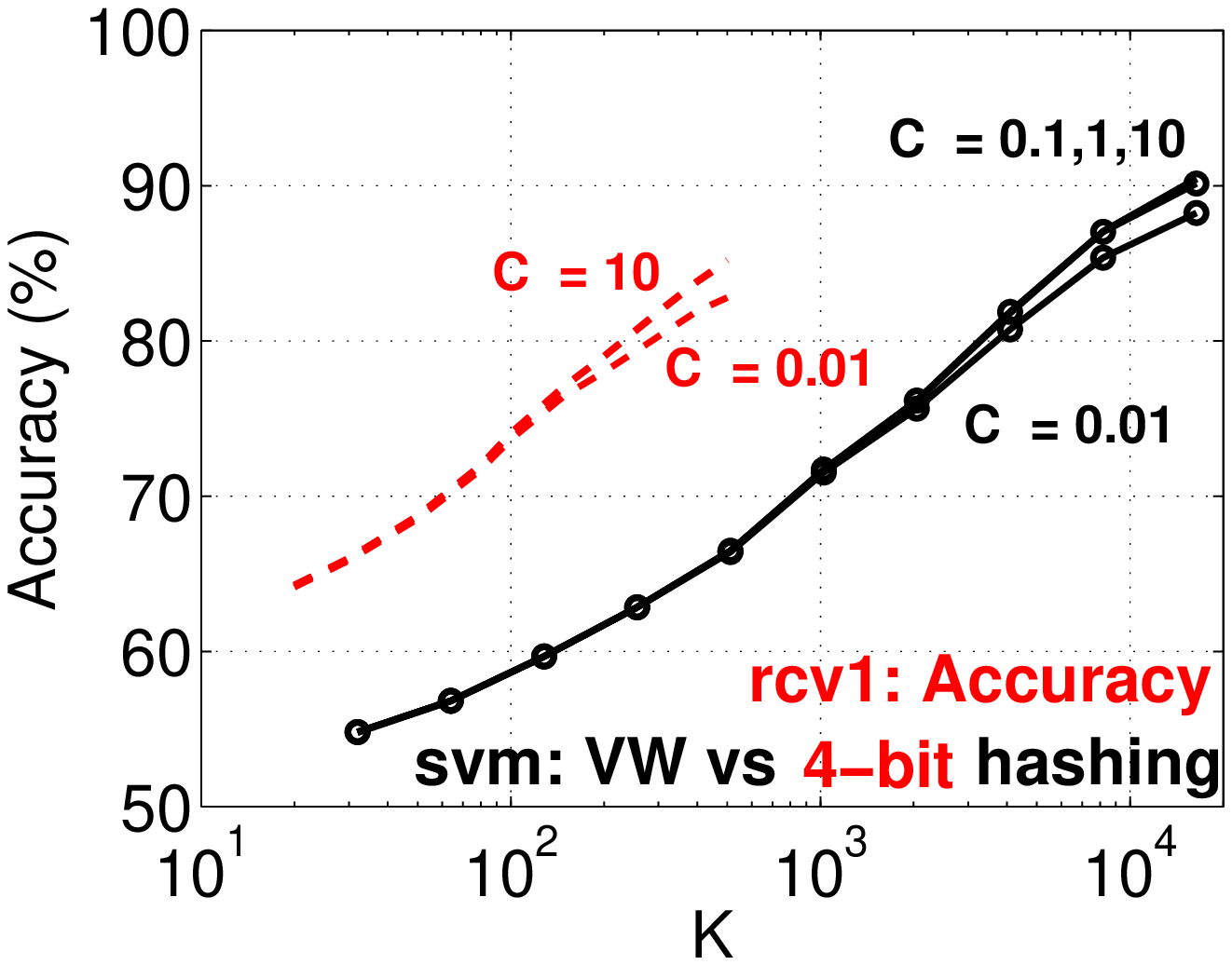}}

\mbox{
\includegraphics[width=1.6in]{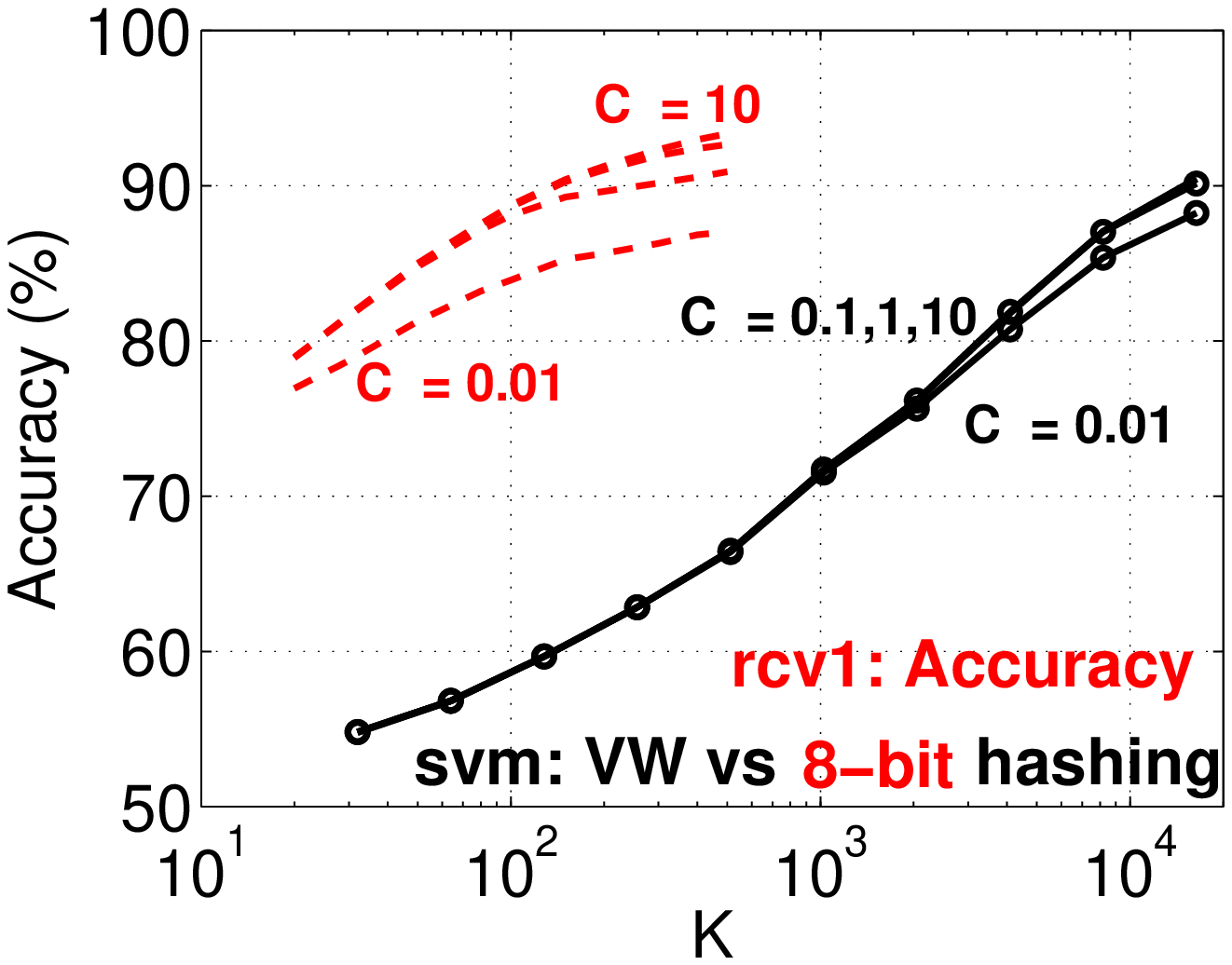}%\hspace{0.2in}
\includegraphics[width=1.6in]{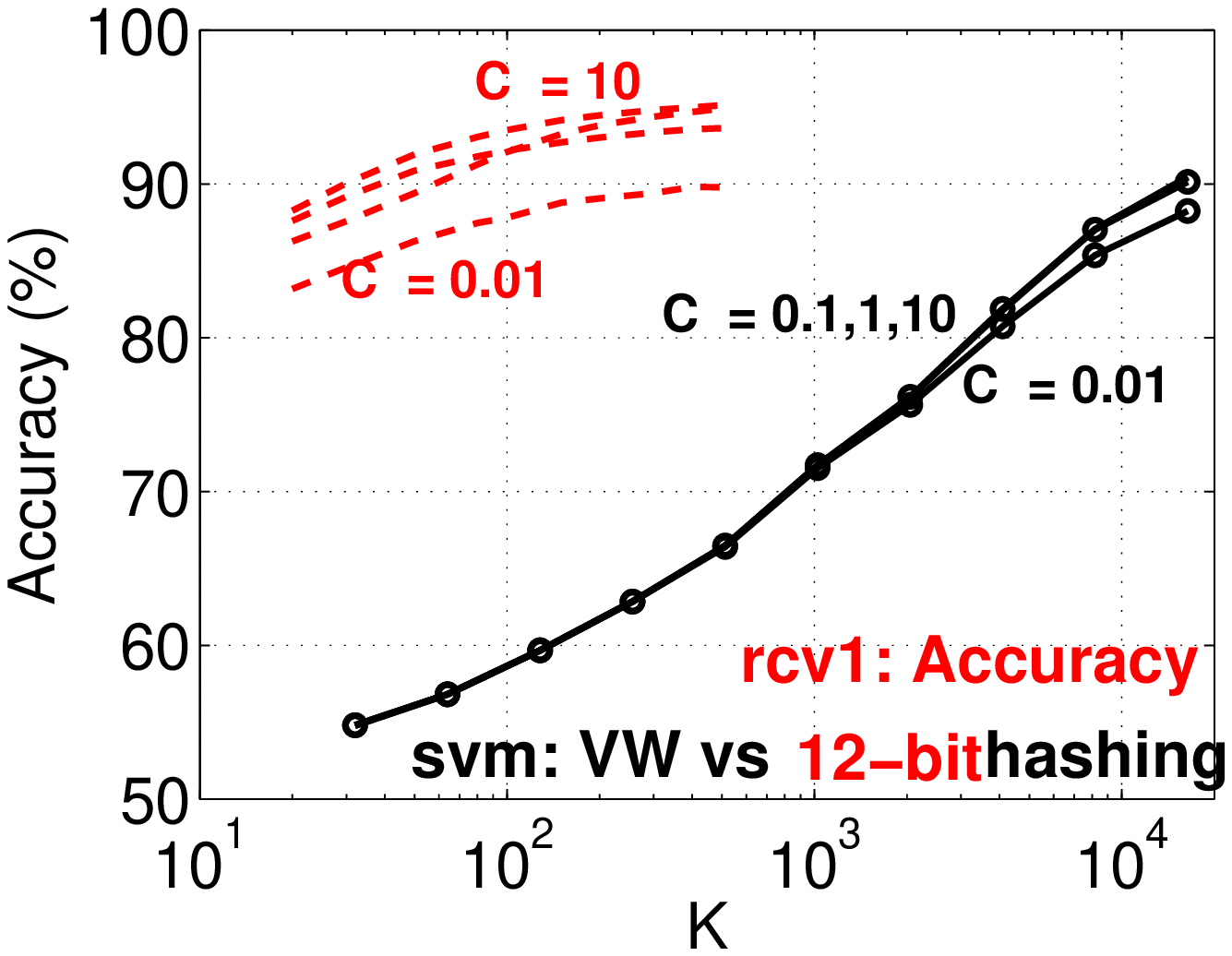}}%\hspace{0.2in}

\end{center}
\vspace{-0.3in}
\caption{\textbf{SVM test accuracy on rcv1} for comparing VW (solid) with $b$-bit minwise hashing (dashed). Each panel plots the same results for VW and results for $b$-bit minwise hashing for a different $b$. We select $C = 0.01, 0.1, 1, 10$. }\label{fig_rcv1_acc_vw}\vspace{-0.1in}
\end{figure}

\begin{figure}[h!]

\begin{center}
\mbox{
\includegraphics[width=1.6in]{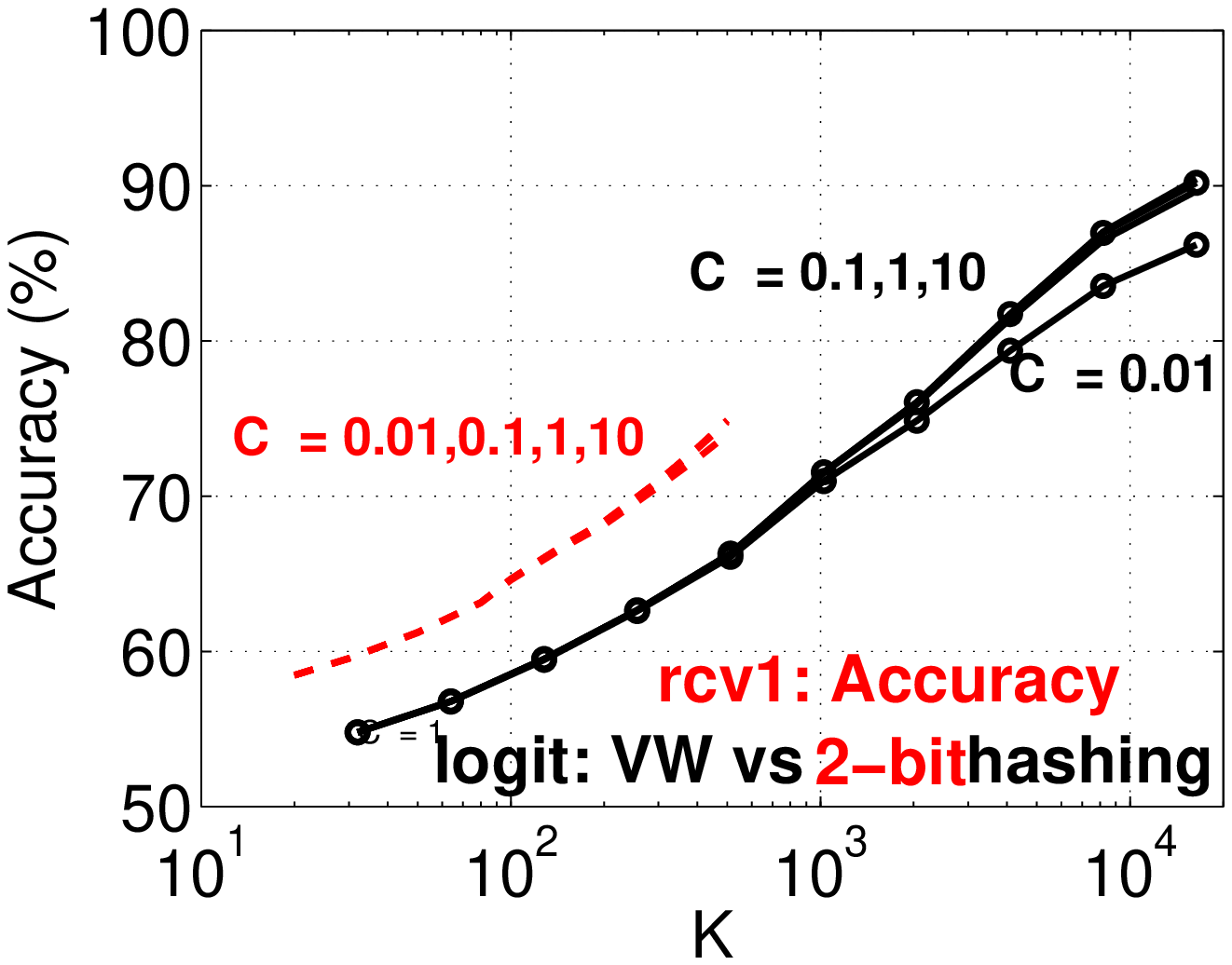}%\hspace{0.2in}
\includegraphics[width=1.6in]{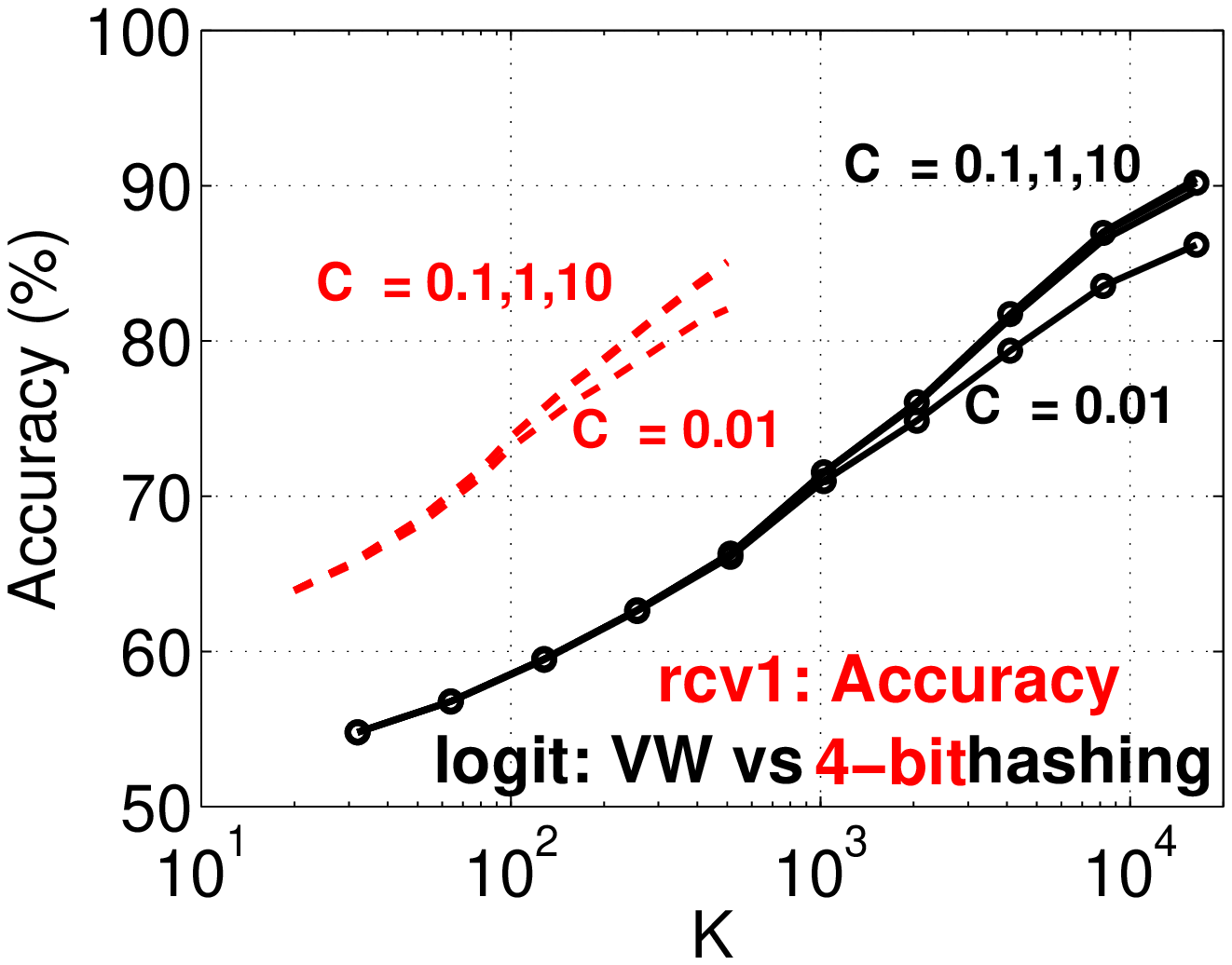}}

\mbox{
\includegraphics[width=1.6in]{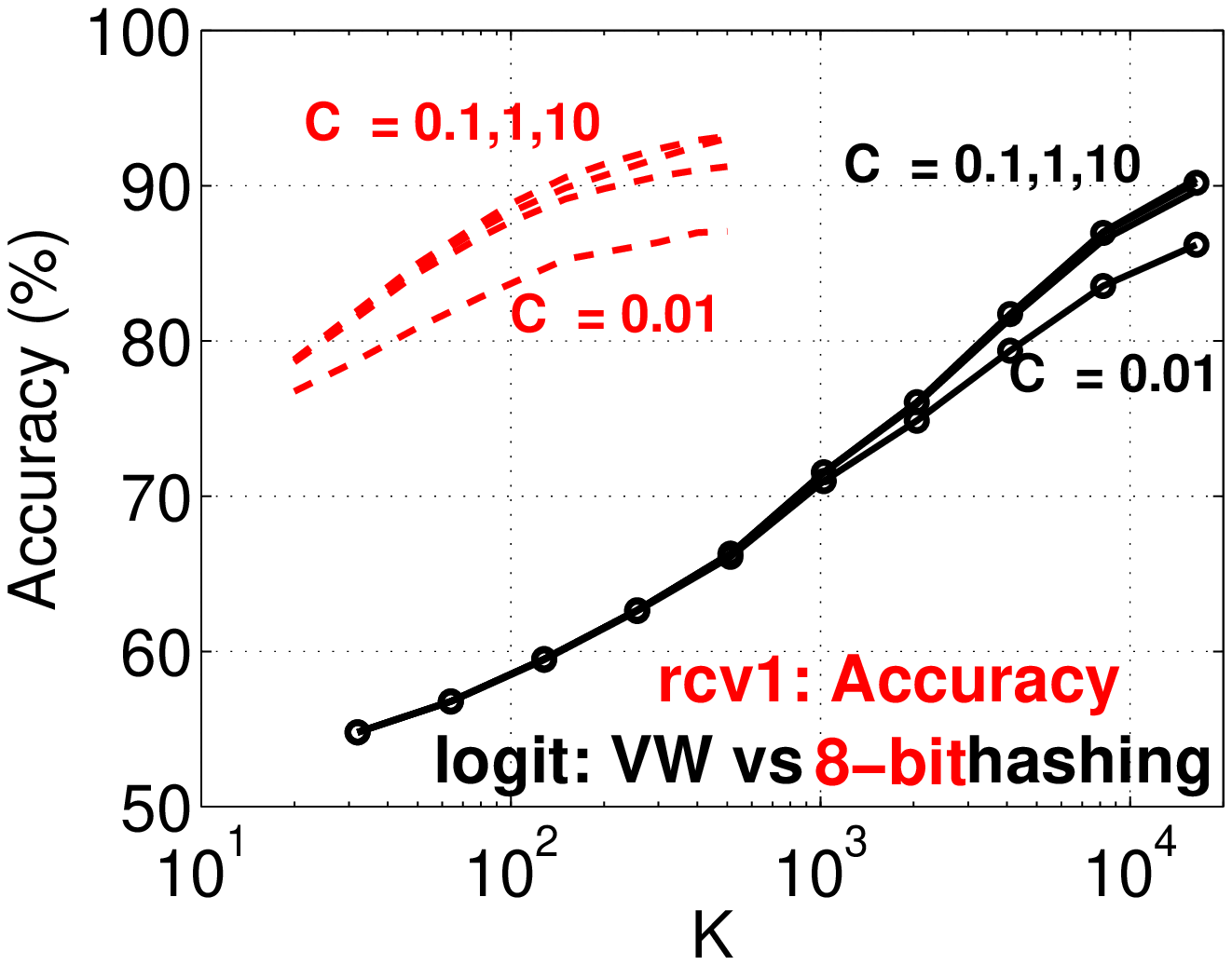}%\hspace{0.2in}
\includegraphics[width=1.6in]{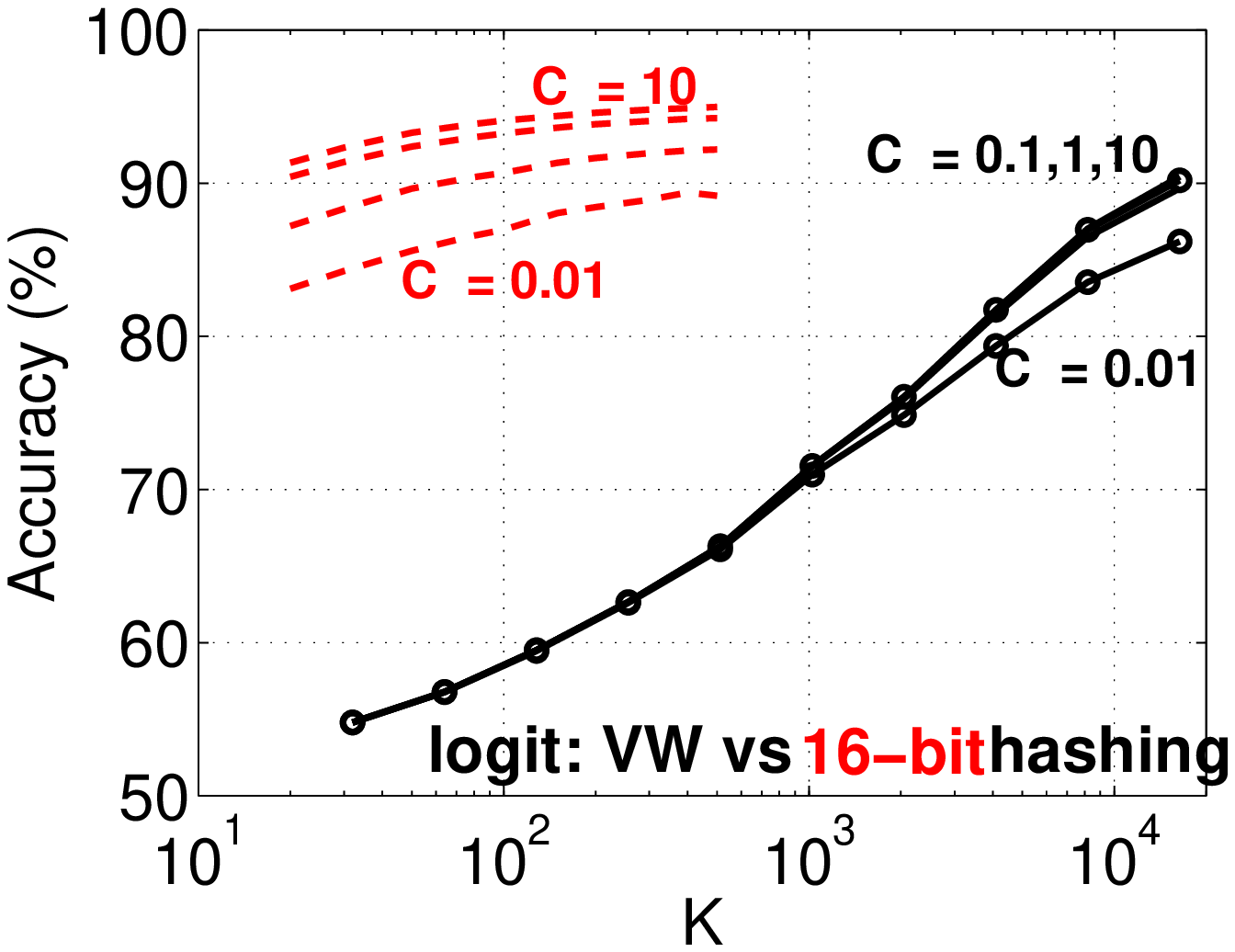}
}

\end{center}

\vspace{-0.3in}

\caption{\textbf{Logistic Regression test accuracy on rcv1} for comparing VW with $b$-bit minwise hashing. }\label{fig_rcv1_acc_vw_logit}\vspace{-0.in}
\end{figure}

Figure~\ref{fig_rcv1_acc_vw} and Figure~\ref{fig_rcv1_acc_vw_logit}  plot the test accuracies for SVM and logistic regression, respectively. In each figure, every panel has the same set of solid curves for VW but a different set of dashed curves for different values of $b$ in $b$-bit minwise hashing. Since the range of $k$ is very large, here we choose to present the test accuracies against $k$. Representative $C$ values (0.01, 0.1, 1, 10) are selected for the presentations.

From Figures~\ref{fig_rcv1_acc_vw} and~\ref{fig_rcv1_acc_vw_logit}, we can see clearly that $b$-bit minwise hashing is substantially more accurate than VW at the same storage. In other words, in order to achieve the same accuracy, VW will require substantially more storage than $b$-bit minwise hashing. %In fact, from the figures, it looks almost like $1$-bit minwise hashing, at the same $k$, can achieve similar test accuracies as VW.

\begin{figure}[h!]

\begin{center}
\mbox{
\includegraphics[width=1.6in]{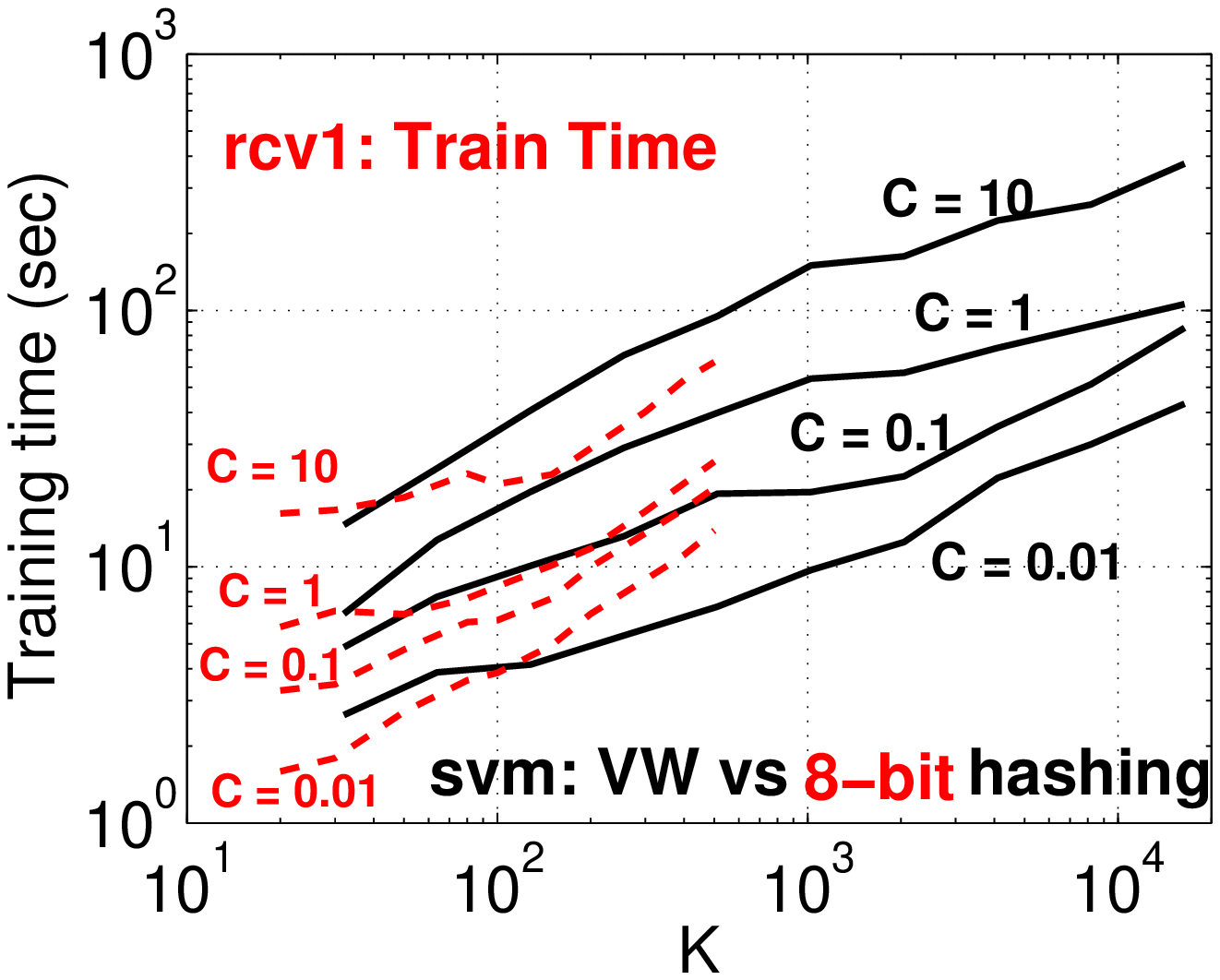}%\hspace{0.2in}
\includegraphics[width=1.6in]{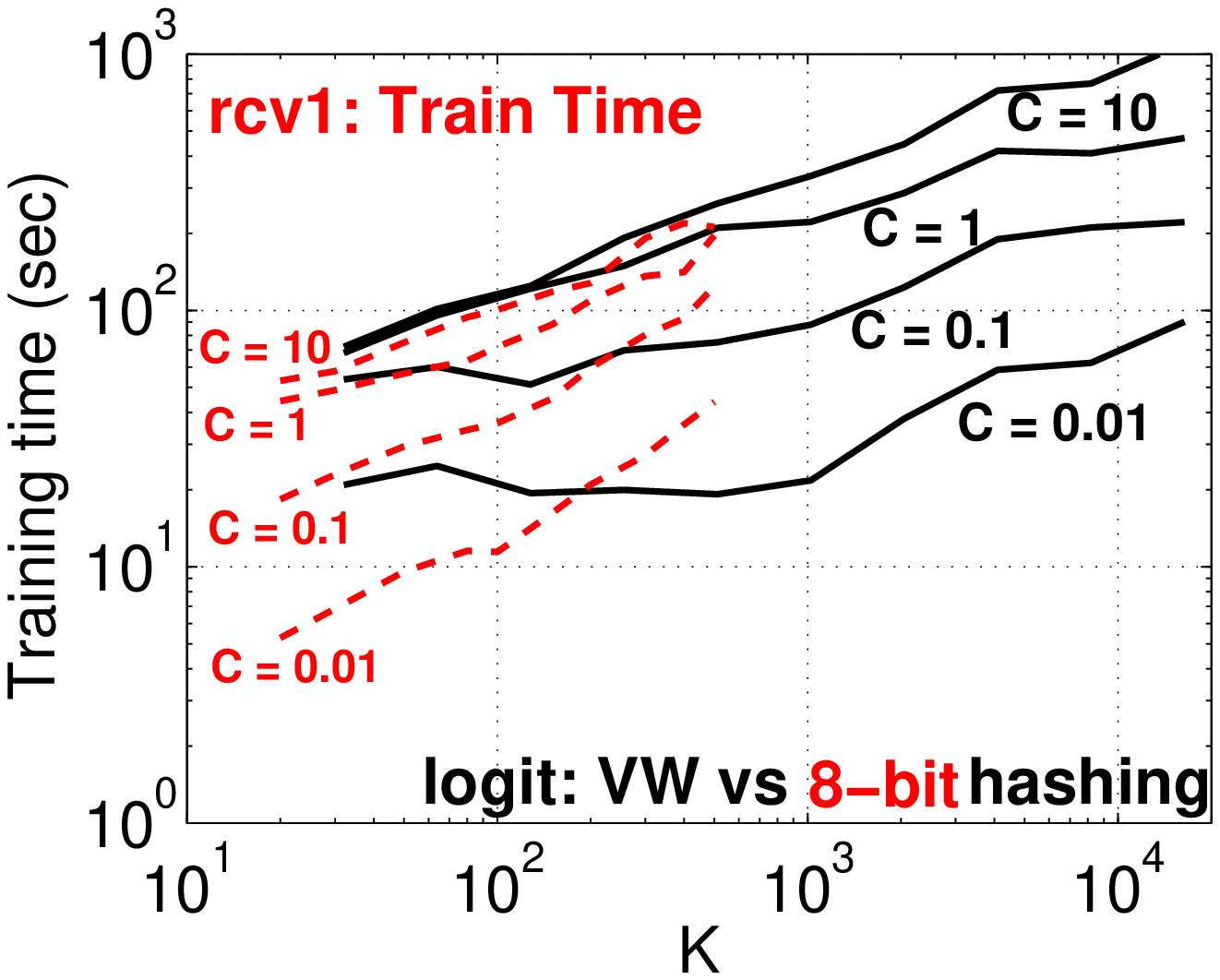}}
\end{center}

\vspace{-0.3in}

\caption{\textbf{Training time for SVM (left) and logistic regression (right) on rcv1} for comparing VW with $8$-bit minwise hashing. }\label{fig_rcv1_train_vw}\vspace{-0.1in}
\end{figure}

Figure~\ref{fig_rcv1_train_vw} presents the training times for comparing VW with $8$-bit minwise hashing. In this case, we can see that even at the same $k$, 8-bit hashing may have some computational advantages compared to VW. Of course, as it is clear that VW will require a much larger $k$ in order to achieve the same accuracies as 8-bit minwise hashing, we know that the advantage of $b$-bit minwise hashing in terms of training time reduction is also enormous.

Our comparison focuses on the VW hashing algorithm, not the VW online learning platform. The prior work~\cite{Proc:Chang_KDD11} experimented with the VW online learning platform on the {\em webspam} dataset and reported an accuracy of $98.42\%$ (compared to $>99.5\%$ in our experiments with $b$-bit hashing) after 597 seconds of training.

\section{Online Learning}

Batch learning algorithms (e.g., the LIBLINEAR package) face a challenge when the data do not fit in memory. In the context of search, the training datasets often far exceed the memory capacity of a single machine. One common solution (e.g.,~\cite{Proc:Yu_KDD10,Proc:Chang_KDD11}) is to partition the data into  blocks, which are repeatedly loaded into memory, to update the model coefficients. However,  this does not solve the computational bottleneck because loading the data blocks for many iterations consumes a large number of disk I/Os. $b$-bit minwise hashing provides a simple solution for high-dimensional data by substantially reducing the data size.

Another increasing popular solution is online learning, which requires only loading one feature vector at a time.  Here, we follow the notational convention used in the online learning literature (and the SGD code~\cite{URL:Bottou_SGD}). Given a training set $\{\mathbf{x}_i,y_i\}_{i=1}^n$, we consider the  following linear SVM optimization problem:
\begin{align}\label{eqn_SVM_online}
\min_{\mathbf{w}}\ \ \frac{\lambda}{2}\mathbf{w^Tw} + \frac{1}{n} \sum_{i=1}^n \max \left\{1 - y_i\mathbf{w^Tx_i},\ 0\right\}.
\end{align}
That is, we replace the  parameter $C$ in (\ref{eqn_SVM}) by $\lambda = \frac{1}{nC}$.

The stochastic gradient descent (SGD)~\cite{Proc:Bottou_COMPSTAT10,URL:Bottou_SGD} is a very popular algorithm for online learning. We modified Bottou's SGD code~\cite{URL:Bottou_SGD} to load one data point at a time. Basically, when the new data point $\{x_t,y_t\}$ arrives, the weights $\mathbf{w}$ are updated according to
\begin{align}
\mathbf{w}  \leftarrow \mathbf{w} -\eta_t\times\left\{\begin{array}{ll}
\lambda \mathbf{w} & \text{if } y_t\mathbf{w}x_t >1 \\
\lambda\mathbf{w} - y_t\mathbf{x}_t &\text{otherwise}
\end{array}\right.
\end{align}
where $\eta_t$ is the learning rate. In Bottou's SGD code, $\eta_t$ is initially set by a careful calibration step using a (small) subset of the examples and is updated at every new data point.

\begin{figure}[h!]
\begin{center}
\mbox{
\includegraphics[width=1.6in]{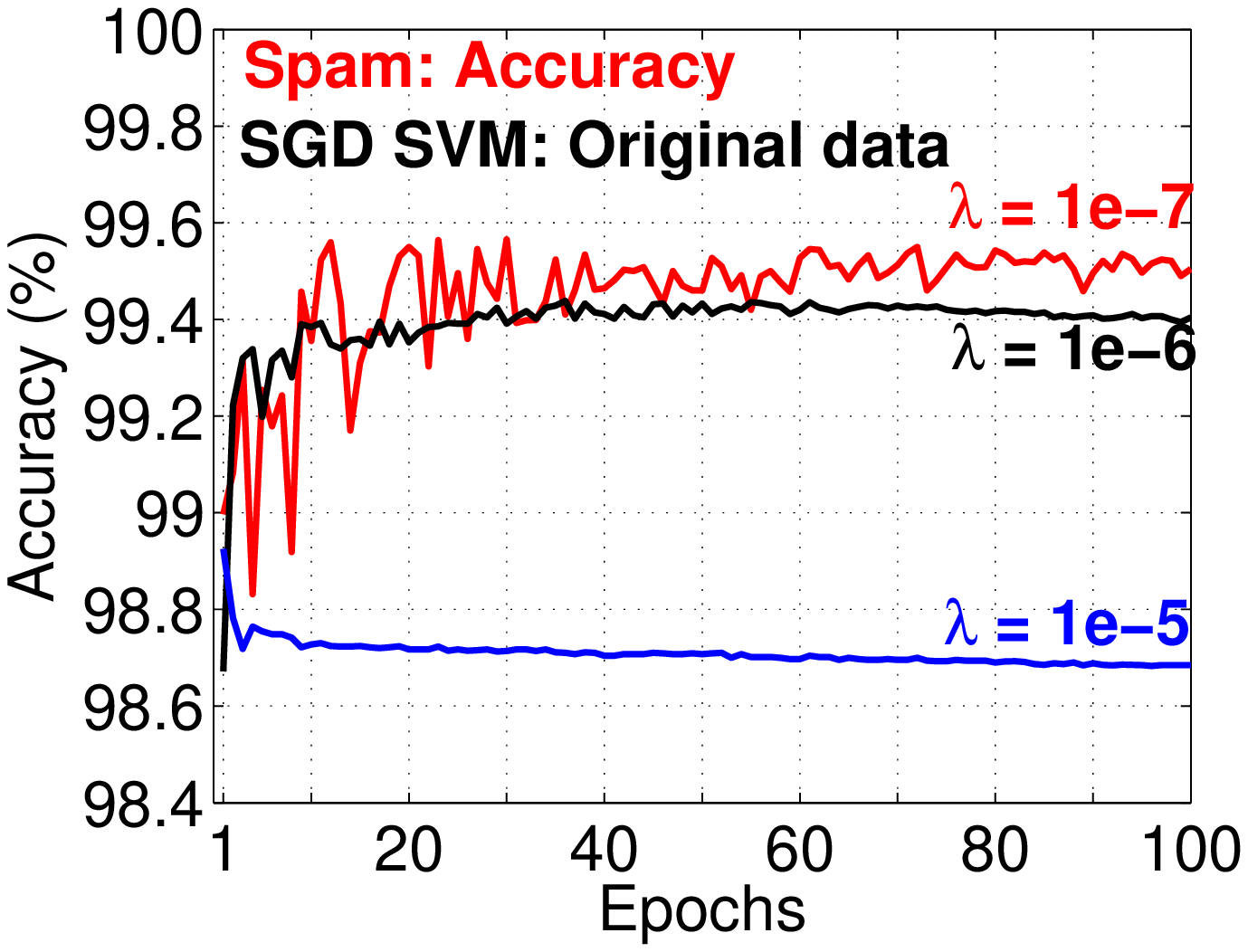}%\hspace{0.1in}
\includegraphics[width=1.6in]{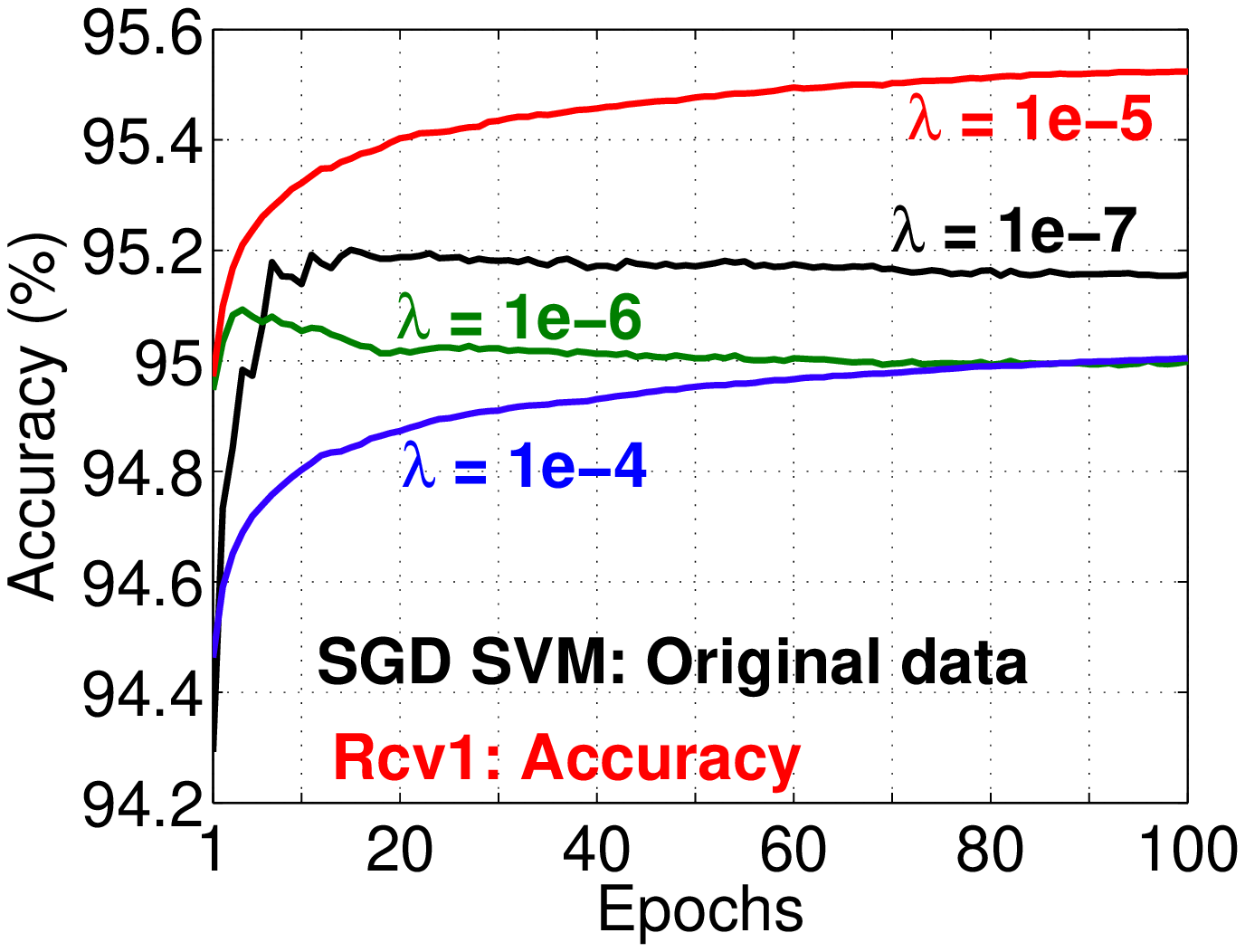}}
\end{center}
\vspace{-0.3in}
\caption{SGD SVM test accuracies on the original {\em webspam} (left panel) and {\em rcv1} (right panel) datasets. The results are somewhat sensitive to $\lambda$. At the best $\lambda$ values, the test accuracies increase with increasing number of epochs.}\label{fig_SGD_base}\vspace{-0.1in}
\end{figure}

It is often observed that the test accuracy  improves with increasing number of epoches (one epoch means one pass of the data). For example, Figure~\ref{fig_SGD_base} shows that we need about 60 epochs for both {\em webspam} and {\em rcv1} datasets  to reach (fairly) stable predictions.

\subsection{SGD SVM Results on Webspam}

\begin{figure}[h!]
\begin{center}
\mbox{
\includegraphics[width=1.6in]{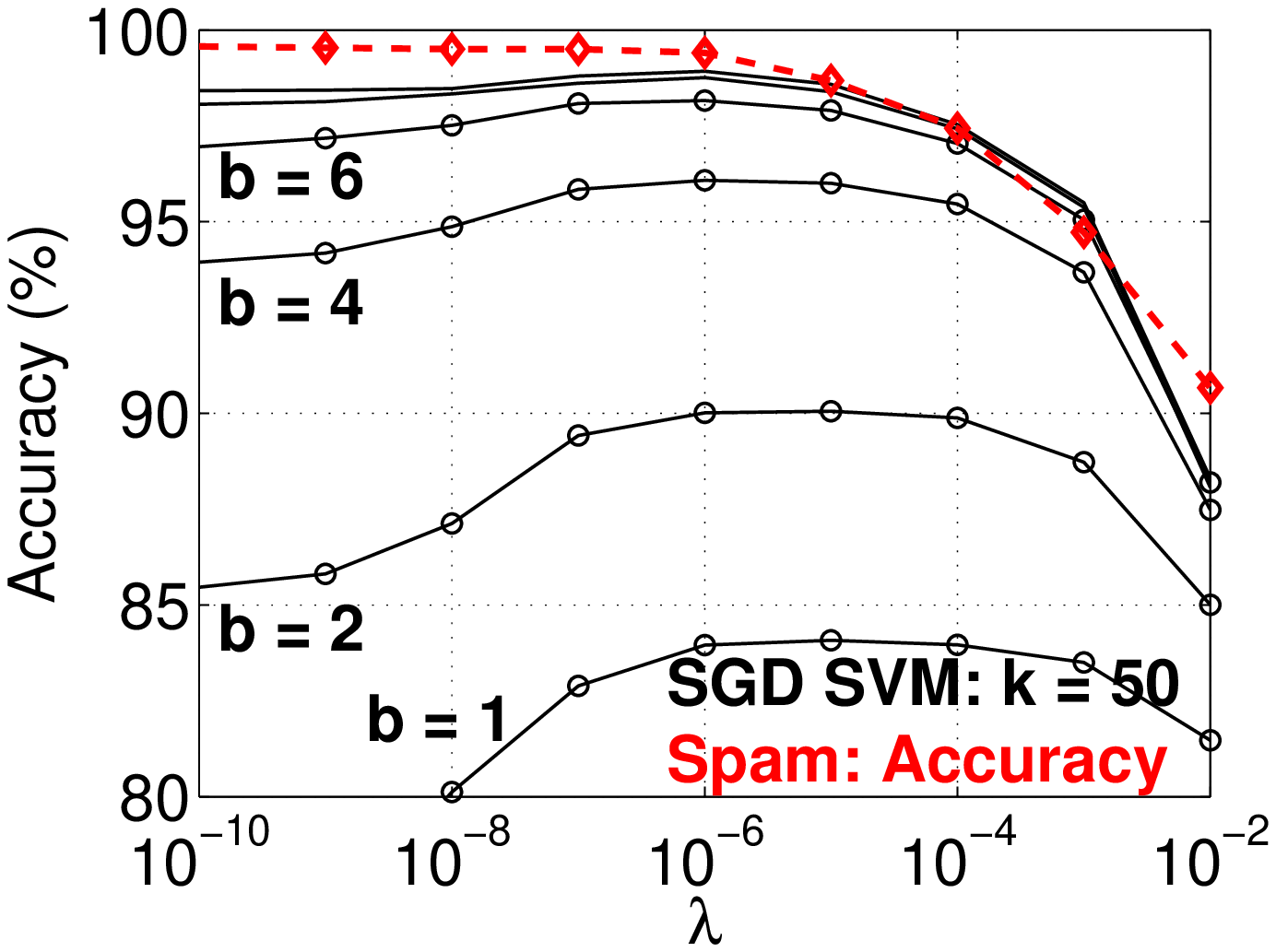}%\hspace{0.1in}
\includegraphics[width=1.6in]{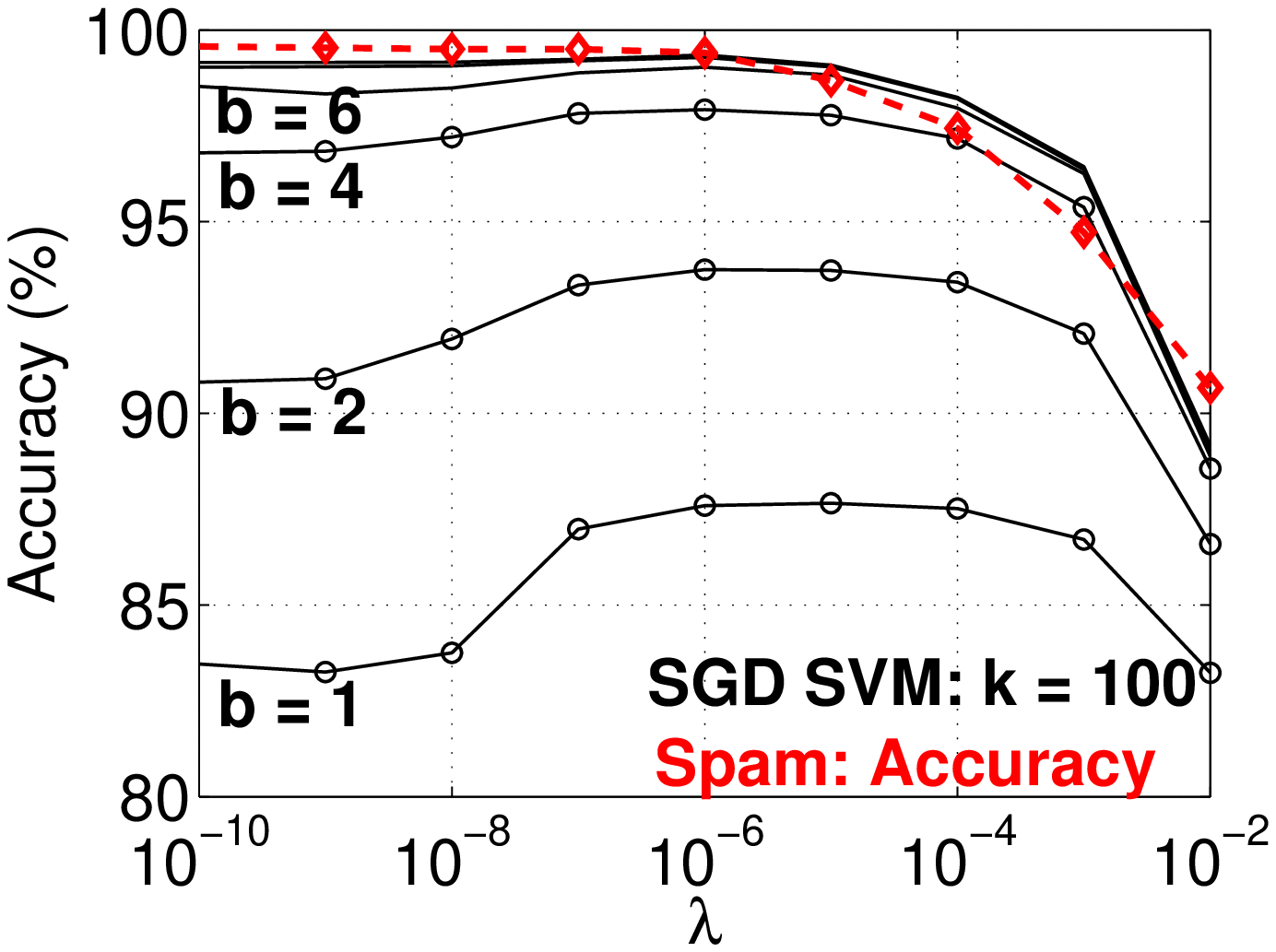}}

\mbox{
\includegraphics[width=1.6in]{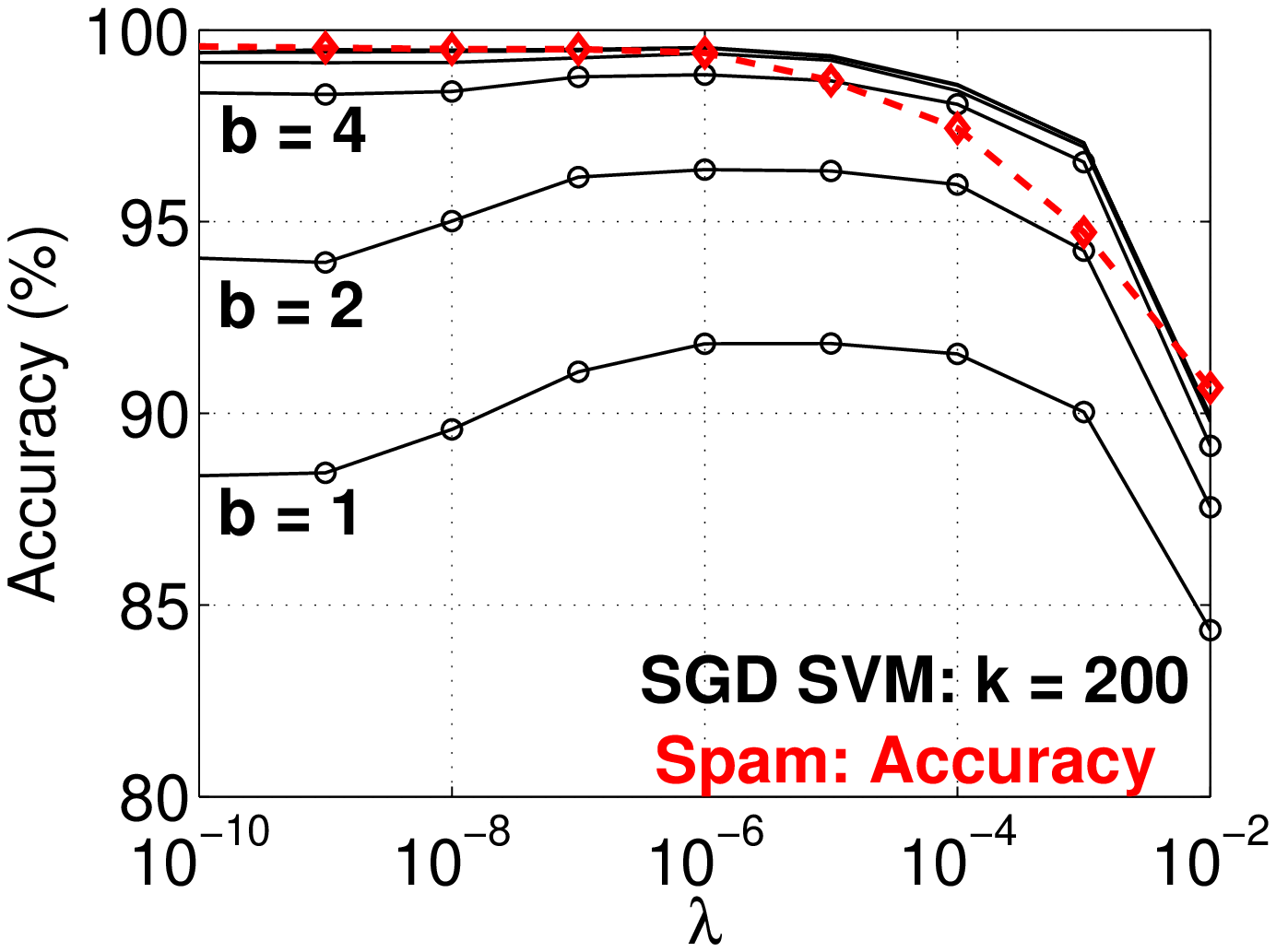}%\hspace{0.1in}
\includegraphics[width=1.6in]{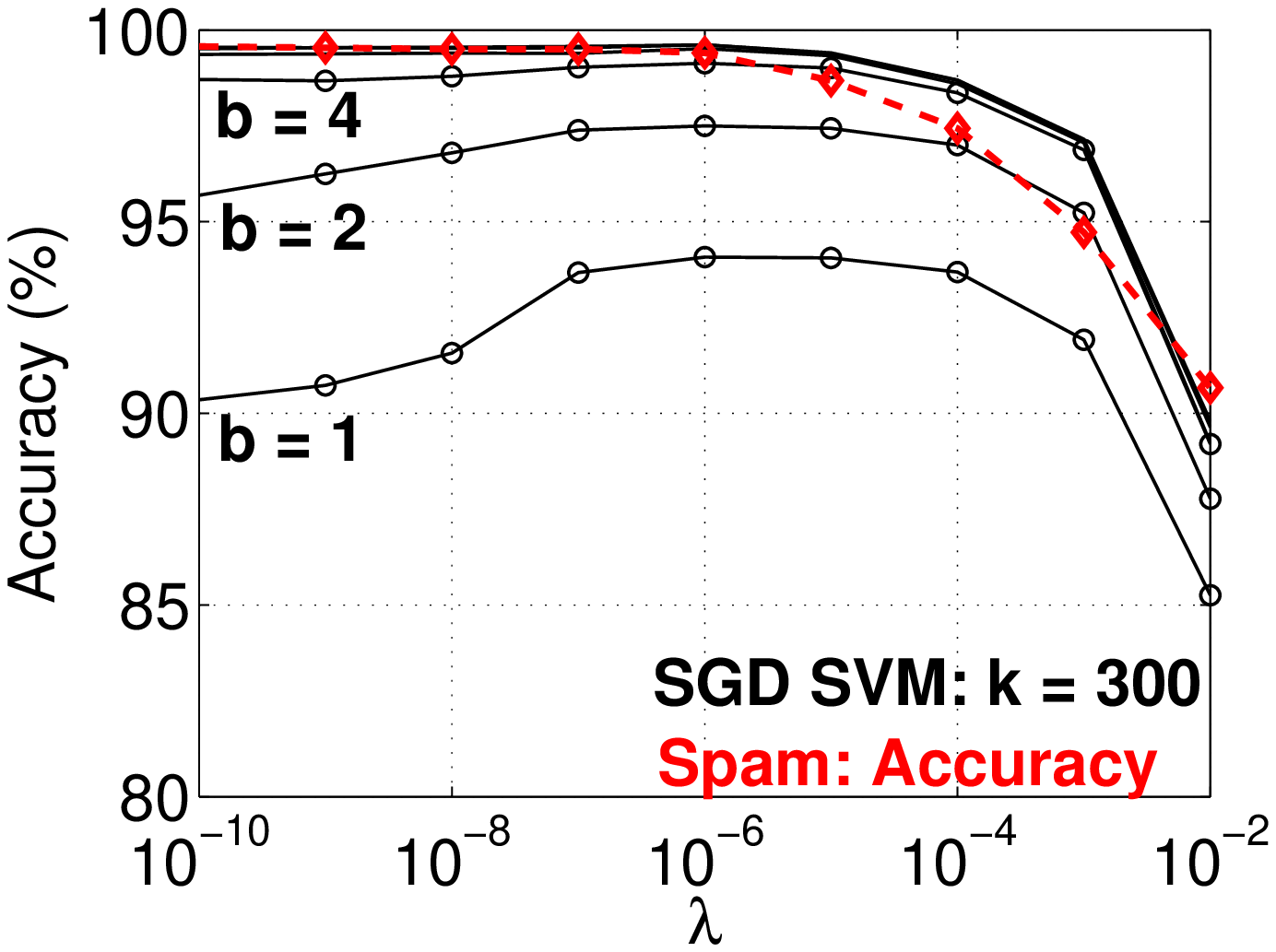}}

\mbox{
\includegraphics[width=1.6in]{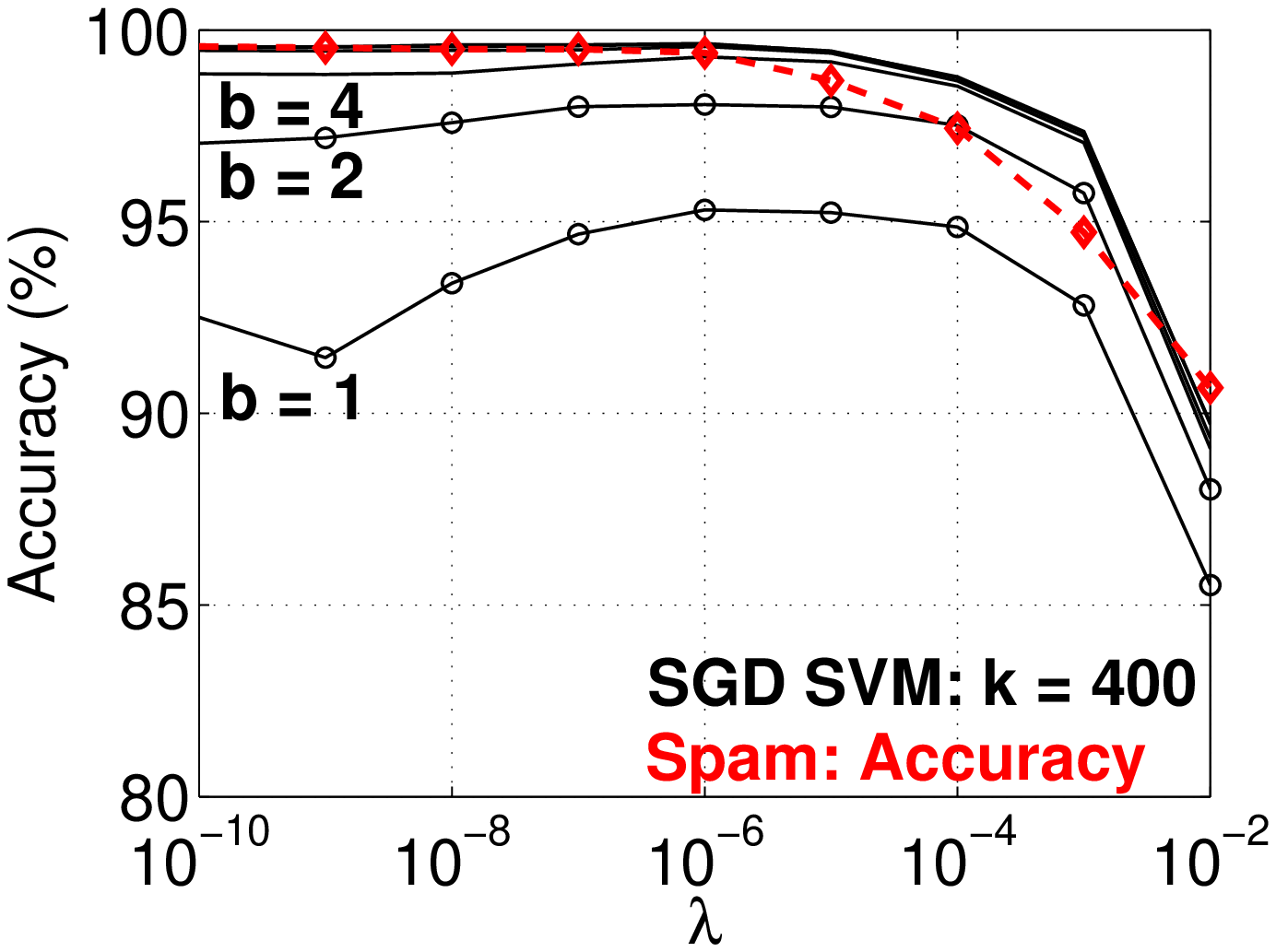}%\hspace{0.1in}
\includegraphics[width=1.6in]{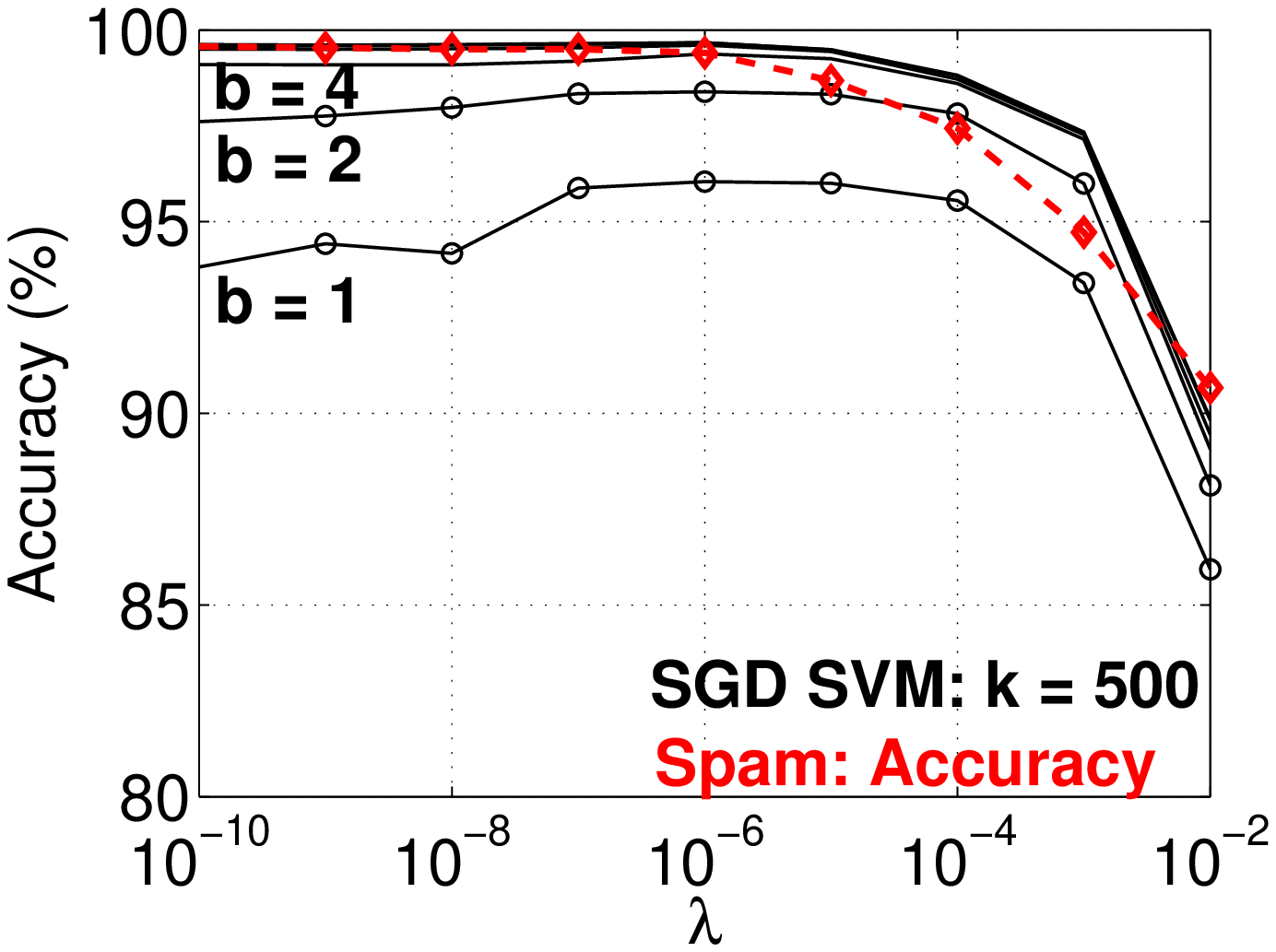}}

\end{center}

\vspace{-0.3in}

\caption{Test accuracies of SGD SVM on {\em webspam}  at the 100th epoch, for both the original data and the $b$-bit hashed data. When $k\geq 200$ and $b\geq 8$, $b$-bit minwise hashing achieves similar accuracies as using the original data.}\label{fig_spam_SGD}
\end{figure}

\begin{figure}[h!]
\begin{center}
\mbox{
\includegraphics[width=1.6in]{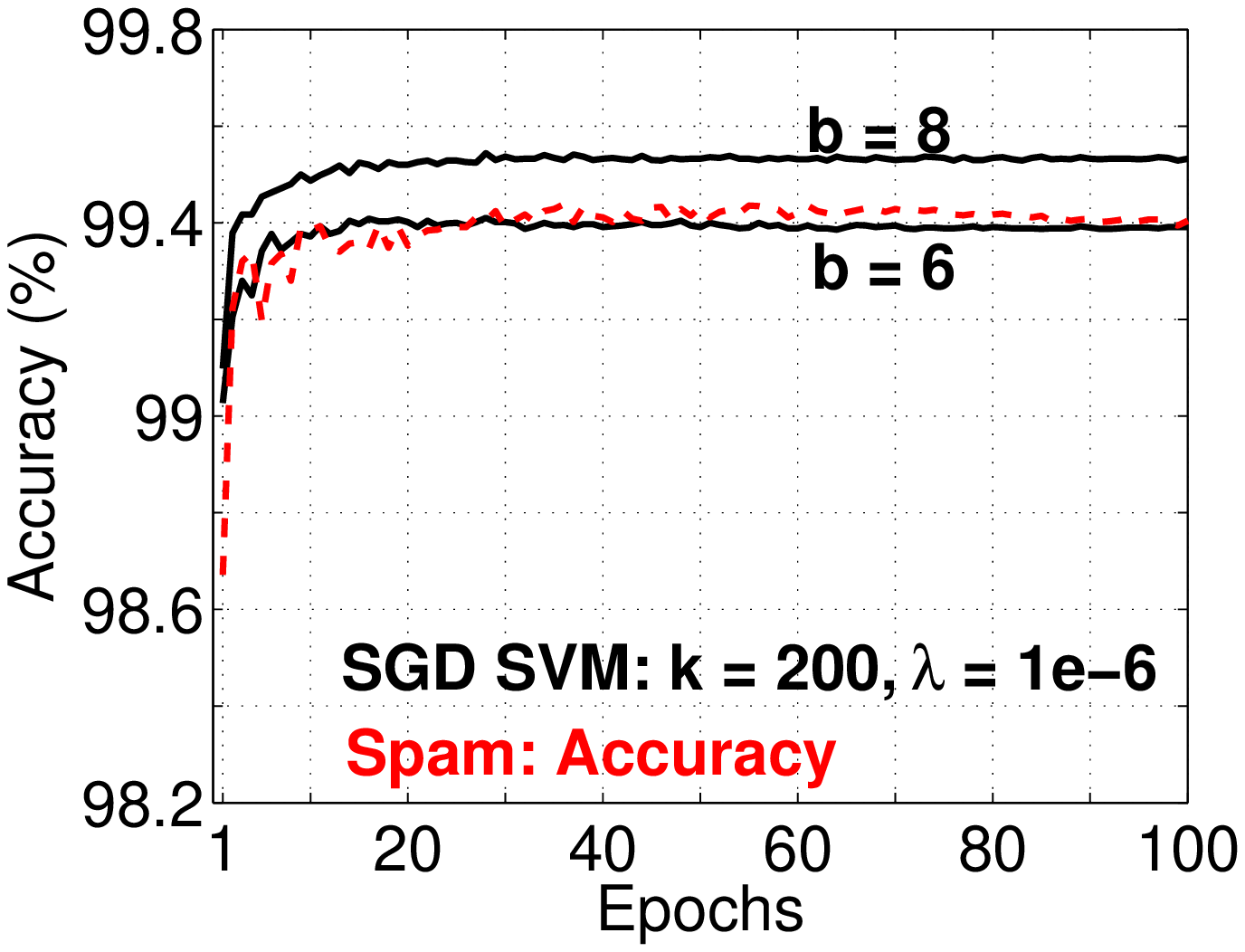}%\hspace{0.1in}
\includegraphics[width=1.6in]{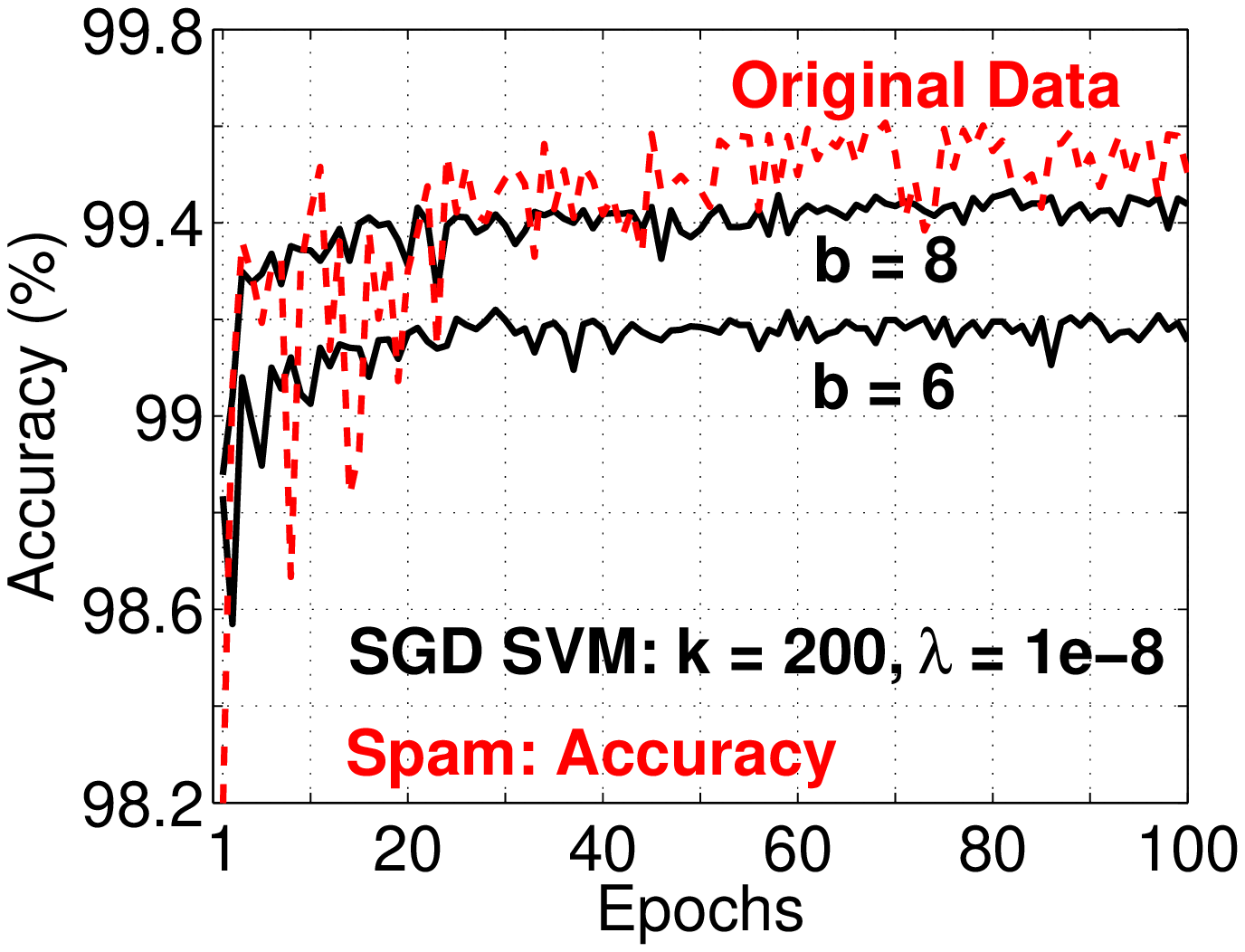}}
\end{center}
\vspace{-0.3in}
\caption{Test accuracies of SGD SVM on {\em webspam}  versus epochs for two selected $\lambda$ values. }\label{fig_spam_SGD_Hist}\vspace{-0.1in}
\end{figure}

Figure~\ref{fig_spam_SGD} presents the test accuracies versus the regularization parameter $\lambda$, at the last (100th) epoch.  When $b\geq 8$ and $k\geq 200$, using $b$-bit hashing can achieve similar test accuracies as using the original data. Figure~\ref{fig_spam_SGD_Hist} illustrates  the test accuracies versus epochs, for two selected $\lambda$ values. Perhaps 20 epochs are sufficient for reaching a sufficient accuracy using $b$-bit hashing.

\begin{figure}[h!]
\begin{center}
\mbox{
\includegraphics[width=1.6in]{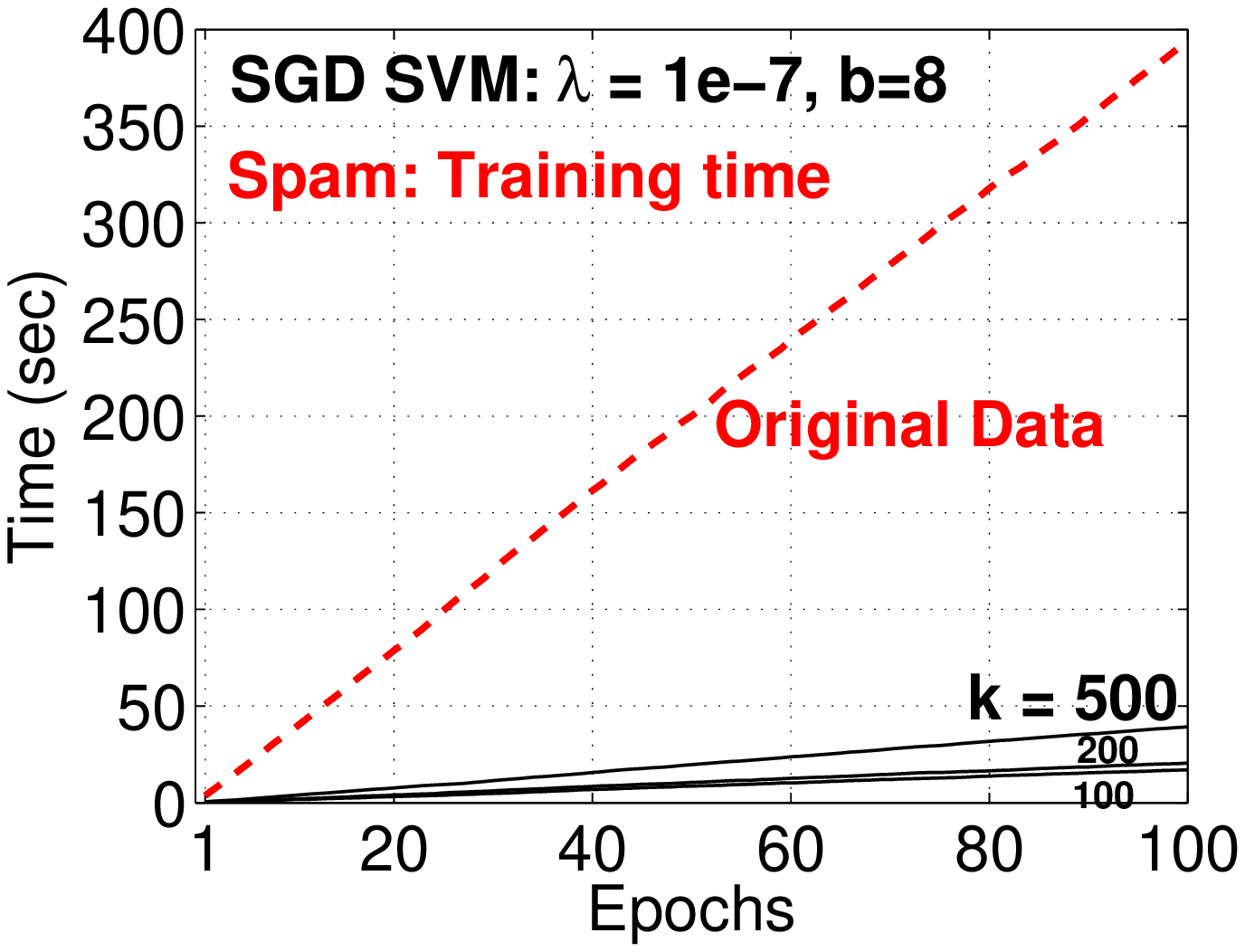}%\hspace{0.1in}
\includegraphics[width=1.6in]{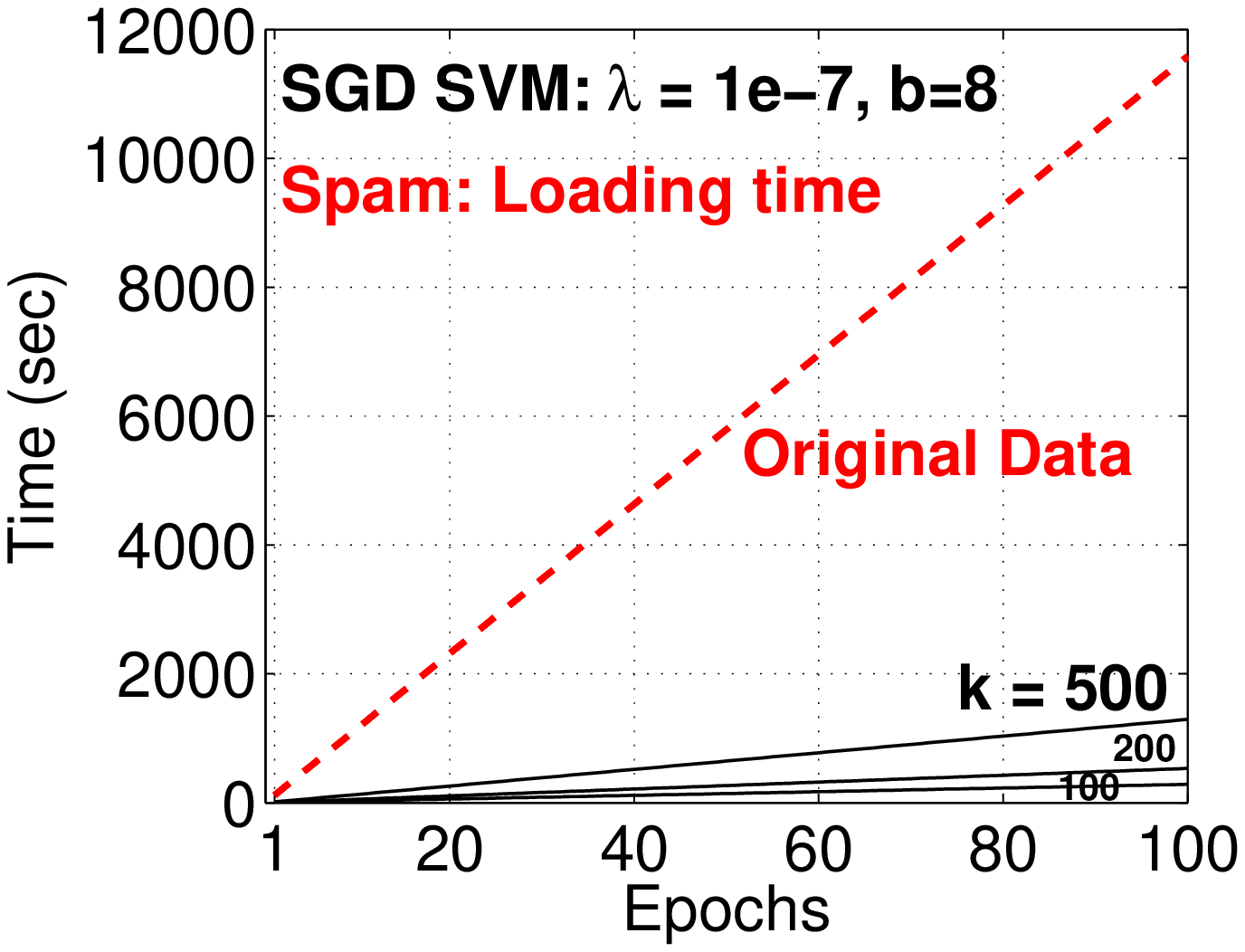}}
\end{center}

\vspace{-0.3in}

\caption{Training time  and data loading time on {\em webspam} versus epochs, for both the original data (dashed curves) and $8$-bit hashed data (solid curves, for $k=500,200,100$).  $b$-bit hashing substantially reduces the training  and loading times. }\label{fig_spam_SGD_time}
\end{figure}

Figure~\ref{fig_spam_SGD_time} plots the training time and loading time for both using the original data and using $8$-bit minwise hashing. On average, using the original data takes about 10 times more time than using $8$-bit hashed data, as reflected in Table~\ref{tab_time_ratio}. Also, clearly the data loading time dominates the cost.

Because the data loading time dominates the cost of online learning, in our SGD experiments, we always first converted the data into binary format as opposed to the LIBSVM format used in our batch learning experiments. All the reported data loading times in this section were based on binary data. We would like to thank Leon Bottou for the highly helpful communications in this matter.

%\vspace{-0.1in}

\subsection{SGD SVM Results on Rcv1}
\begin{figure}[h!]
\begin{center}
\mbox{
\includegraphics[width=1.6in]{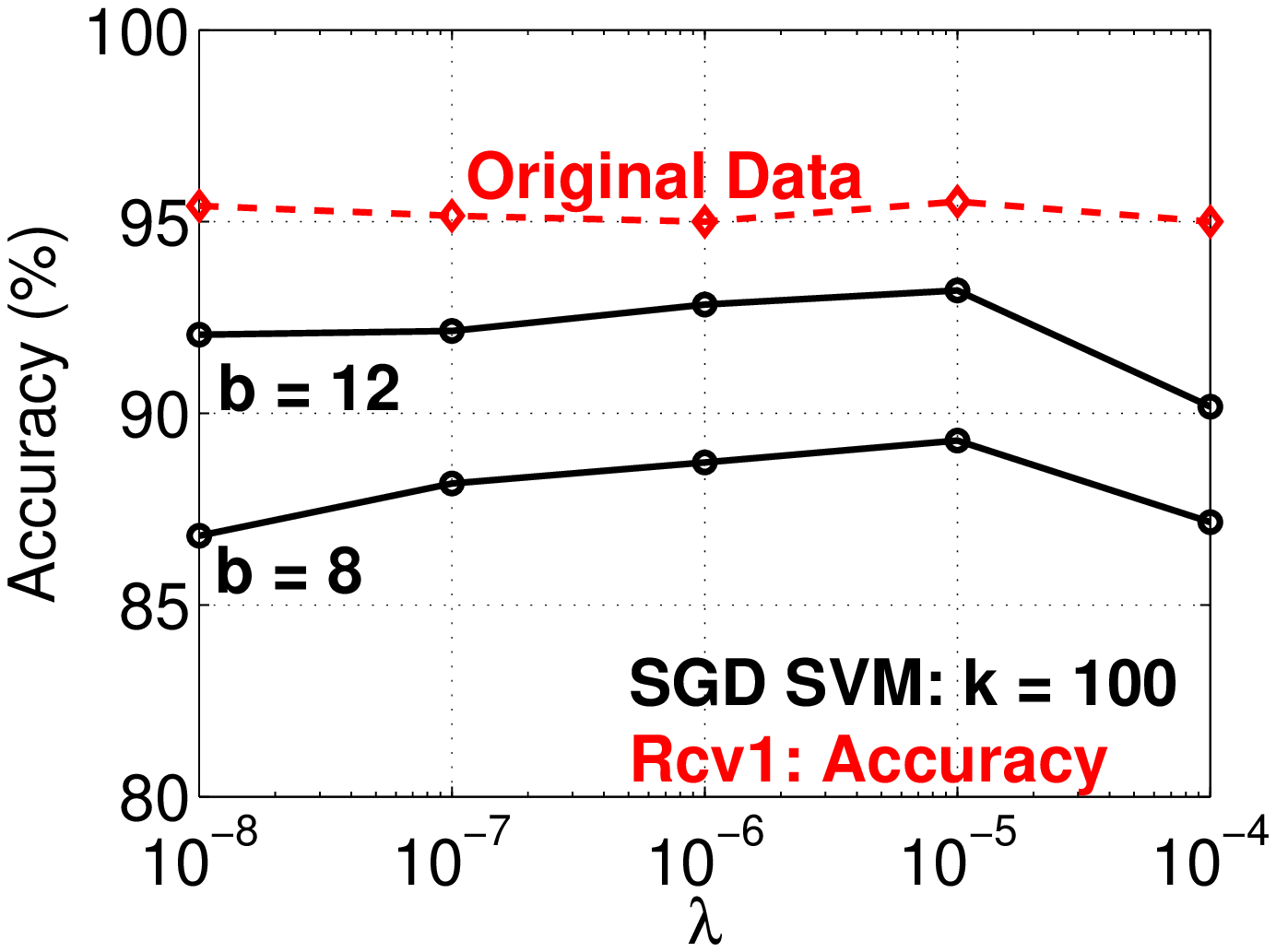}%\hspace{0.1in}
\includegraphics[width=1.6in]{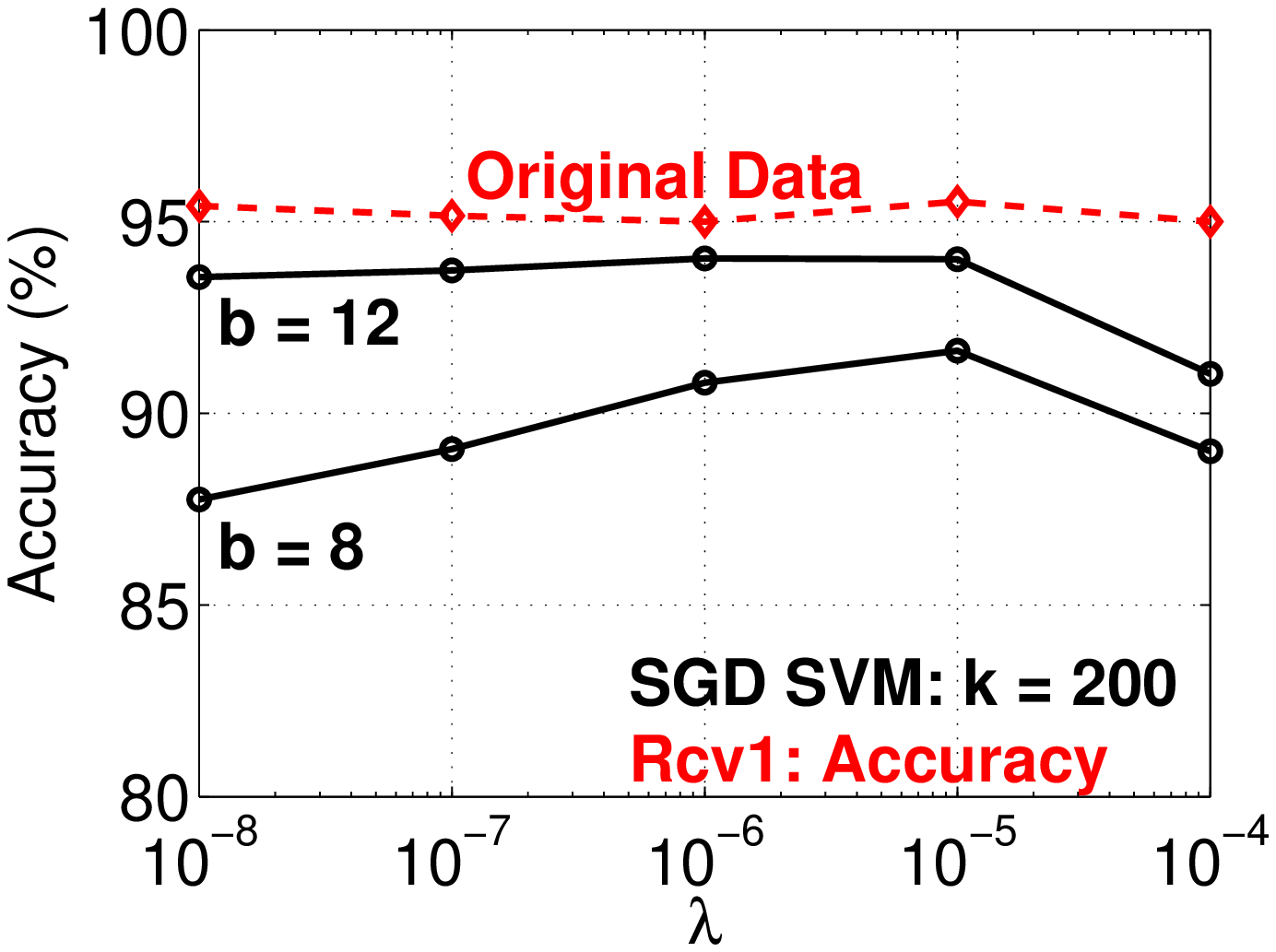}}

\mbox{
\includegraphics[width=1.6in]{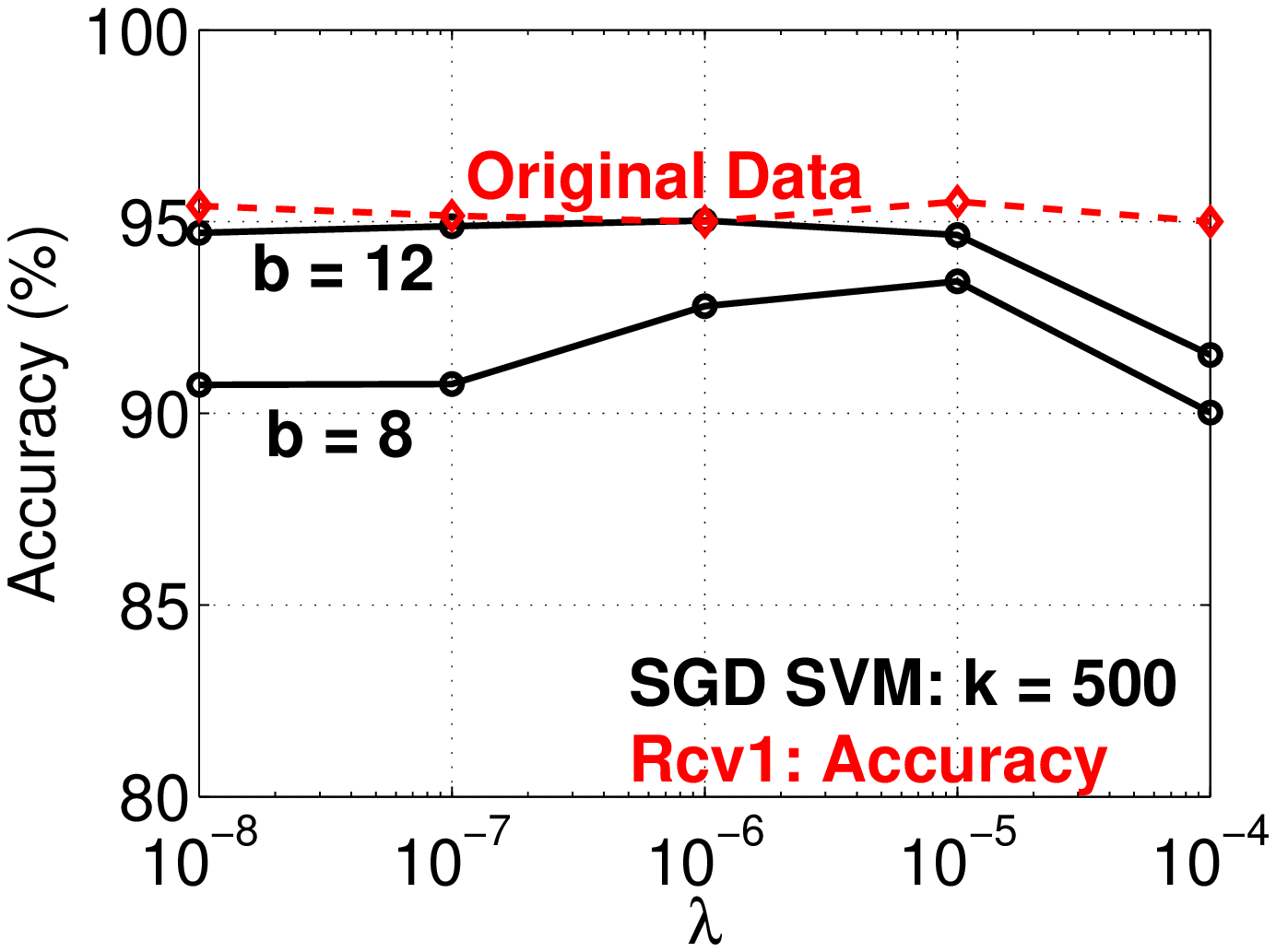}%\hspace{0.1in}
\includegraphics[width=1.6in]{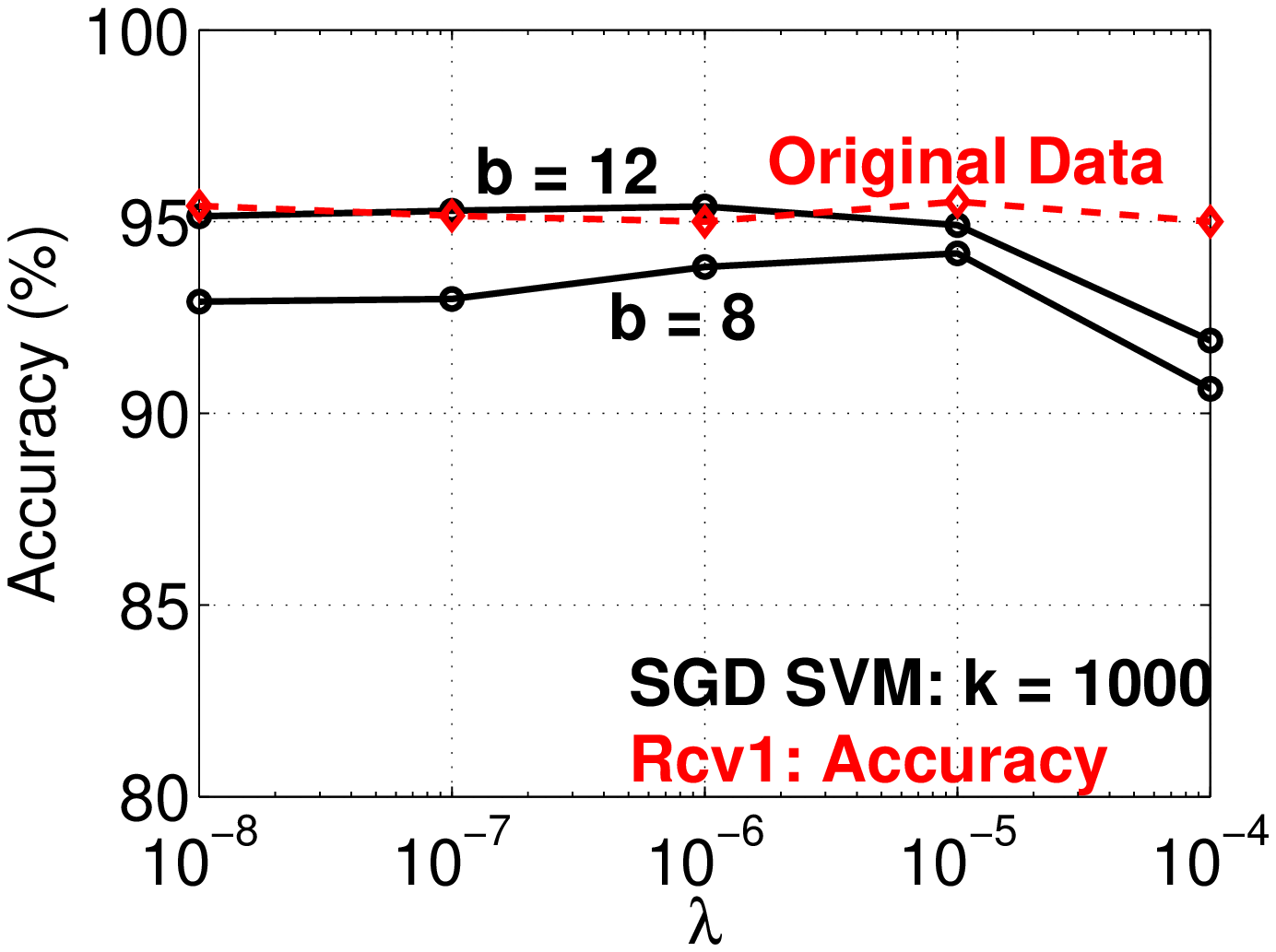}}

\end{center}

\vspace{-0.3in}

\caption{ Test accuracies of SGD SVM on {\em rcv1}  at the 100th epoch, for both the original data and the $b$-bit ($b=8$ and $b=12$) hashed data. When $k\geq 500$ and $b= 12$, $b$-bit minwise hashing achieves similar accuracies as using the original data.}\label{fig_rcv1_SGD}\vspace{-0.in}
\end{figure}

\begin{figure}[h!]
\begin{center}
\mbox{
\includegraphics[width=1.6in]{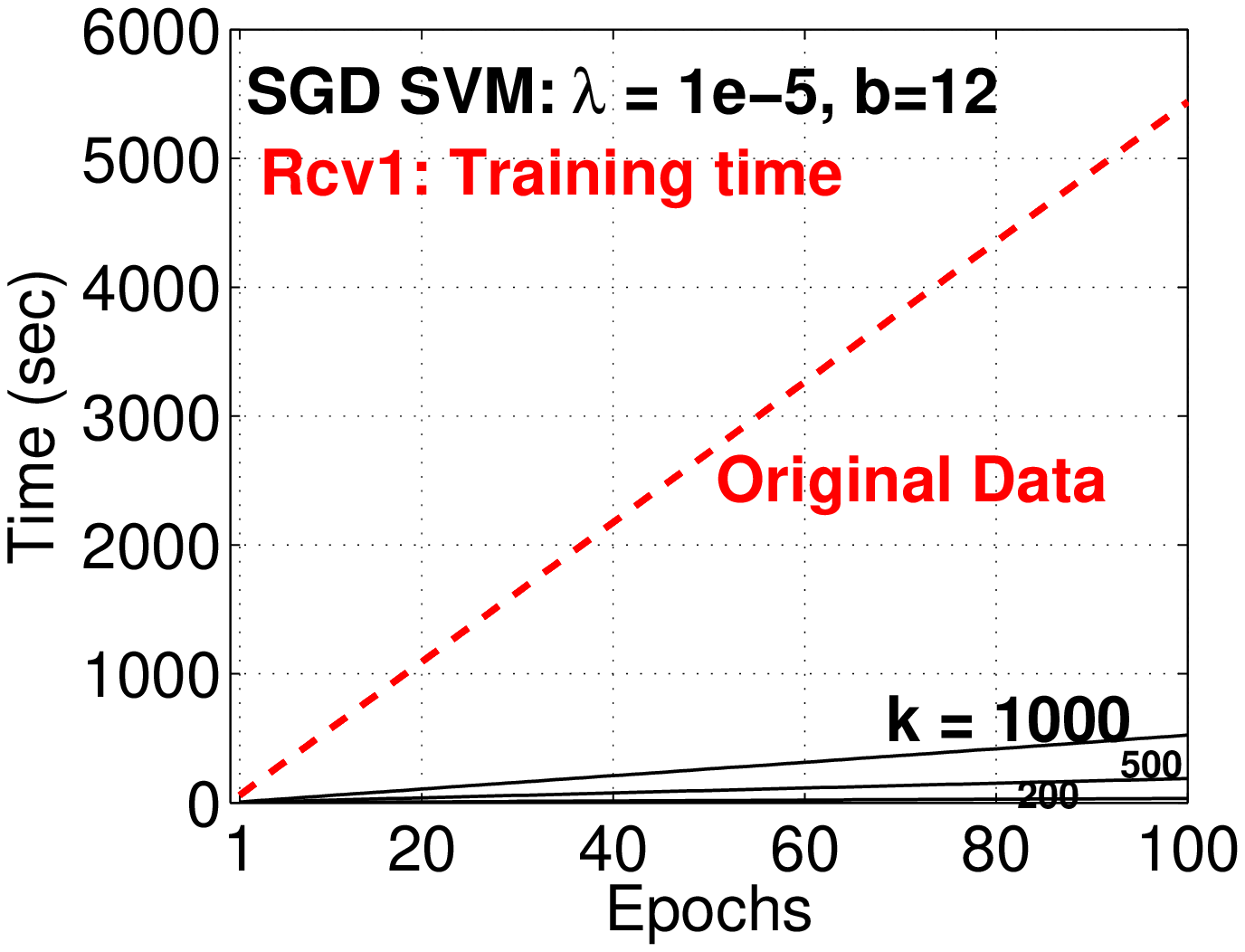}%\hspace{0.1in}
\includegraphics[width=1.6in]{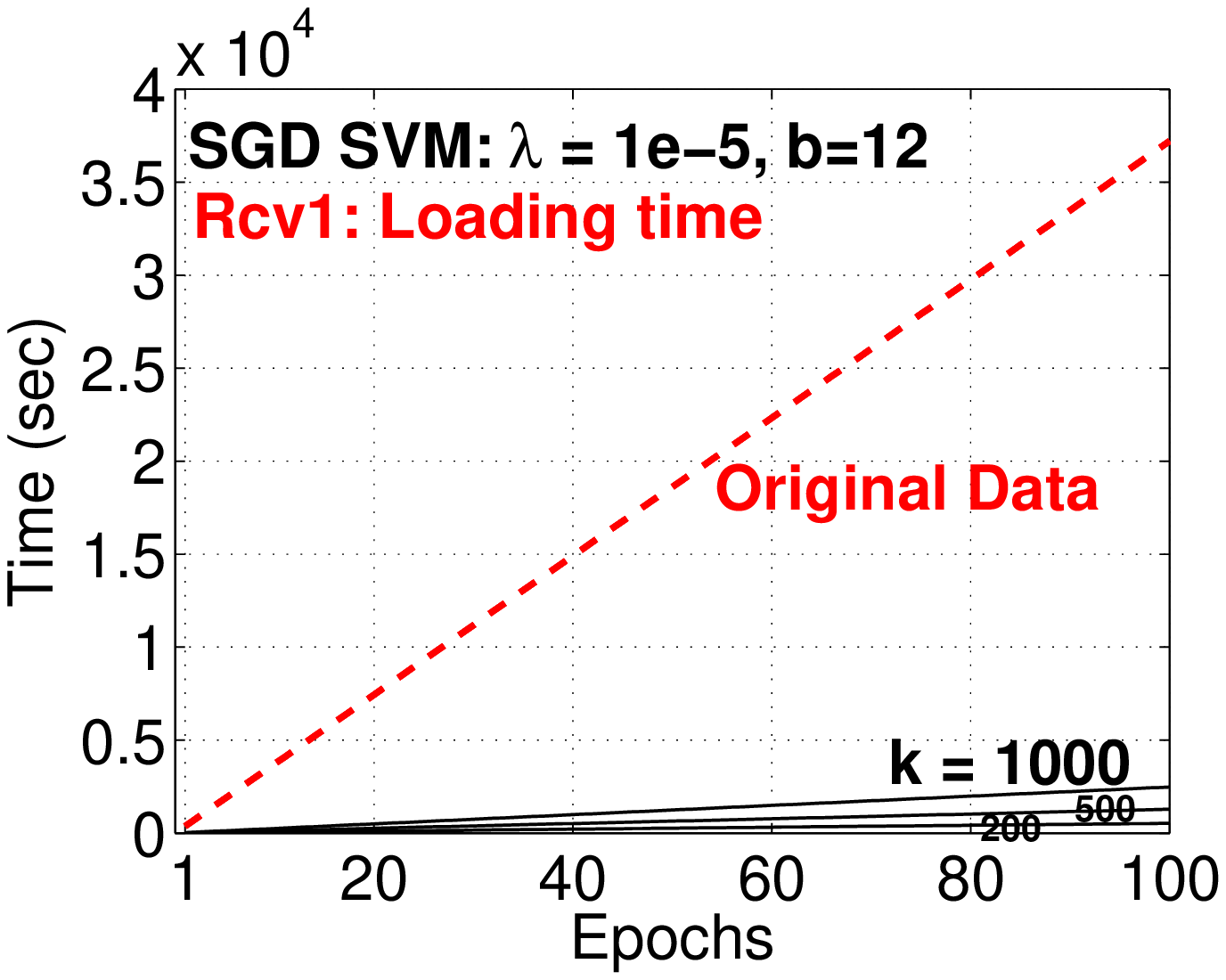}}
\end{center}

\vspace{-0.3in}

\caption{Training time  and data loading time on {\em rcv1} versus epochs, for both the original data (dashed red curves) and $12$-bit hashed data (solid curves, for $k=1000,500,200$).  $b$-bit hashing substantially reduces the training and loading times. }\label{fig_rcv1_SGD_time}
\end{figure}

Figure~\ref{fig_rcv1_SGD} presents the test accuracies of SGD SVM on {\em rcv1}  at the 100th epoch, for both the original data and the $b$-bit ($b=8$ and $b=12$) hashed data. When $k\geq 500$ and $b= 12$, $b$-bit minwise hashing achieves similar accuracies as using the original data. Figure~\ref{fig_rcv1_SGD_time} presents the training time and data loading time. As explicitly calculated in Table~\ref{tab_time_ratio}, using the original data costs 30 times more time than using 12-bit minwise hashing. Again, the data loading time dominates the cost.

\begin{table}[h]
\caption{Time ratios for {\em webspam} and {\em rcv1}, as in Figures~\ref{fig_spam_SGD_time} and~\ref{fig_rcv1_SGD_time}, averaged over 100 epochs. For example, the entry 10.05 means training with the original {\em webspam} data on average requires 10.05 times as much time as using $8$-bit minwise hashing.}
\begin{center}{
\begin{tabular}{lrr}
\hline \hline
Dataset & Training time ratio &Loading time ratio\\\hline
Webspam & 10.05& 8.95\\\hline
Rcv1 &28.91 &29.07\\
\hline\hline
\end{tabular}
}
\end{center}
\label{tab_time_ratio}\vspace{-0.15in}
\end{table}

%\newpage

\subsection{Averaged SGD (ASGD) }

\begin{figure}[h!]
\begin{center}
\mbox{
\includegraphics[width=1.6in]{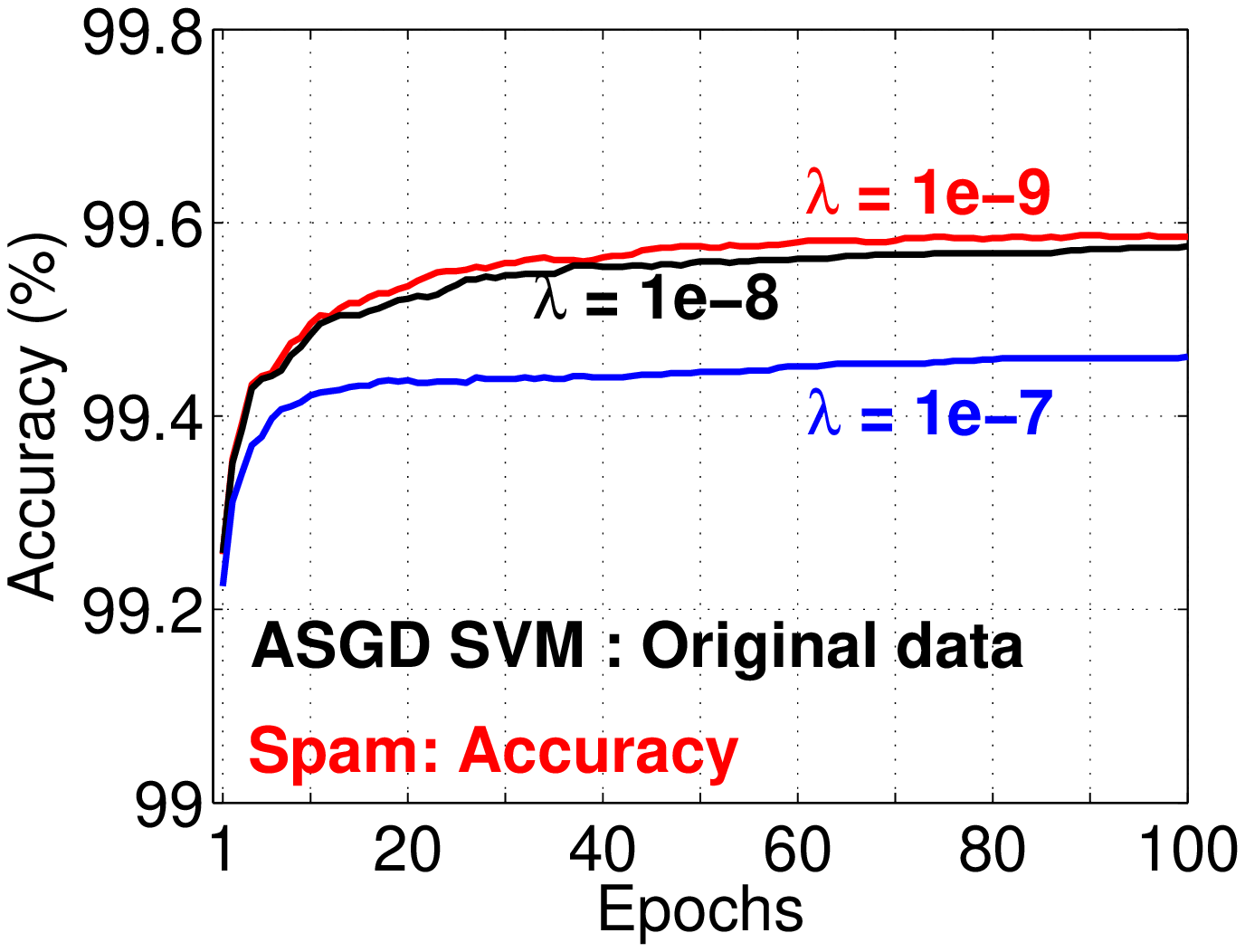}%\hspace{0.1in}
\includegraphics[width=1.6in]{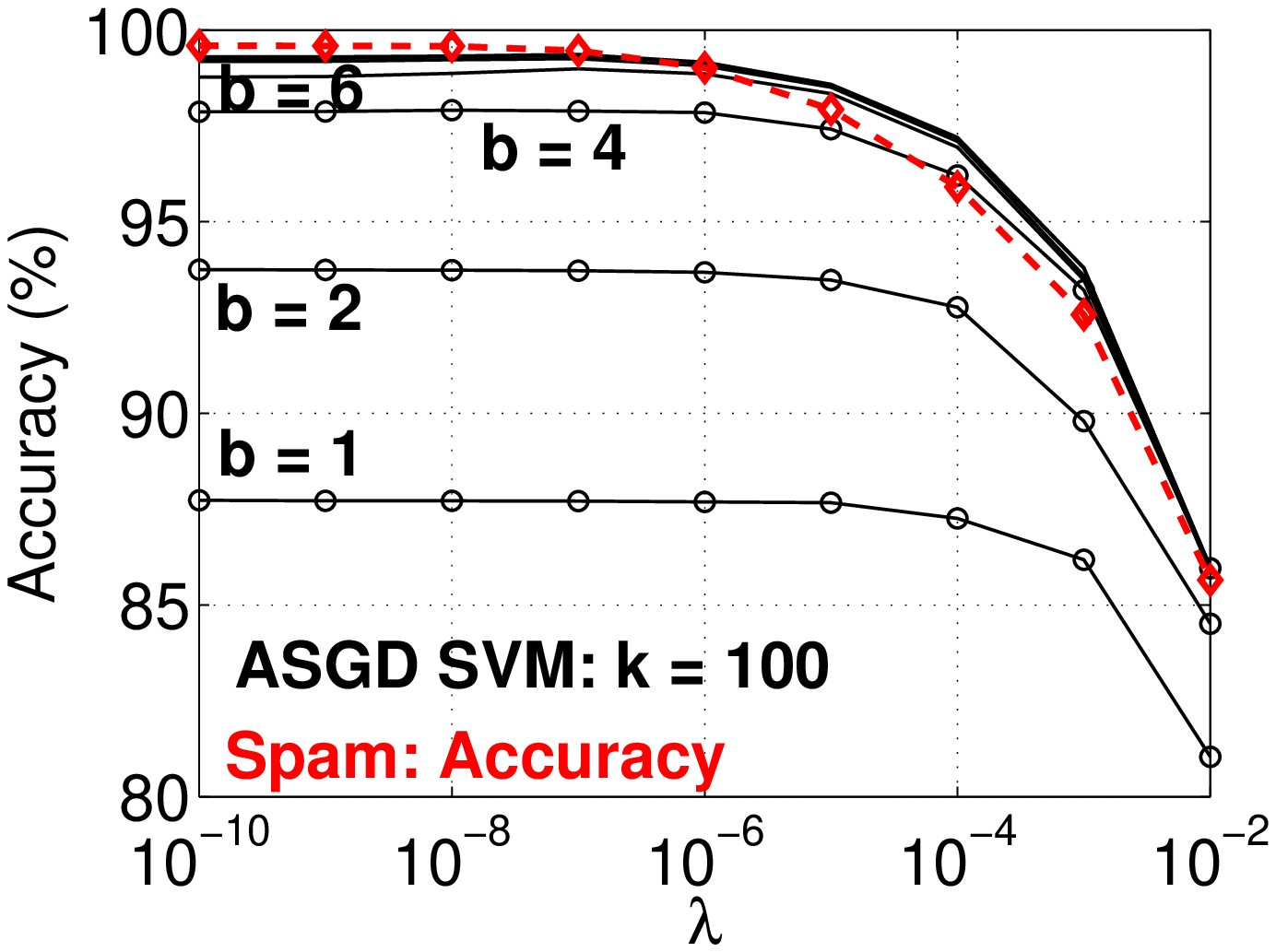}}

\mbox{
\includegraphics[width=1.6in]{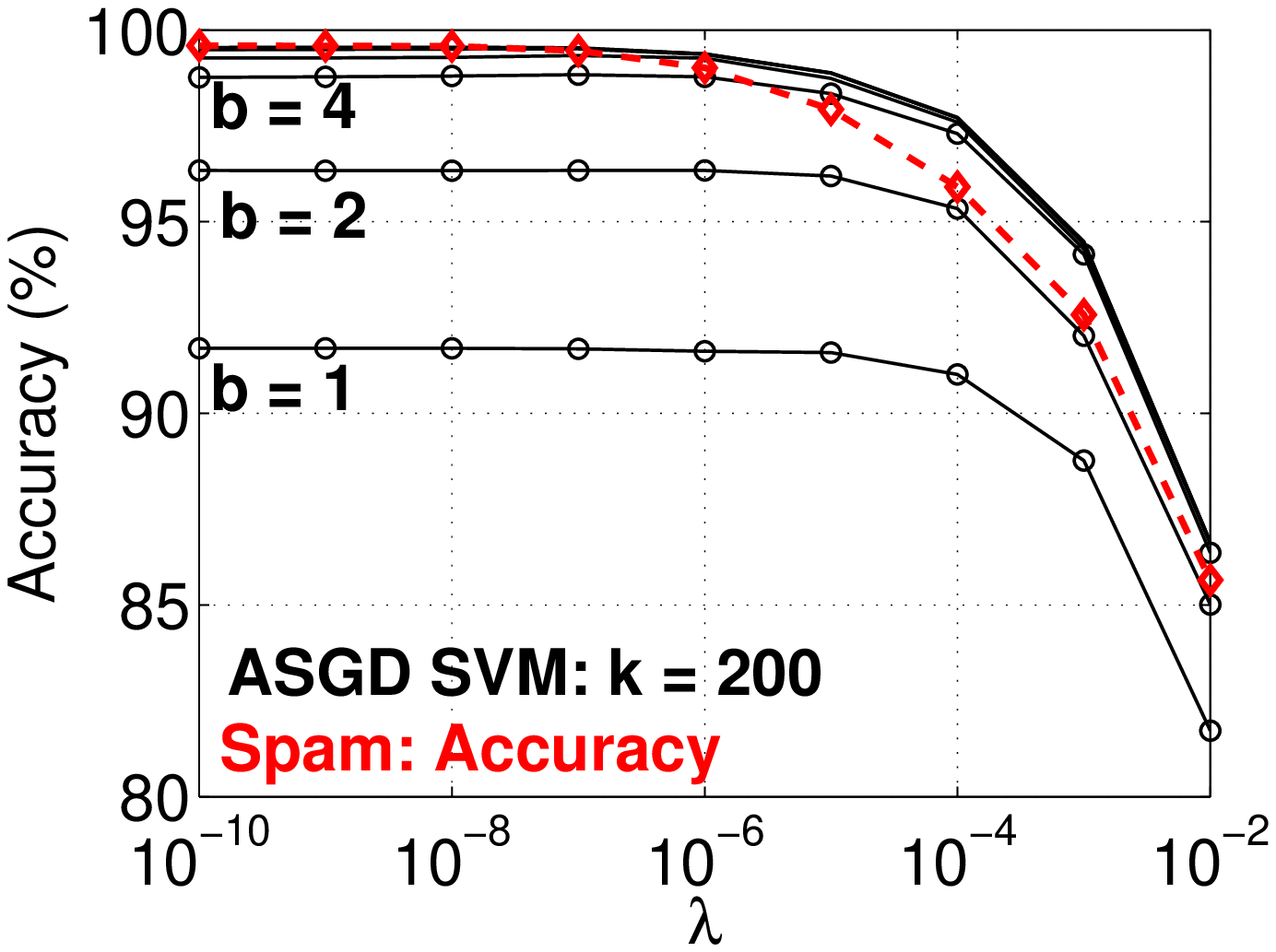}%\hspace{0.1in}
\includegraphics[width=1.6in]{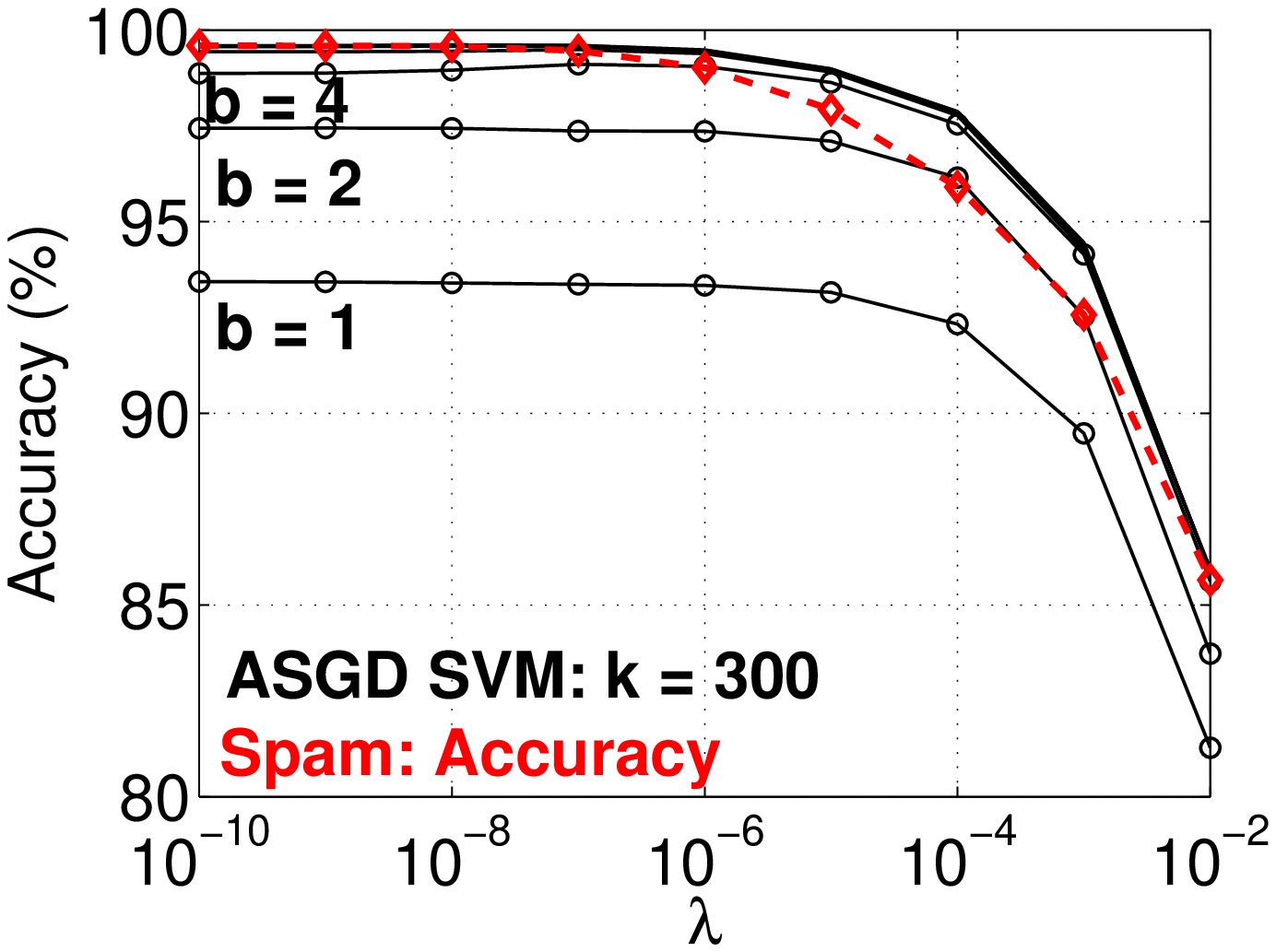}}

\mbox{
\includegraphics[width=1.6in]{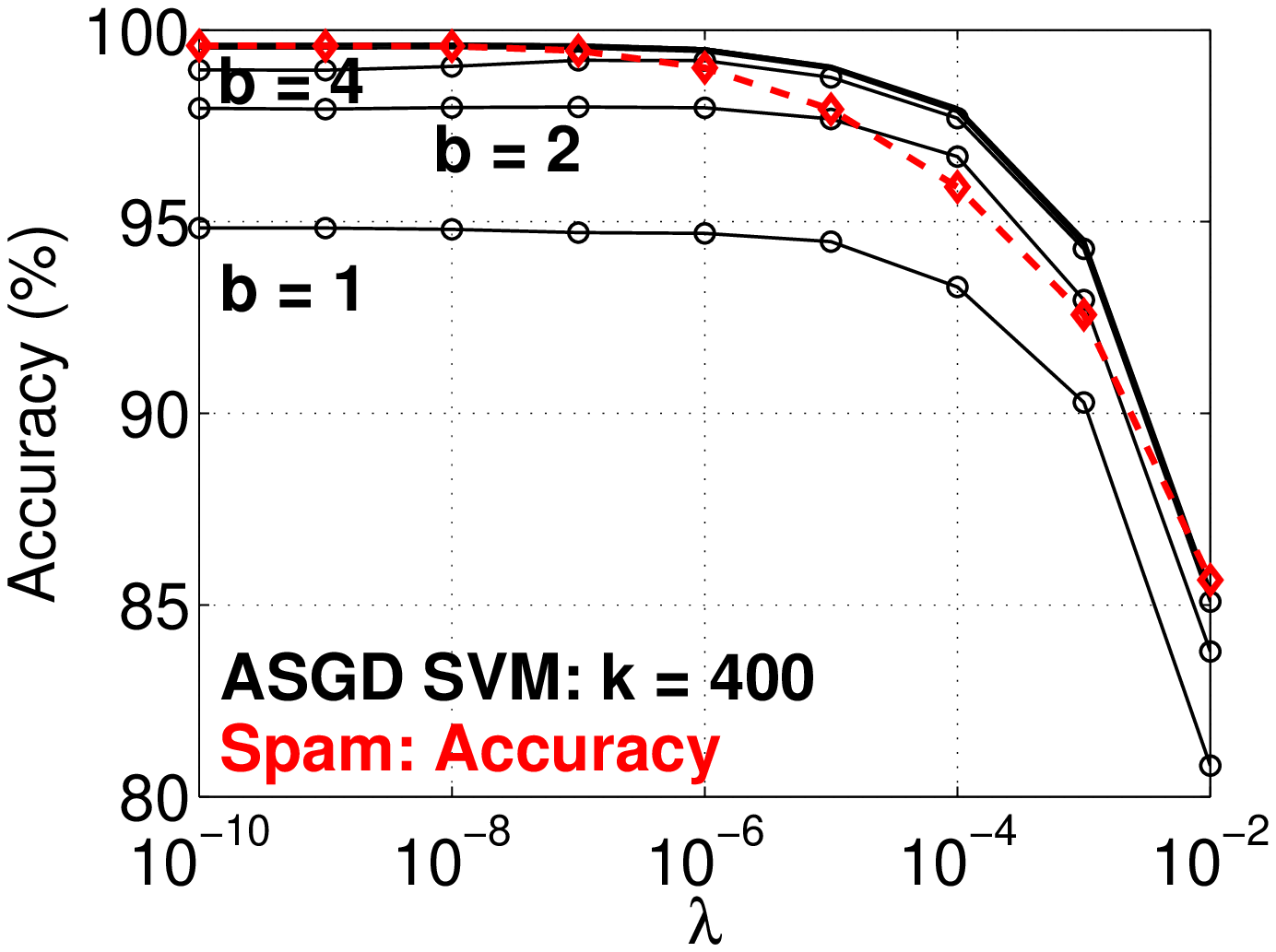}%\hspace{0.1in}
\includegraphics[width=1.6in]{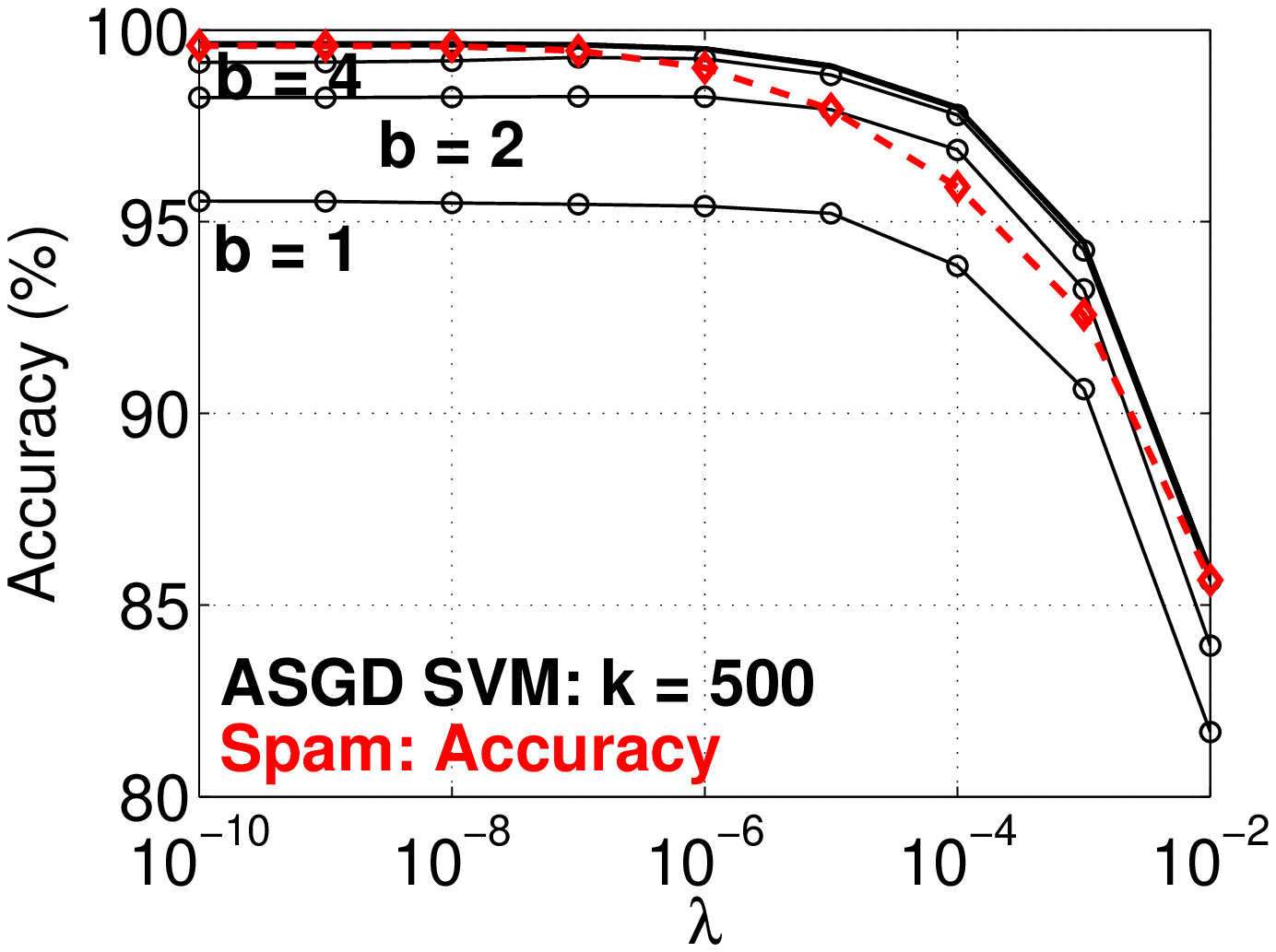}}

\end{center}

\vspace{-0.25in}

\caption{Test accuracies of ASGD SVM on the {\em webspam} dataset. The left upper panel is only for the original data (accuracies versus epochs). All other panels are the test accuracies versus $\lambda$, at the 100th epoch.}\label{fig_spam_ASGD}
\end{figure}

In the course of writing this paper in 2011, Leon Bottou kindly informed us that there was a  recent major upgrade of the SGD code, which implemented ASGD (Averaged SGD)~\cite{Report:Wu_arXiv11}. Thus,  we also provide  experiments of ASGD on the {\em webspam} dataset as shown in Figure~\ref{fig_spam_ASGD}.

Compared with the SGD results, it appears that ASGD does have some noticeable improvements over SGD. Nevertheless, ASGD still needs more than 1 epoch (perhaps 10 to 20) to approach the best accuracy. Also, $b$-bit hashing continues to perform very well in terms of accuracy and training time reduction.

%\clearpage
%\newpage

\section{Conclusion} \label{Conclusion}

($b$-bit) Minwise Hashing is a standard technique for similarity computation which has also recently been shown~\cite{Proc:HashLearning_NIPS11} to be a valuable data reduction technique in (batch) machine learning, where it can reduce both the computational overhead as well as the required infrastructure and energy consumption by orders of magnitude, at often negligible reduction in learning accuracy.

However, the use of $b$-bit minwise hashing on truly large learning datasets, which frequently occur in the context of search, requires study of a number of related challenges. First, datasets with very large numbers of features make it impossible to use pre-computed permutation matrices for the permutation step, due to prohibitive storage requirements. Second, for very large data, the initial pre-processing phase during which minhash signatures are computed, consumes significant resources. And finally, while the technique has been successfully applied in the context of batch learning (on a fairly small dataset), its efficacy in the context of online learning algorithms (which are becoming increasingly popular in the context of web search and advertising) has not been shown.\\

In the context of duplicate detection (which normally concerns only highly similar pairs of documents) using minwise hashing with 64 bits per hashed value, the prior studies (e.g.,~\cite{Proc:Broder}) demonstrated that  it would be  sufficient to use about $k\approx 200$ permutations. However, $b$-bit minwise hashing (for small values of $b$) does require more permutations than the original minwise hashing, as explained in~\cite{Article:Li_Konig_CACM11}, for example, by increasing $k$ by a factor of 3 when using $b=1$ and the resemblance threshold is  $R=0.5$. In the context of machine learning and $b$-bit minwise hashing, we have also found that in some datasets $k$ has to be fairly large, e.g., $k=500$ or even more. This is because machine learning algorithms use all similarities, not just highly similar pairs. \\

In this paper, we addressed all of these challenges to adoption of $b$-bit minwise hashing in the context of Web-scale learning tasks. Regarding the first challenge, we were able to show that the use of 2- and 4-universal hash functions matches the accuracy of fully random permutations, while enabling the use of ($b$-bit) minwise hashing even for data with extremely large numbers of features.

Regarding the 2nd challenge, we were able to formulate an implementation of the minwise hashing algorithm that effectively leverages
the properties of current GPU architectures, in particular their massive parallelism and SIMD instruction processing, while minimizing the impact of their constraints (most notably slow modulo operations and the limited bandwidth available for main memory data transfer). We observed that
the new GPU-based implementation resulted in speed-ups of between 20-80$\times$ for the minhash computation, thereby making the
data loading time itself (and not the preprocessing) the new bottleneck.

Finally, we were able to show that,  similarly to batch learning, $b$-bit minwise hashing can dramatically reduce the resource requirements for online learning as well, with little reduction in accuracy. However, for online learning, the reduction was mainly due to the reduction in data loading time, which becomes a major factor when online learning requires a large number of epochs to converge.  A side-effect of the use of models leveraging b-bit minwise hashing is that the space requirements of the resulting model itself are also dramatically reduced, which is important in the context of web search, where an incoming search query may trigger various models for query classification, vertical selection, etc. and all of these models compete for memory on the user-facing servers.

%\newpage
\section*{Acknowledgement}

We thank Leon Bottou for very helpful communications.

%This work is partially supported by NSF, ONR, and DARPA.

{\scriptsize
%\bibliographystyle{plain}
%\bibliography{../../bib/IEEEabrv,../../bib/mybibfile}

\begin{thebibliography}{10}

\bibitem{Url:Lanford_Learning0511}
{\em
  \url{http://cacm.acm.org/blogs/blog-cacm/108385-research-directions-for-mach%
ine-learning-and-algorithms/fulltext}}.

\bibitem{Url:NVIDIA_CUDA}
{\em
  \url{http://developer.download.nvidia.com/compute/cuda/2_3/toolkit/docs/NVID%
IA_CUDA_Programming_Guide_2.3.pdf}}.

\bibitem{Bilenko:2005:APN:1106326.1106331}
Mikhail Bilenko, Sugato Basu, and Mehran Sahami.
\newblock Adaptive product normalization: Using online learning for record
  linkage in comparison shopping.
\newblock In {\em ICDM}, 2005.

\bibitem{URL:Bottou_SGD}
Leon Bottou.
\newblock http://leon.bottou.org/projects/sgd.

\bibitem{Proc:Bottou_COMPSTAT10}
Leon Bottou.
\newblock Large-scale machine learning with stochastic gradient descent.
\newblock In {\em COMPSTAT}, 2010.

\bibitem{Proc:Broder}
Andrei~Z. Broder.
\newblock On the resemblance and containment of documents.
\newblock In {\em the Compression and Complexity of Sequences}, pages 21--29,
  Positano, Italy, 1997.

\bibitem{Proc:Broder_STOC98}
Andrei~Z. Broder, Moses Charikar, Alan~M. Frieze, and Michael Mitzenmacher.
\newblock Min-wise independent permutations (extended abstract).
\newblock In {\em STOC}, pages 327--336, Dallas, TX, 1998.

\bibitem{Proc:Broder_WWW97}
Andrei~Z. Broder, Steven~C. Glassman, Mark~S. Manasse, and Geoffrey Zweig.
\newblock Syntactic clustering of the web.
\newblock In {\em WWW}, pages 1157 -- 1166, Santa Clara, CA, 1997.

\bibitem{Proc:Carter_STOC77}
J.~Lawrence Carter and Mark~N. Wegman.
\newblock Universal classes of hash functions (extended abstract).
\newblock In {\em STOC}, pages 106--112, 1977.

\bibitem{Proc:Chang_KDD11}
Kai-Wei Chang and Dan Roth.
\newblock Selective block minimization for faster convergence of limited memory
  large-scale linear models.
\newblock In {\em KDD}, pages 699--707, 2011.

\bibitem{Proc:Cherkasova_KDD09}
Ludmila Cherkasova, Kave Eshghi, Charles B.~Morrey III, Joseph Tucek, and
  Alistair~C. Veitch.
\newblock Applying syntactic similarity algorithms for enterprise information
  management.
\newblock In {\em KDD}, pages 1087--1096, Paris, France, 2009.

\bibitem{Ciaramita:2008:OLC:1367497.1367529}
Massimiliano Ciaramita, Vanessa Murdock, and Vassilis Plachouras.
\newblock Online learning from click data for sponsored search.
\newblock In {\em {W}{W}{W} {C}onference}, 2008.

\bibitem{Dietzfelbinger:1996:UHK:646511.695324}
Martin Dietzfelbinger.
\newblock Universal hashing and k-wise independent random variables via integer
  arithmetic without primes.
\newblock In {\em Proceedings of the 13th Annual Symposium on Theoretical
  Aspects of Computer Science}, pages 569--580, 1996.

\bibitem{Article:Dietzfelbinger97}
Martin Dietzfelbinger, Torben Hagerup, Jyrki Katajainen, and Martti Penttonen.
\newblock A reliable randomized algorithm for the closest-pair problem.
\newblock {\em Journal of Algorithms}, 25(1):19--51, 1997.

\bibitem{Article:Fan_JMLR08}
Rong-En Fan, Kai-Wei Chang, Cho-Jui Hsieh, Xiang-Rui Wang, and Chih-Jen Lin.
\newblock Liblinear: A library for large linear classification.
\newblock {\em Journal of Machine Learning Research}, 9:1871--1874, 2008.

\bibitem{Proc:Fetterly_WWW03}
Dennis Fetterly, Mark Manasse, Marc Najork, and Janet~L. Wiener.
\newblock A large-scale study of the evolution of web pages.
\newblock In {\em WWW}, pages 669--678, Budapest, Hungary, 2003.

\bibitem{Article:Forman09}
George Forman, Kave Eshghi, and Jaap Suermondt.
\newblock Efficient detection of large-scale redundancy in enterprise file
  systems.
\newblock {\em SIGOPS Oper. Syst. Rev.}, 43(1):84--91, 2009.

\bibitem{DBLP:conf/wsdm/GantiKL10}
Venkatesh Ganti, Arnd~Christian K{\"o}nig, and Xiao Li.
\newblock Precomputing search features for fast and accurate query
  classification.
\newblock In {\em WSDM}, pages 61--70, 2010.

\bibitem{He:2009:RQC:1620585.1620588}
Bingsheng He, Mian Lu, Ke~Yang, Rui Fang, Naga~K. Govindaraju, Qiong Luo, and
  Pedro~V. Sander.
\newblock Relational {Q}uery {C}oprocessing on {G}raphics {P}rocessors.
\newblock {\em ACM Trans. Database Syst.}, 34:21:1--21:39, December 2009.

\bibitem{Indyk:2001:SAM:370968.370980}
Piotr Indyk.
\newblock A {S}mall {A}pproximately {M}in-wise {I}ndependent {F}amily of {H}ash
  {F}unctions.
\newblock {\em J. Algorithms}, 38, 2001.

\bibitem{Proc:Joachims_KDD06}
Thorsten Joachims.
\newblock Training linear svms in linear time.
\newblock In {\em KDD}, pages 217--226, Pittsburgh, PA, 2006.

\bibitem{Kim:2010:FFA:1807167.1807206}
Changkyu Kim, Jatin Chhugani, Nadathur Satish, Eric Sedlar, Anthony~D. Nguyen,
  Tim Kaldewey, Victor~W. Lee, Scott~A. Brandt, and Pradeep Dubey.
\newblock Fast: fast architecture sensitive tree search on modern cpus and
  gpus.
\newblock In {\em SIGMOD}, pages 339--350, 2010.

\bibitem{L'Huillier:2010:OPC:1809400.1809421}
Gaston L'Huillier, Richard Weber, and Nicolas Figueroa.
\newblock Online {P}hishing {C}lassification using {A}dversarial {D}ata
  {M}ining and {S}ignaling {G}ames.
\newblock {\em SIGKDD Explor. Newsl.}, 11:92--99, May 2010.

\bibitem{Report:Li_EM}
Ping Li.
\newblock Image classification with hashing on locally and gloablly expanded
  features.
\newblock Technical report.

\bibitem{Proc:Li_Hastie_Church_KDD06}
Ping Li, Trevor~J. Hastie, and Kenneth~W. Church.
\newblock Very sparse random projections.
\newblock In {\em KDD}, pages 287--296, Philadelphia, PA, 2006.

\bibitem{Article:Li_Konig_CACM11}
Ping Li and Arnd~Christian K\"onig.
\newblock Theory and applications b-bit minwise hashing.
\newblock {\em Commun. ACM}, 2011.

\bibitem{Proc:HashLearning_NIPS11}
Ping Li, Anshumali Shrivastava, Joshua Moore, and Arnd~Christian K\"onig.
\newblock Hashing algorithms for large-scale learning.
\newblock In {\em NIPS}, Vancouver, BC, 2011.

\bibitem{Proc:Manku_WWW07}
Gurmeet~Singh Manku, Arvind Jain, and Anish~Das Sarma.
\newblock {D}etecting {N}ear-{D}uplicates for {W}eb-{C}rawling.
\newblock In {\em WWW}, Banff, Alberta, Canada, 2007.

\bibitem{Proc:Pandey_WWW09}
Sandeep Pandey, Andrei Broder, Flavio Chierichetti, Vanja Josifovski, Ravi
  Kumar, and Sergei Vassilvitskii.
\newblock Nearest-neighbor caching for content-match applications.
\newblock In {\em WWW}, pages 441--450, Madrid, Spain, 2009.

\bibitem{Patrascu:2010:KRL:1880918.1880996}
Mihai P\u{a}tra\c{s}cu and Mikkel Thorup.
\newblock On the k-independence required by linear probing and minwise
  independence.
\newblock In {\em ICALP}, pages 715--726, 2010.

\bibitem{Sculley:2007:ROS:1277741.1277813}
D.~Sculley and Gabriel~M. Wachman.
\newblock Relaxed online {S}{V}{M}s for {S}pam {F}iltering.
\newblock In {\em SIGIR}, 2007.

\bibitem{Proc:Shalev-Shwartz_ICML07}
Shai Shalev-Shwartz, Yoram Singer, and Nathan Srebro.
\newblock Pegasos: Primal estimated sub-gradient solver for svm.
\newblock In {\em ICML}, pages 807--814, Corvalis, Oregon, 2007.

\bibitem{Article:Shi_JMLR09}
Qinfeng Shi, James Petterson, Gideon Dror, John Langford, Alex Smola, and
  S.V.N. Vishwanathan.
\newblock Hash kernels for structured data.
\newblock {\em Journal of Machine Learning Research}, 10:2615--2637, 2009.

\bibitem{Proc:Thorup_ALENEX10}
Mikkel Thorup and Yin Zhang.
\newblock Tabulation based 5-universal hashing and linear probing.
\newblock In {\em ALENEX}, pages 62--76, 2010.

\bibitem{GoogleBlog}
Simon Tong.
\newblock Lessons learned developing a practical large scale machine learning
  system.
\newblock
  http://googleresearch.blogspot.com/2010/04/lessons-learned-developing-practi%
cal.html, 2008.

\bibitem{Article:Urvoy08}
Tanguy Urvoy, Emmanuel Chauveau, Pascal Filoche, and Thomas Lavergne.
\newblock Tracking web spam with html style similarities.
\newblock {\em ACM Trans. Web}, 2(1):1--28, 2008.

\bibitem{Proc:Weinberger_ICML2009}
Kilian Weinberger, Anirban Dasgupta, John Langford, Alex Smola, and Josh
  Attenberg.
\newblock Feature hashing for large scale multitask learning.
\newblock In {\em ICML}, pages 1113--1120, 2009.

\bibitem{Report:Wu_arXiv11}
Wei Wu.
\newblock Towards optimal one pass large scale learning with averaged
  stochastic gradient descent.
\newblock Technical report, 2011.

\bibitem{Proc:Yu_KDD10}
Hsiang-Fu Yu, Cho-Jui Hsieh, Kai-Wei Chang, and Chih-Jen Lin.
\newblock Large linear classification when data cannot fit in memory.
\newblock In {\em KDD}, pages 833--842, 2010.

\end{thebibliography}

}

\appendix

\vspace{-0.1in}
\section{Resemblance Estimation Using Simple Hash Functions}

In this section we study the effect of using 2U/4U hashing function in place of (fully) random permutation matrices on the accuracy of resemblance estimation via b-bit minwise hashing. This will provide us a better understanding why the learning results (for SVM and logistic regression)
using $b$-bit minwise hashing are not noticeably affected much by replacing the
fully random permutation matrix with 2U/4U hash functions. As we shall see, as long as the original data
are not too dense, using 2U/4U hash functions will not result in
loss of estimation accuracy. As we observed that results from 2U and 4U are
essentially indistinguishable, we only report the 2U experiments.

The task we study here is the estimation of word associations. The dataset, extracted from commercial Web crawls, consists of 9 pairs of sets (18 English words). Each set consists of the document IDs which contain the word at least once. Table~\ref{tab_10pairs} summarizes the data.

\vspace{-0.1in}
\begin{table}[h]
\caption{\small Data information of the 9 pairs of English words. For example, ``KONG'' and ``HONG'' correspond to the two sets of document IDs which contained  word ``KONG'' and  word ``HONG'' respectively.\vspace{-0.05in}
 }
\begin{center}{\scriptsize
\begin{tabular}{l l r r  l }
\hline \hline
Word 1 & Word 2 &$f_1$  &$f_2$  &$R$ \\\hline
KONG & HONG &948 &940 &0.925 \\
RIGHTS & RESERVED  &12234      &11272 &0.877\\
OF & AND &37339   &36289 &0.771\\
GAMBIA &KIRIBATI &206   &186  &0.712\\
%UNITED &STATES &4079 &3981 &0.591\\
SAN &FRANCISCO &3194 &1651 &0.476\\
CREDIT & CARD &2999 &2697 &0.285\\
TIME & JOB &37339   &36289 &0.128\\
LOW  & PAY  &2936        &2828 &0.112\\
A  & TEST &39063        &2278&0.052\\
\hline\hline
\end{tabular}
}
\end{center}
\label{tab_10pairs}\vspace{-0.15in}
\end{table}

\begin{figure}[h!]
\begin{center}
\mbox{
\includegraphics[width=1.2in]{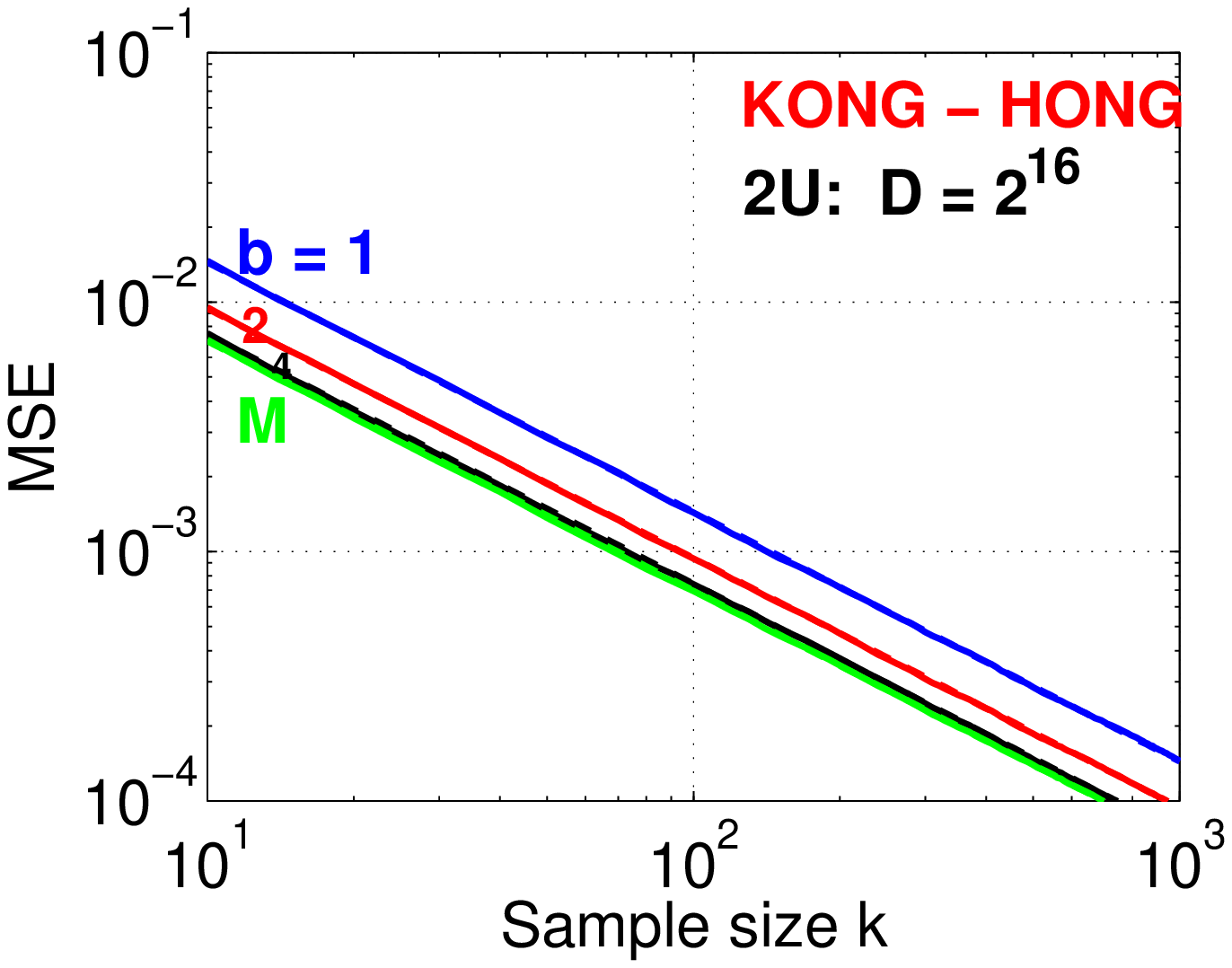}\hspace{-0.1in}
\includegraphics[width=1.2in]{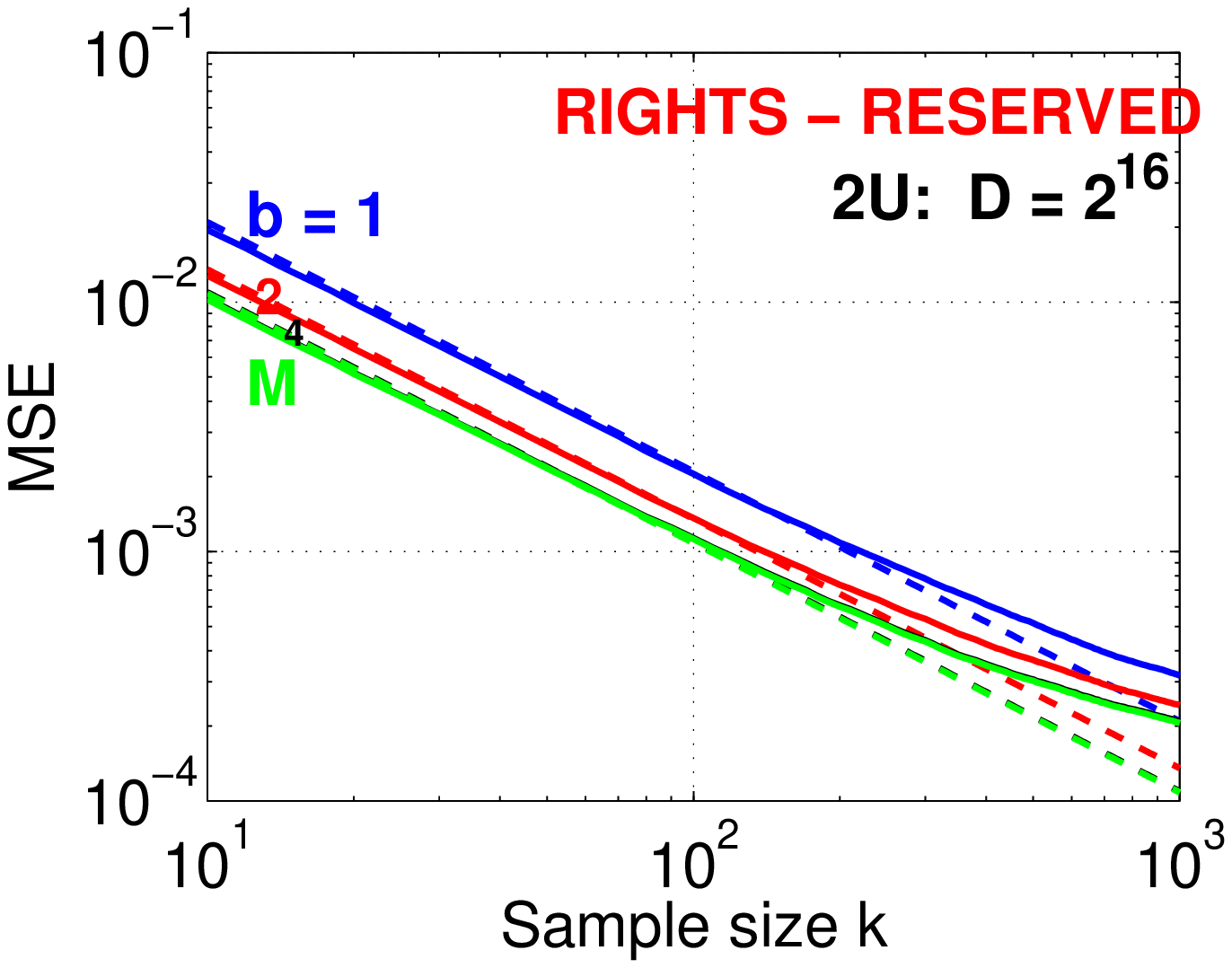}\hspace{-0.1in}
\includegraphics[width=1.2in]{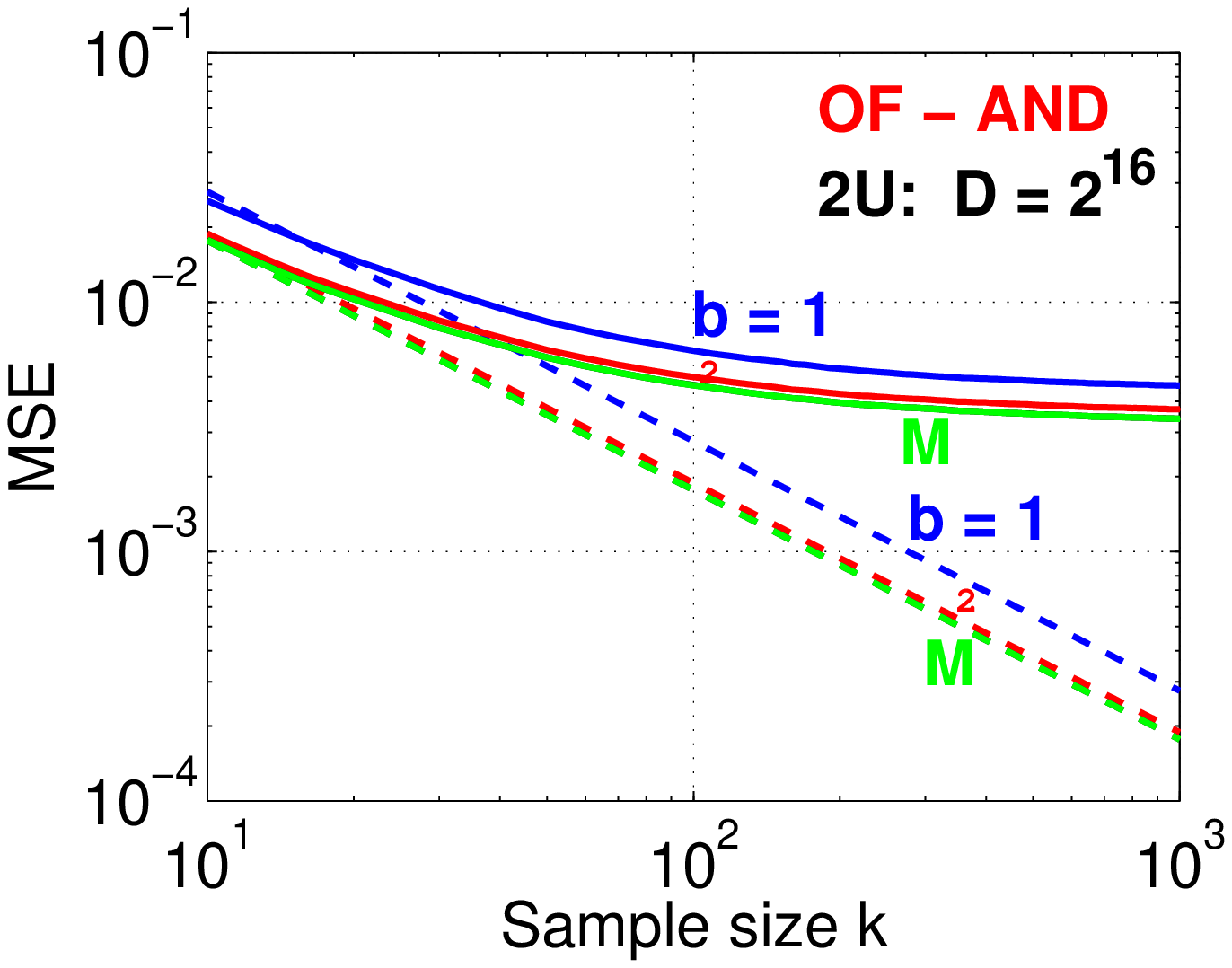}}

\mbox{
\includegraphics[width=1.2in]{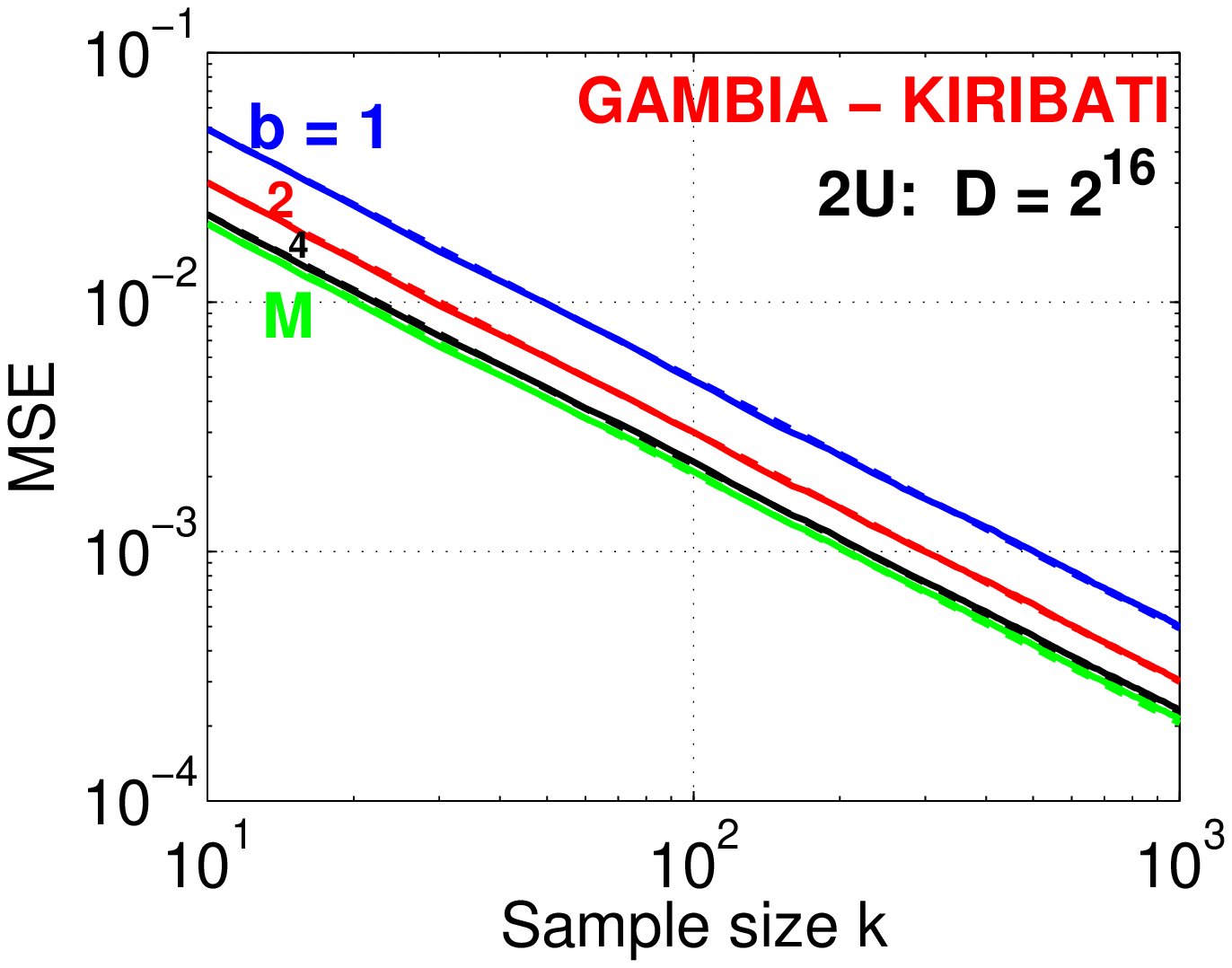}\hspace{-0.1in}
\includegraphics[width=1.2in]{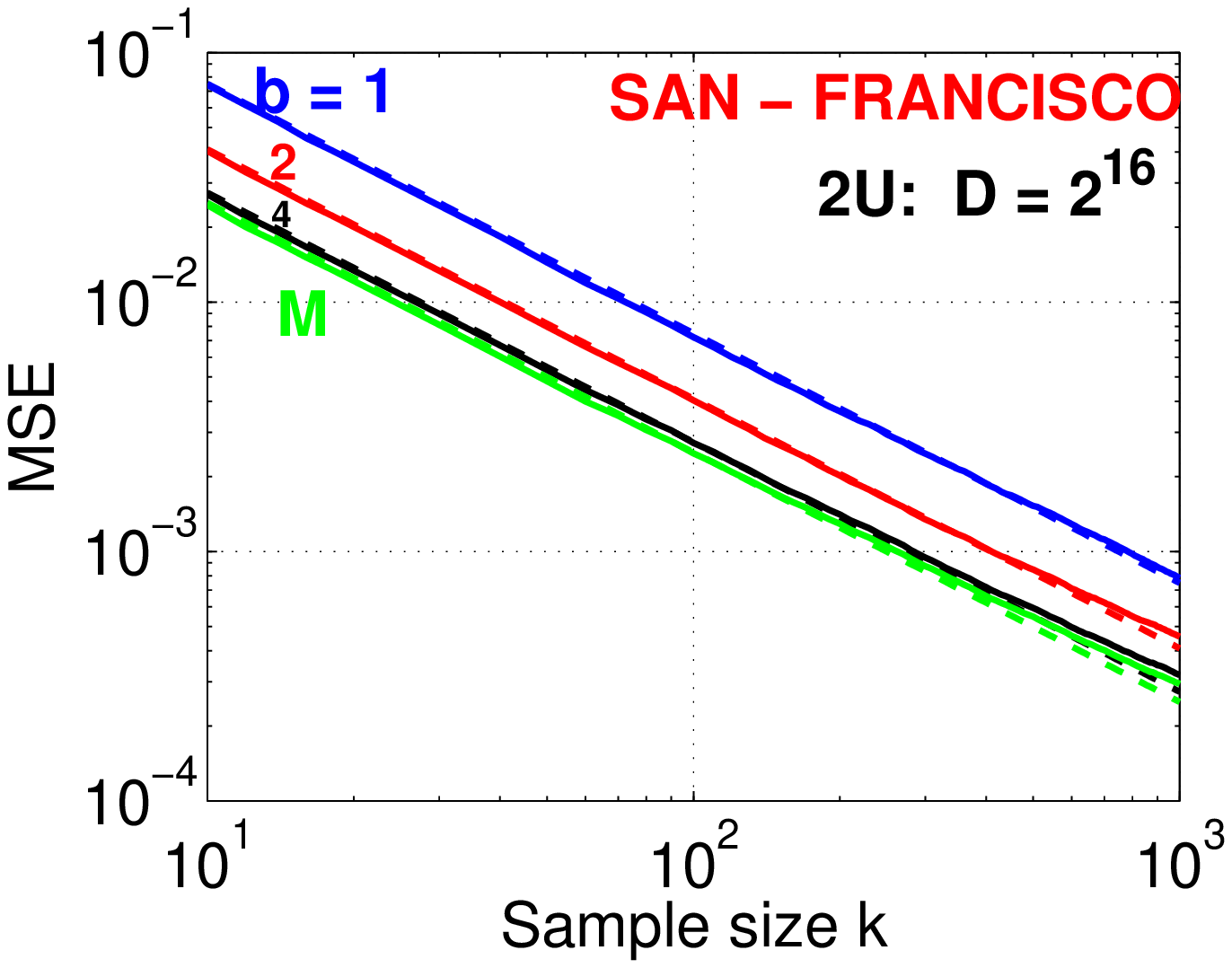}\hspace{-0.1in}
\includegraphics[width=1.2in]{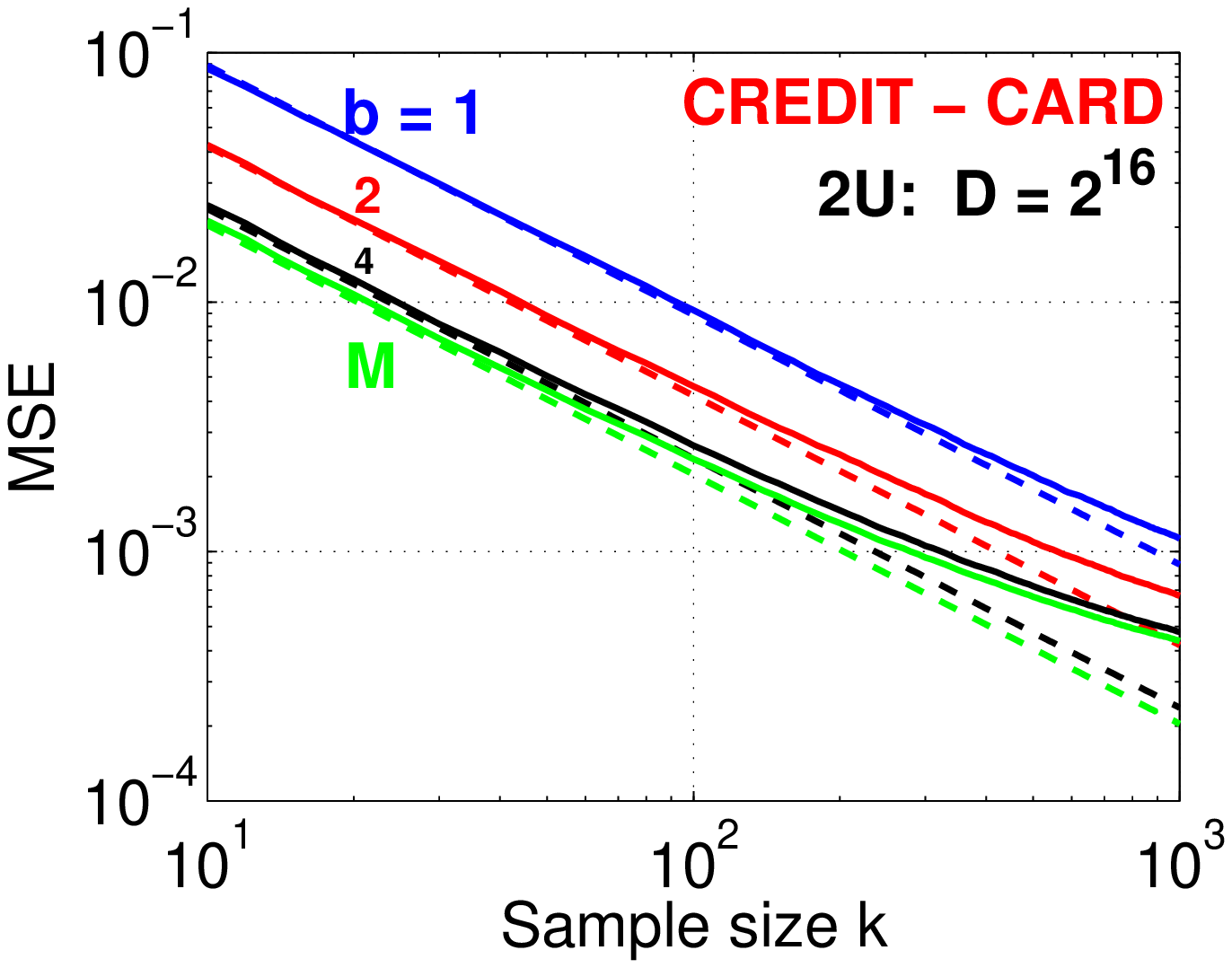}}

\mbox{
\includegraphics[width=1.2in]{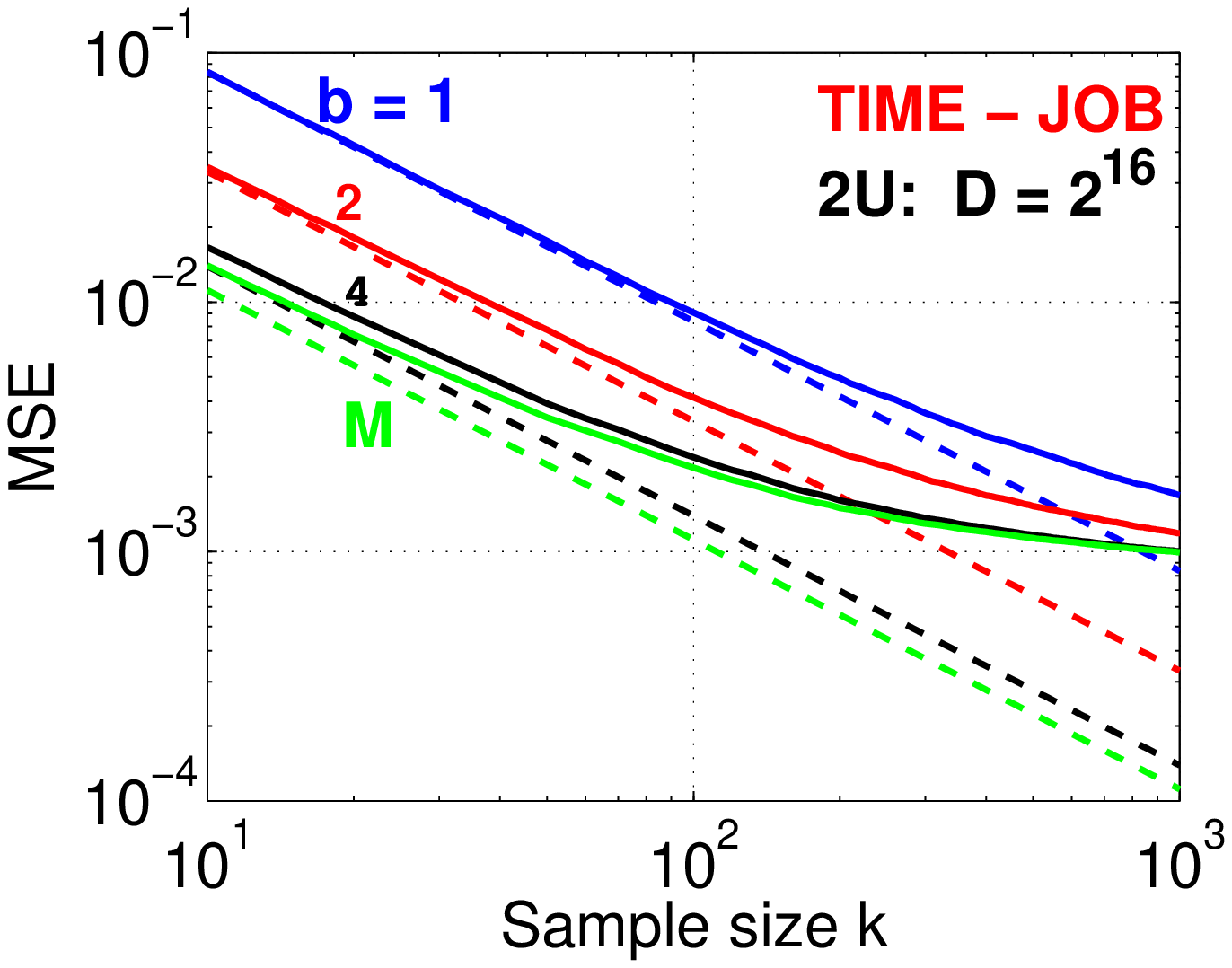}\hspace{-0.1in}
\includegraphics[width=1.2in]{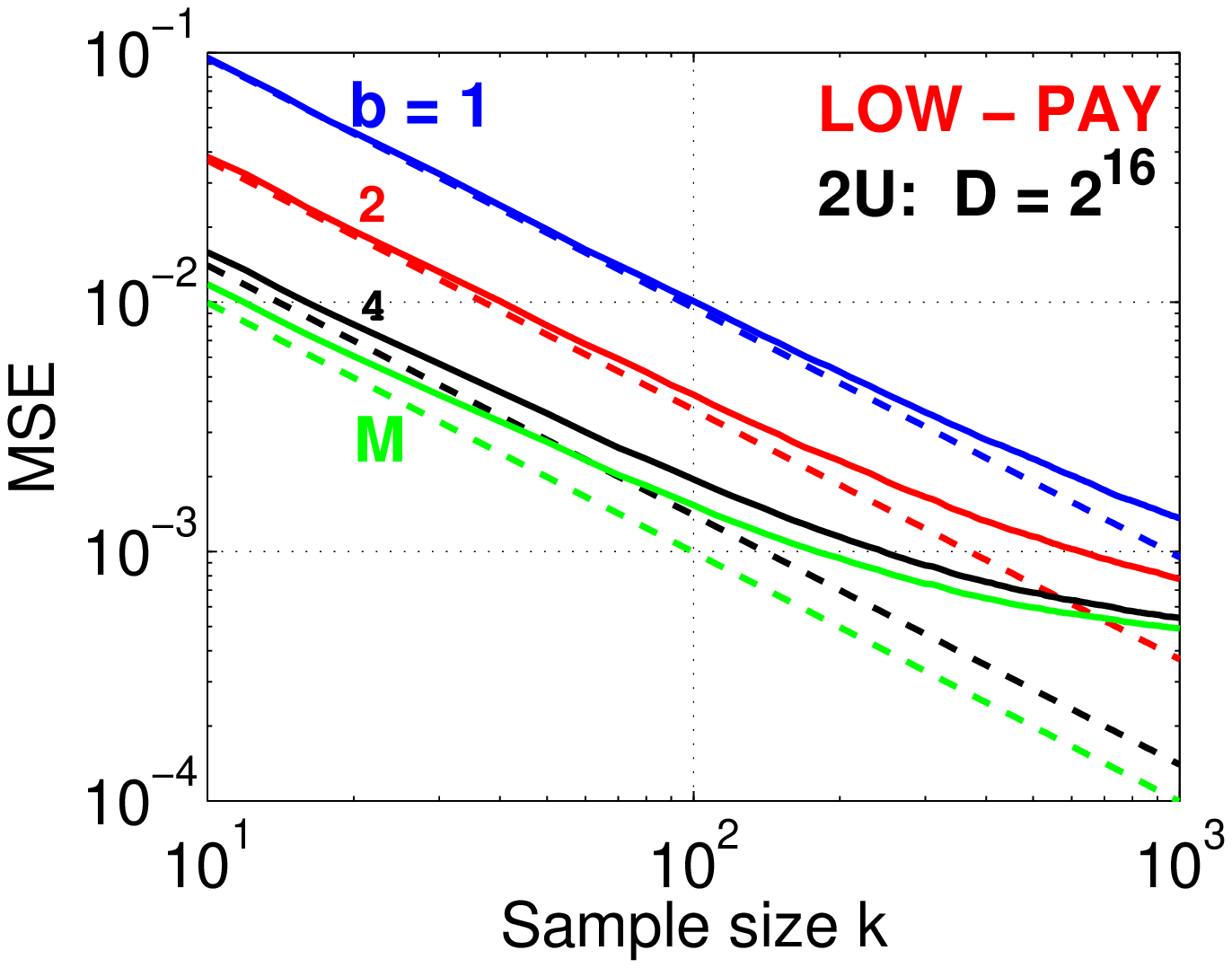}\hspace{-0.1in}
\includegraphics[width=1.2in]{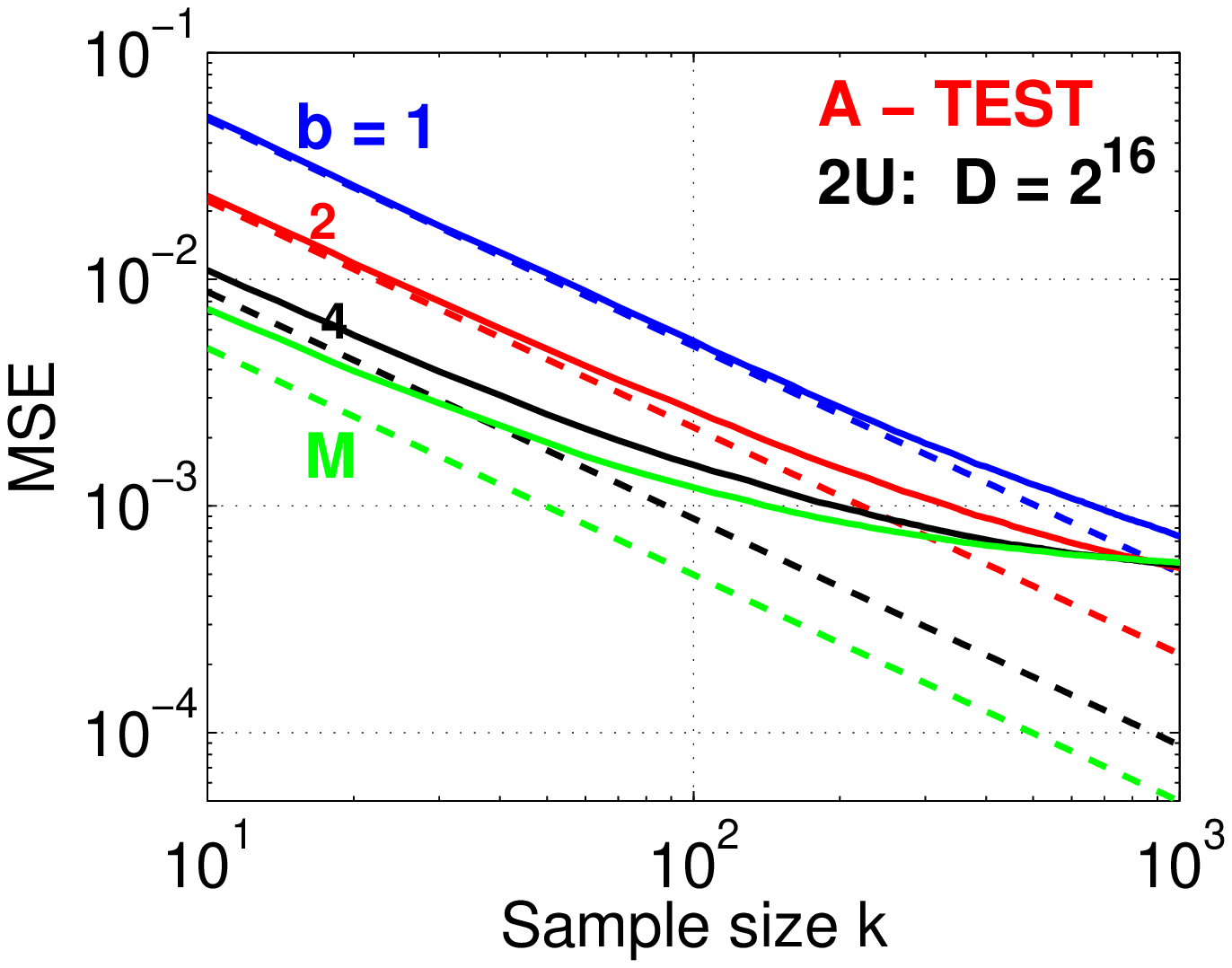}}

\end{center}

\vspace{-0.3in}

\caption{Mean square errors (MSEs) of the resemblance estimates using (\ref{eqn_R_b}) and 2U hashing with $D=2^{16}$, on the 9 English word vector pairs in Table~\ref{tab_10pairs}. We  present $b=1,2,4$ and the original minwise hashing (i.e., ``M''). The dashed curves are the theoretical variances (Eq. (11) in~\cite{Article:Li_Konig_CACM11}). Ideally the solid and dashed curves should overlap (e.g., KONG-HONG). Due to limited randomness, when the data are fairy dense (e.g., OF-AND), the empirical estimates deviate from  theoretical predictions.} \label{fig_2U_D16}\vspace{-0.15in}
\end{figure}

\begin{figure}[h!]
\begin{center}
\mbox{
\includegraphics[width=1.2in]{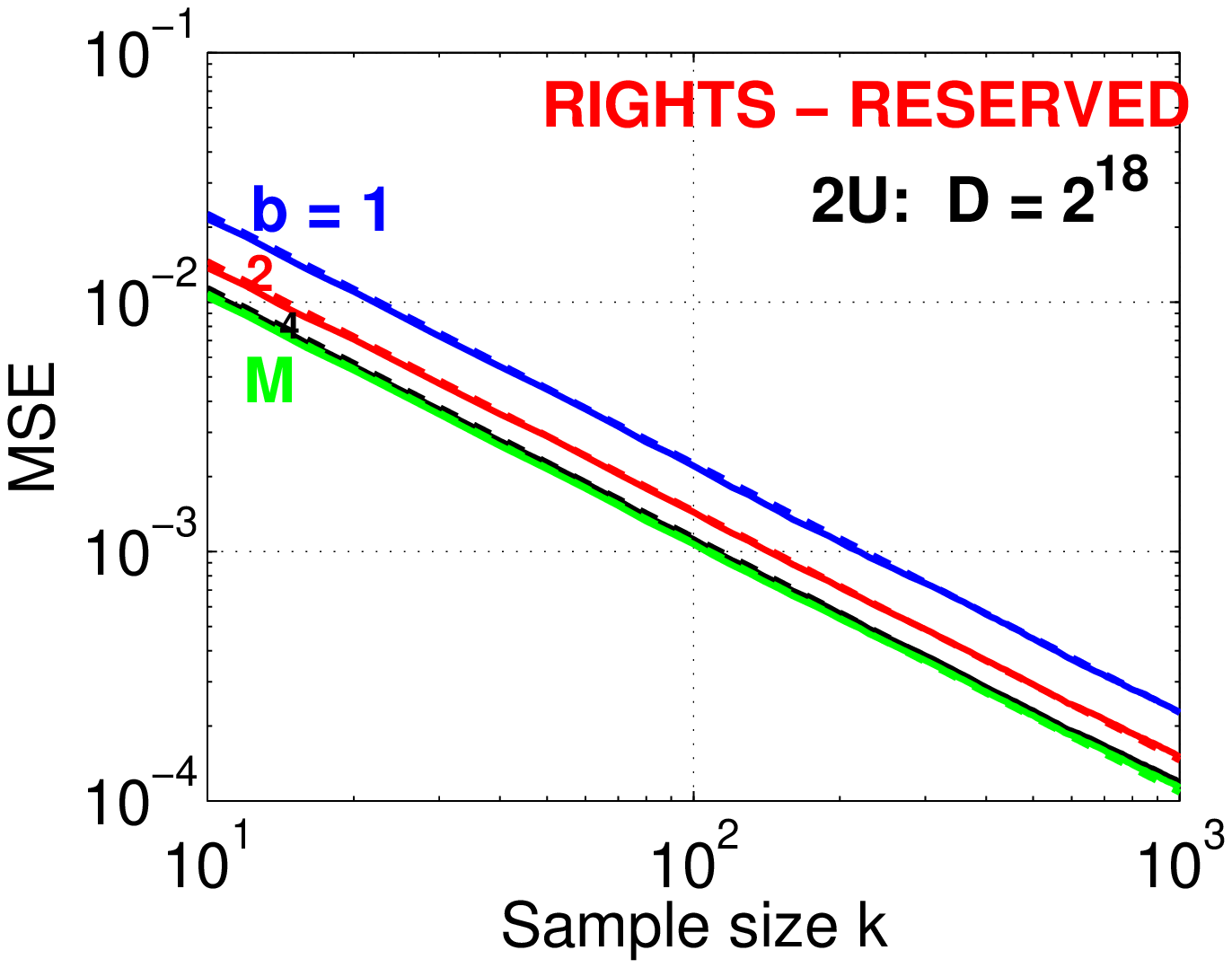}\hspace{-0.1in}
\includegraphics[width=1.2in]{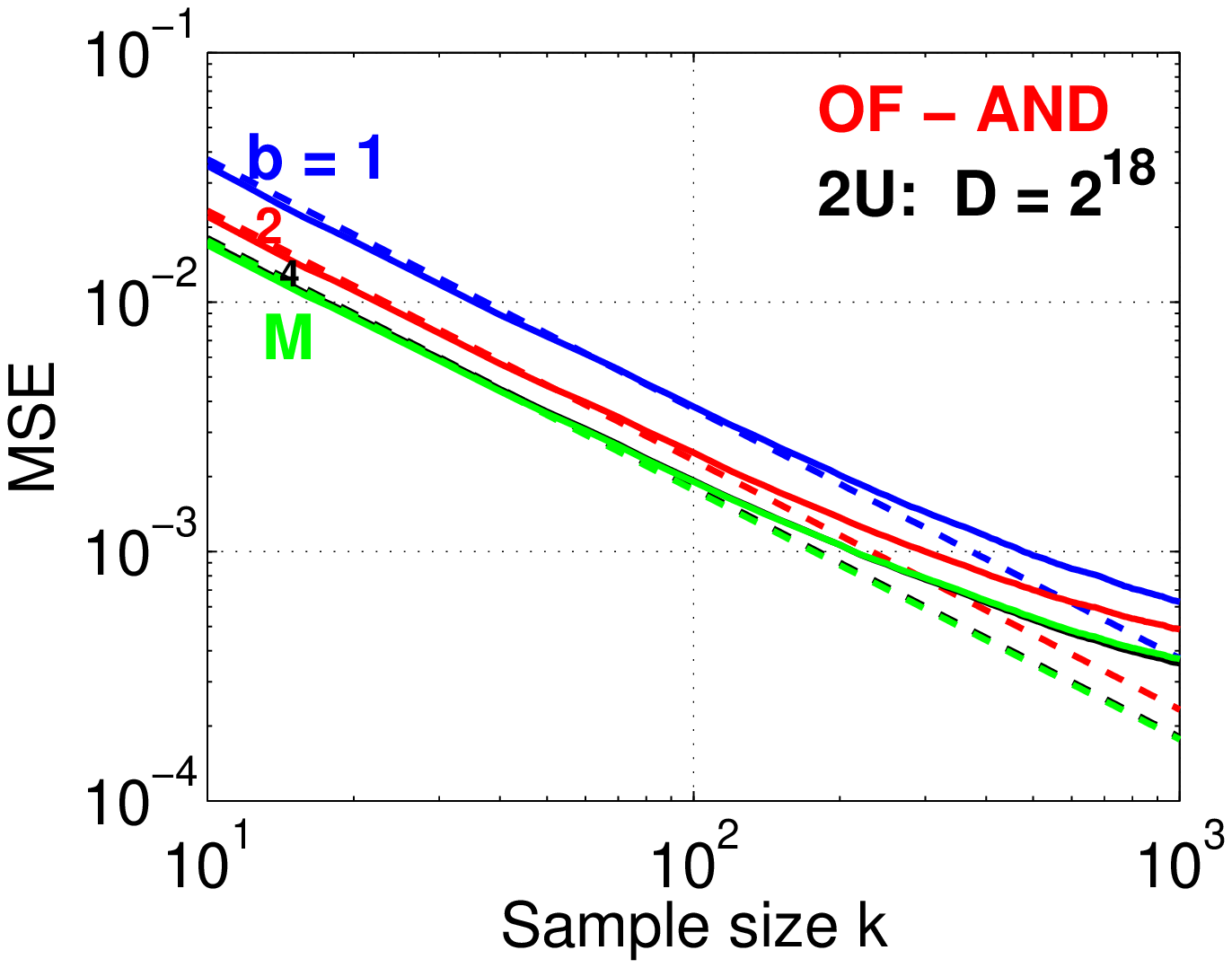}\hspace{-0.1in}
\includegraphics[width=1.2in]{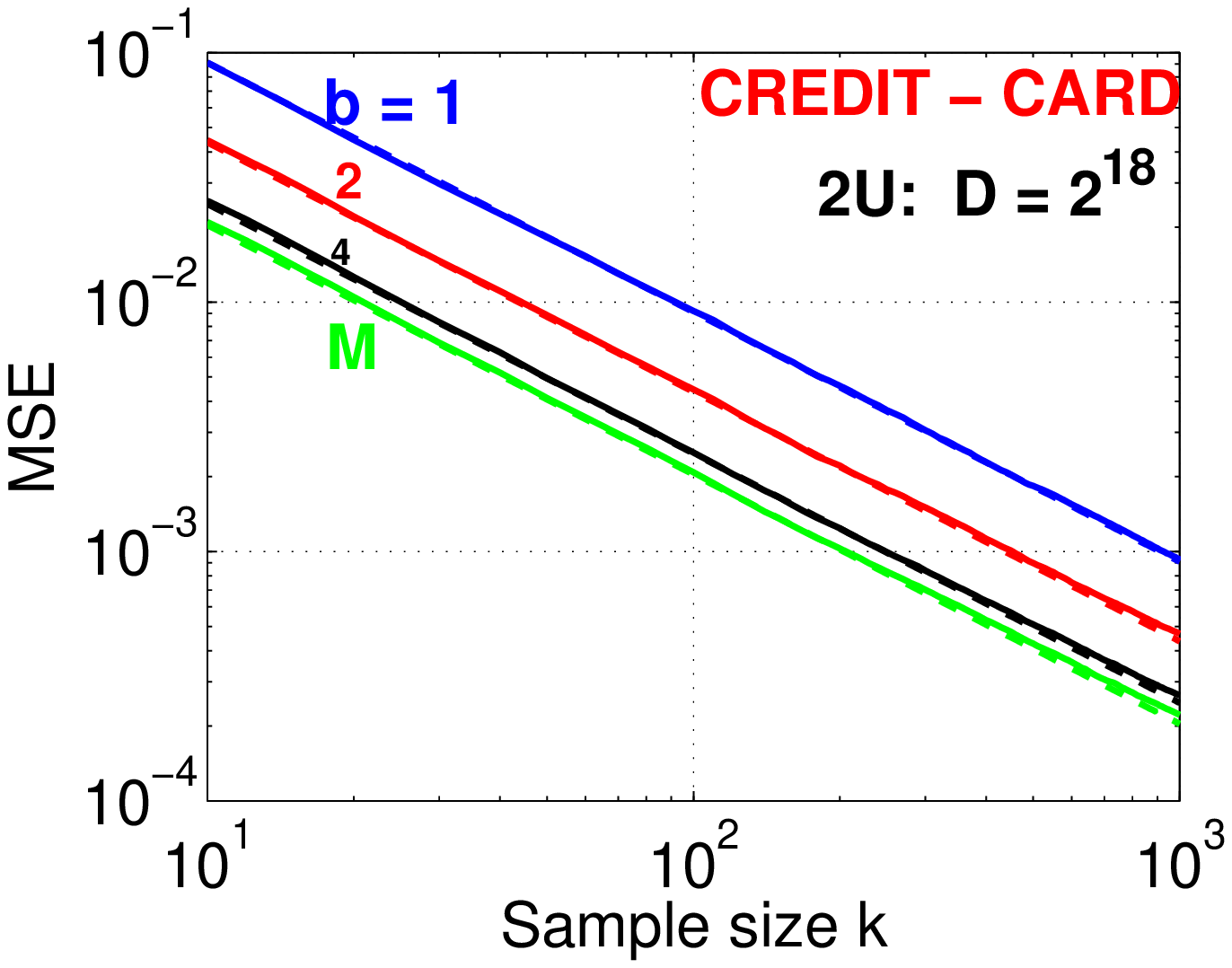}}

\mbox{
\includegraphics[width=1.2in]{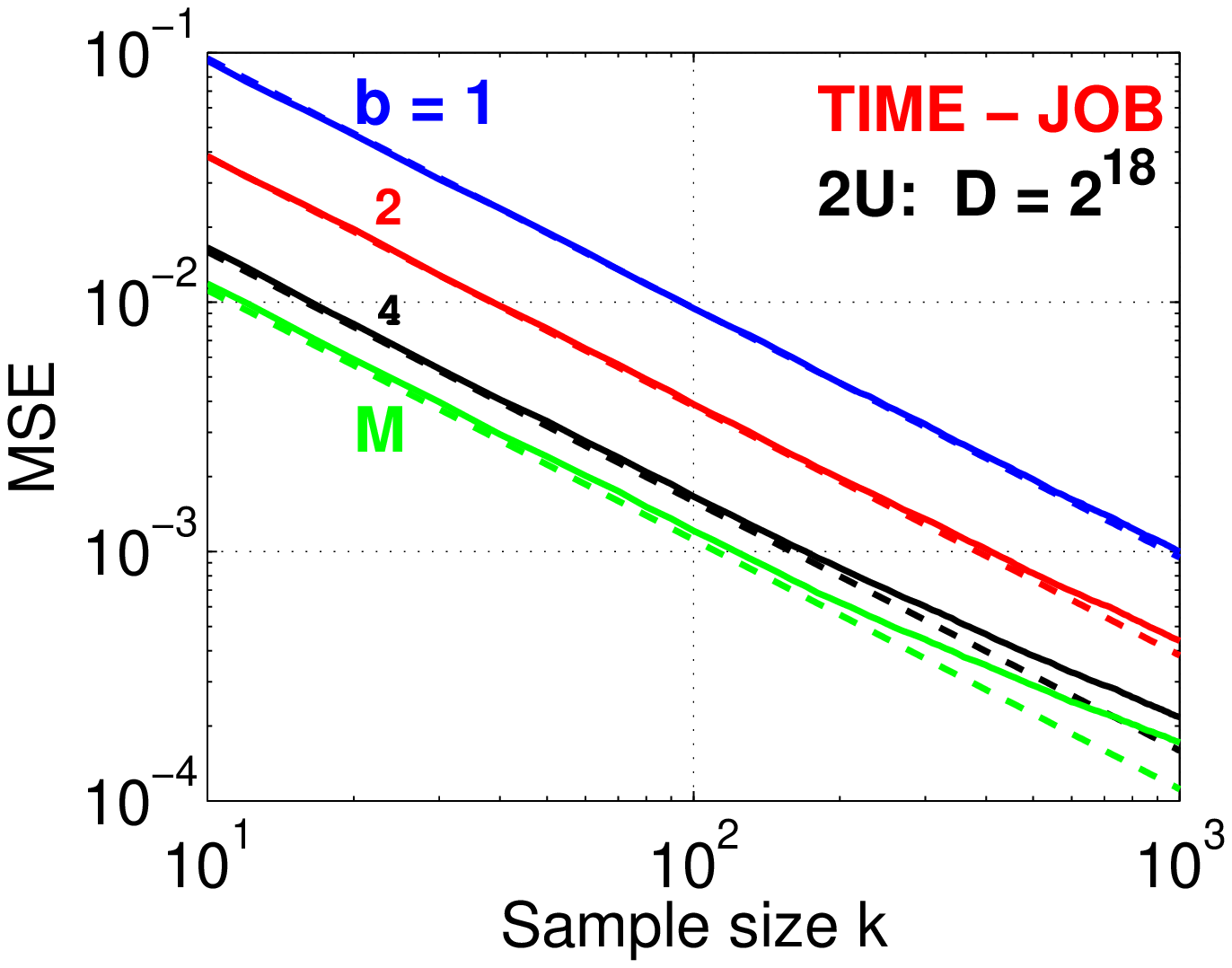}\hspace{-0.1in}
\includegraphics[width=1.2in]{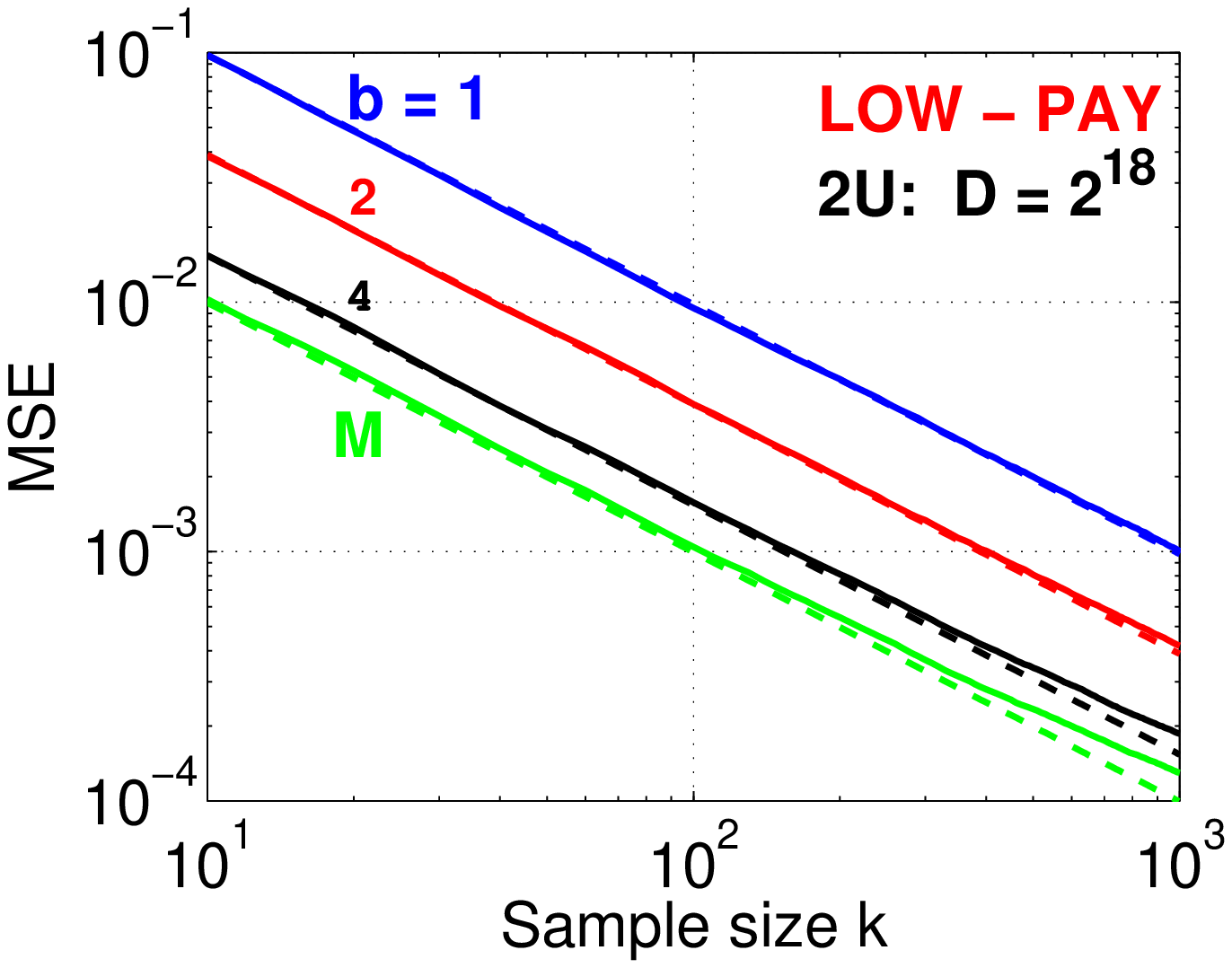}\hspace{-0.1in}
\includegraphics[width=1.2in]{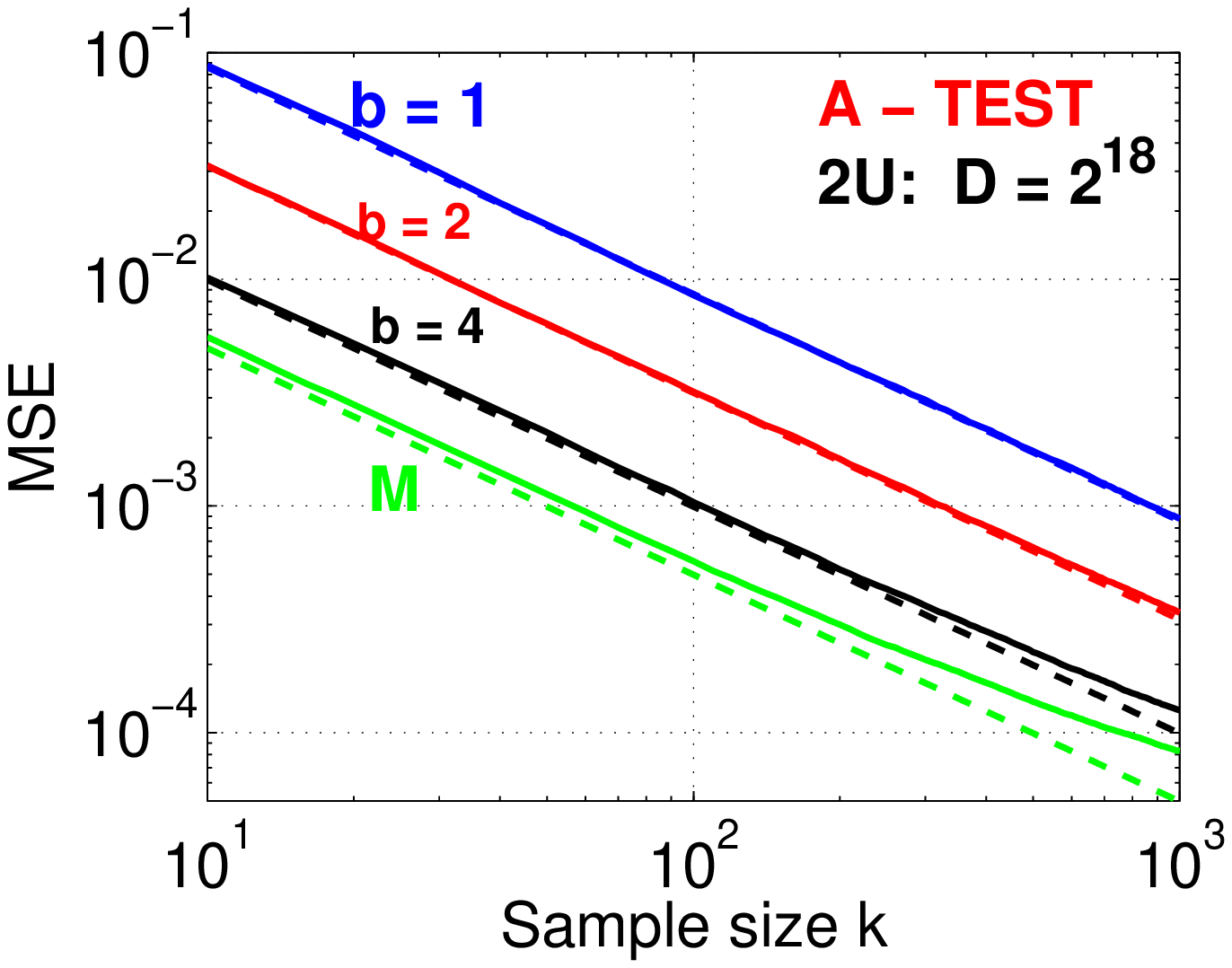}}
\end{center}
\vspace{-0.3in}

\caption{Mean square errors (MSEs) of the resemblance estimates using (\ref{eqn_R_b}) and 2U hashing with $D=2^{18}$, on 6 English word vector pairs which do not perform too well with $D=2^{16}$ in Figure~\ref{fig_2U_D16}. We can see that the results become much better.}\label{fig_2U_D18}\vspace{-0.05in}
\end{figure}

\begin{figure}[h!]
\begin{center}

\mbox{
\includegraphics[width=1.2in]{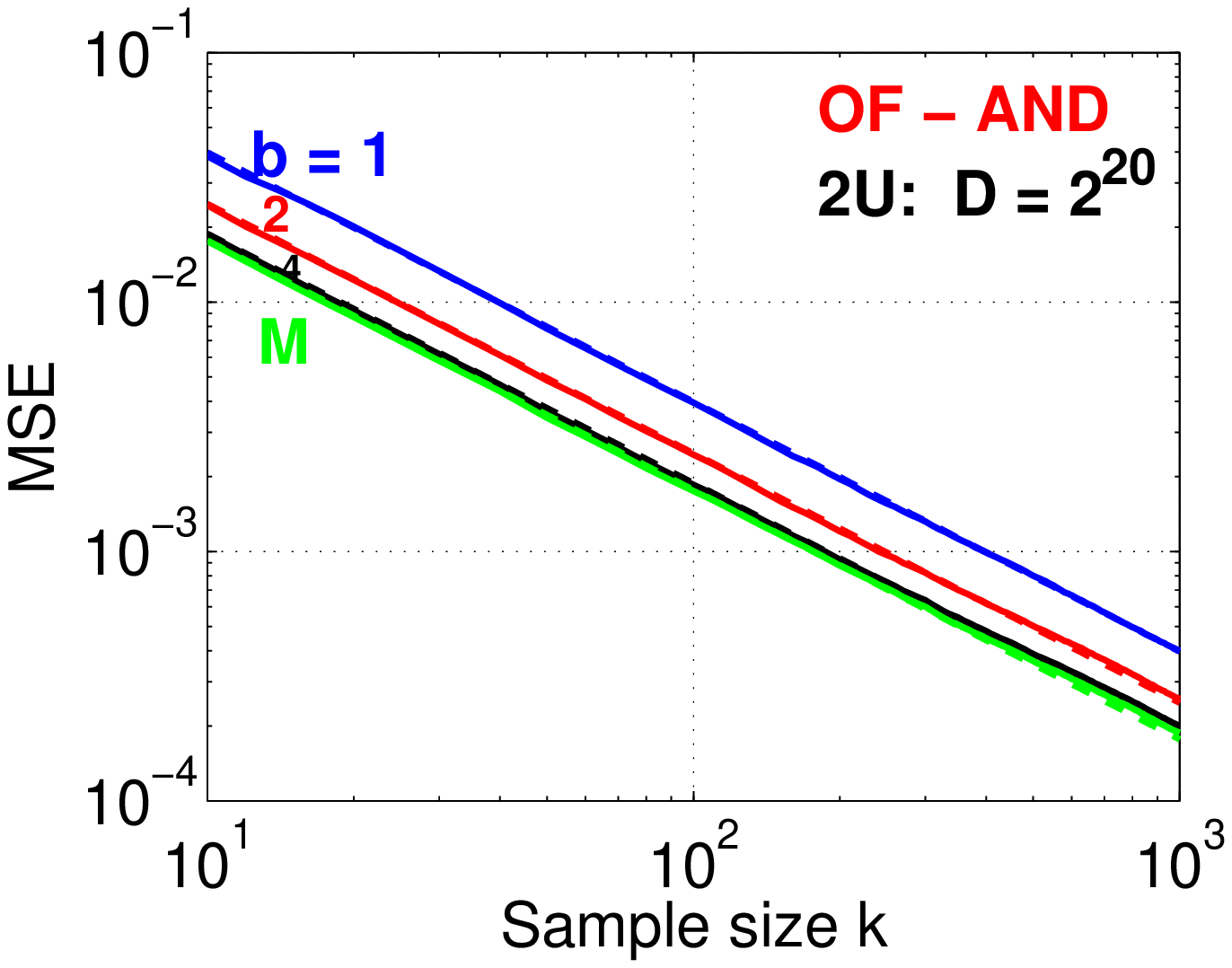}\hspace{-0.1in}
\includegraphics[width=1.2in]{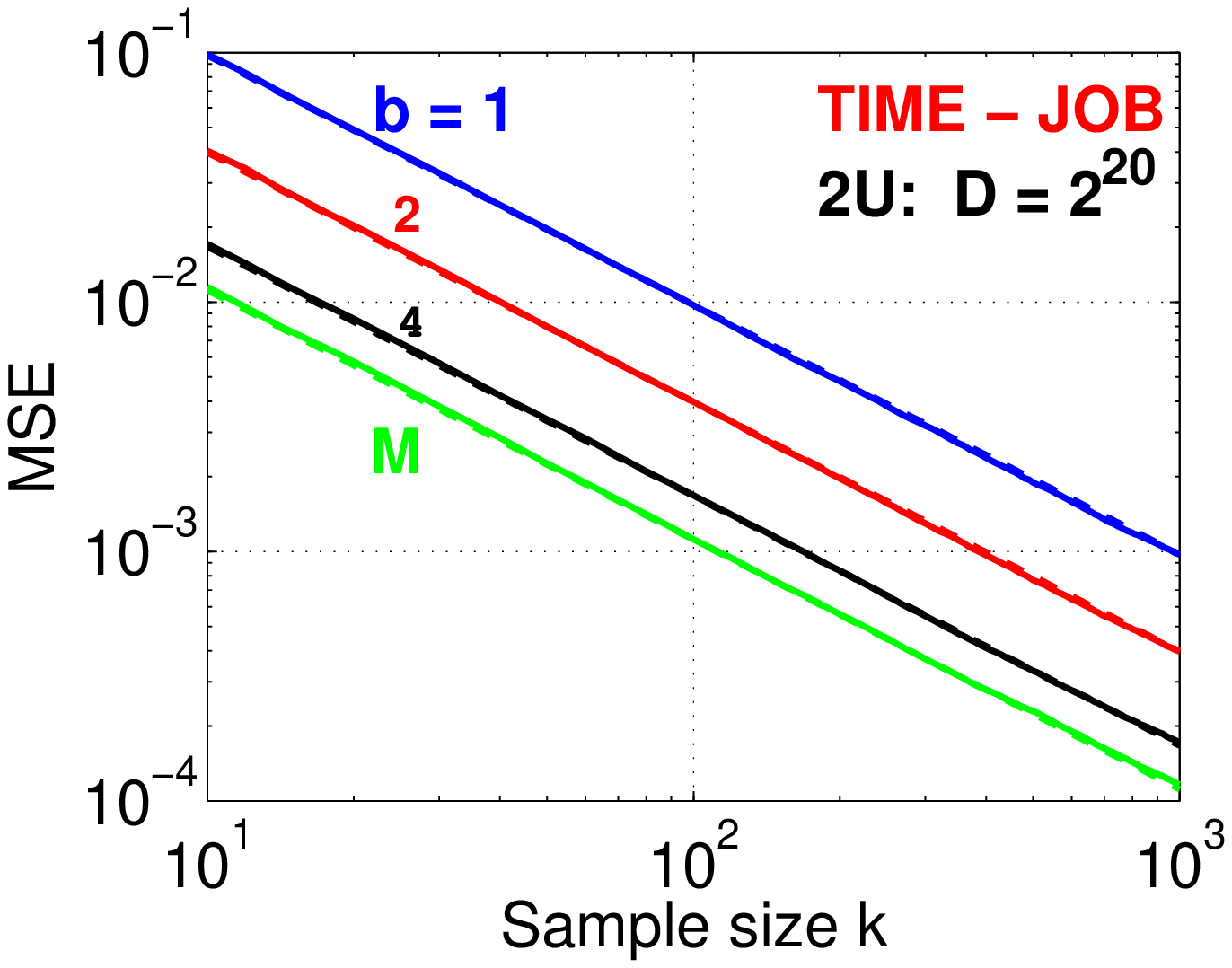}\hspace{-0.1in}
\includegraphics[width=1.2in]{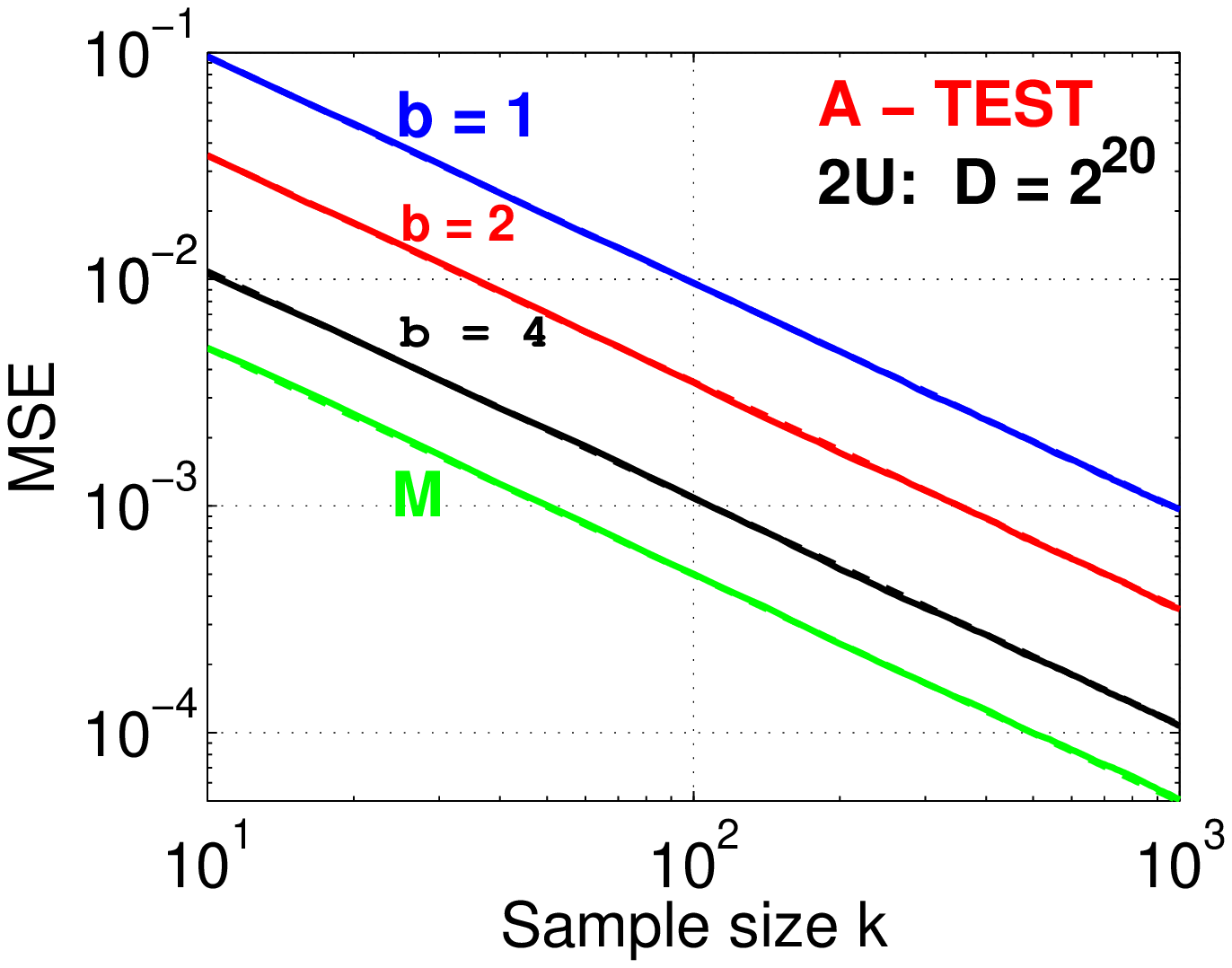}}

\end{center}
\vspace{-0.3in}

\caption{Mean square errors (MSEs) of the resemblance estimates using (\ref{eqn_R_b}) and 2U hashing with $D=2^{20}$, on 3 English word vector pairs which do not perform too well with $D=2^{18}$ in Figure~\ref{fig_2U_D18}. We can see now all the dashed curves (theoretical) match the solid curves (empirical) now. }\label{fig_2U_D20}\vspace{-0.1in}
\end{figure}

We implemented both  2U hash (\ref{eqn_h_s}) and 4U hash schemes, for $D = 2^{16}, 2^{18}, 2^{20}, 2^{22}, 2^{24}, 2^{26}, 2^{28}, 2^{30}, 2^{32}$.  Note that $D\geq 2^{16}$ is necessary for this dataset. After sufficient number of repetitions, we computed the simulated mean square error (MSE = Var + Bias$^2$) for each case, to compare with the theoretical variance (Eq. (11) in~\cite{Article:Li_Konig_CACM11}), which was derived by assuming perfect random permutations. Ideally, the empirical MSEs and the theoretical variances should overlap. Indeed, we observe this is always the case when $D\geq 2^{20}$. This is the reason why we only plot the results for $D\leq 2^{20}$ in Figures~\ref{fig_2U_D16} to~\ref{fig_2U_D20}. In fact, as shown in Figure~\ref{fig_2U_D16}, when the data are not dense (e.g., KONG-HONG, GABMIA-KIRIBATI, SAN-FRANCISCO), using 2U can achieve very similar results as using perfect random permutations, even at the smallest $D=2^{16}$.

\vspace{0.1in}

\noindent\textbf{Practical Implication}:\hspace{0.02in} In practice, we expect the data vectors to be very sparse for a large number of applications,
especially the many search-related tasks where features correspond to the presence/absence of text $n$-grams.
For these tasks, the large number of distinct words (e.g.,~\cite{DBLP:conf/wsdm/GantiKL10} reports 38M distinct 1-grams in an early Wikipedia corpus) and
the much smaller number of terms in individual documents combine to cause this property.
%especially as the standard industry practice is to use $D=2^{64}$.
Therefore, we expect that 2U/4U hash functions will perform well when used for $b$-bit minwise hashing, as verified in the main body of the paper.

%\vspace{-0.1in}

\end{document}